\documentclass{aa}  

\pdfoutput=1
\usepackage{graphicx,rotating}
\usepackage{times}
\usepackage{latexsym}
\usepackage{epsfig}
\usepackage{longtable,lscape,multirow,supertabular}
\usepackage{color}

\usepackage{natbib,afterpage}
\usepackage{color}
\usepackage[english]{babel}
\usepackage{subfigure}
\usepackage{ifpdf}
\usepackage[rightcaption]{sidecap}
\usepackage{multicol}
\usepackage{booktabs}
\usepackage{amssymb,times}
\usepackage{morefloats}
\usepackage{dblfloatfix}
\usepackage{float}

\newif\ifpdf \ifx\pdfoutput\undefined\pdffalse\else\pdftrue\fi
\ifpdf\pdfoutput=1
        \usepackage{graphicx}
        \usepackage{epstopdf}
        \DeclareGraphicsExtensions{.pdf,.png,.gif,.jpg}
        \else \usepackage[dvips]{color,graphicx} \fi

\def\Msun{\hbox{M$_{\odot}$}}               
\def\Lsun{\hbox{L$_{\odot}$}}               
\def\Rstar{\hbox{R$_{\star}$}}              
\def\Tstar{\hbox{T$_{\star}$}}              
\def\Mdot{\hbox{$\dot{M}$}}               
\def\arcsec{\hbox{$^{\prime\prime}$}}

\begin{document} 

\selectlanguage{english}
\newcommand{\red}{\textcolor[rgb]{1,0,0}}
\newcommand{\blue}{\textcolor[rgb]{0,0,1}}

\newlength{\fntxvi} \newlength{\fntxvii}
\newcommand{\chemical}[1]
{{\fontencoding{OMS}\fontfamily{cmsy}\selectfont
\fntxvi\the\fontdimen16\font
\fntxvii\the\fontdimen17\font
\fontdimen16\font=3pt\fontdimen17\font=3pt
$\mathrm{#1}$
\fontencoding{OMS}\fontfamily{cmys}\selectfont
\fontdimen16\font=\fntxvi \fontdimen17\font=\fntxvii}}

\title{ALMA spectral line and imaging survey of a low and a high mass-loss rate AGB star between 335 and 362 GHz}

   \author{
   L. Decin\inst{1}
          \and
          A. M. S. Richards\inst{2}
          \and
          T. Danilovich\inst{1}
          \and
          W. Homan\inst{1}
            \and 
            J.A. Nuth\inst{3}
          }

  \offprints{Leen.Decin@kuleuven.be}

  \institute{
  Instituut voor Sterrenkunde, Katholieke Universiteit Leuven, Celestijnenlaan 200D, 3001 Leuven, Belgium \\ 
  \email{Leen.Decin@kuleuven.be}
  \and
 JBCA, Department Physics and Astronomy, University of Manchester, Manchester M13 9PL, UK
   \and
  NASA/GSFC, Mail Code: 690, Greenbelt, MD 20771, USA
 }
 

   \date{Received date; accepted date}

 
  \abstract
   {Low and intermediate mass stars are known to power strong stellar winds when evolving through the asymptotic giant branch (AGB) phase. Initial mass, luminosity, temperature, and composition determine the pulsation characteristics of the star and the dust species formed in the pulsating photospheric layers. Radiation pressure on these grains triggers the onset of a stellar wind. However, as of today, we still cannot predict the wind mass-loss rates and wind velocities from first principles neither do we know which species are the first to condense in the upper atmospheric regions.}  
   {We aim to characterise the dominant physical, dynamical, and chemical processes in the inner wind region of two archetypical oxygen-rich (C/O$<$1) AGB stars, that is,\ the low mass-loss rate AGB star R~Dor (\Mdot$\sim$1$\times 10^{-7}$\,\Msun/yr) and the high mass-loss rate AGB star IK~Tau (\Mdot$\sim$5$\times 10^{-6}$\,\Msun/yr). The purpose of this study is to observe the key molecular species contributing to the formation of dust grains and to cross-link the observed line brightnesses of several species to the global and local properties of the star and its wind.}
   {A spectral line and imaging survey of IK~Tau and R~Dor was made with ALMA between 335 and 362\,GHz (band~7) at a spatial resolution of $\sim$150\,mas, which corresponds to the locus of the main dust formation region of both targets.}
   {Some two hundred spectral features from 15 molecules (and their isotopologues) were observed, including rotational lines in both the ground and vibrationally excited states (up to v=5 for SiO).
    Detected species include the gaseous precursors of dust grains such as SiO, AlO, AlOH, TiO, and TiO$_2$. We present a spectral atlas for both stars and the parameters of all detected spectral features. A clear dichotomy for the sulphur chemistry is seen: while CS, SiS, SO, and SO$_2$ are abundantly present in IK~Tau, only SO and SO$_2$ are detected in R~Dor. Also other species such as NaCl, NS, AlO, and AlOH display a completely different behaviour. From some selected species, the minor isotopologues can be used to assess the isotopic ratios. The channel maps of many species prove that both large and small-scale inhomogeneities persist in the inner wind of both stars in the form of blobs, arcs, and/or a disk. The high sensitivity of ALMA allows us to spot the impact of these correlated density structures in the spectral line profiles. The spectral lines often display a half width at zero intensity much larger than expected from the terminal velocity, $v_\infty$, previously derived for both objects (36\,km/s versus $v_\infty\sim$17.7\,km/s for IK~Tau and 23\,km/s versus $v_\infty\sim$5.5\,km/s for R~Dor). Both a more complex 3D morphology and a more forceful wind acceleration of the (underlying) isotropic wind can explain this trend. The formation of fractal grains in the region beyond $\sim$400\,mas can potentially account for the latter scenario.
     From the continuum map, we deduce a dust mass of $\sim3.7\times 10^{-7}$\,\Msun\ and $\sim2 \times 10^{-8}$\,\Msun\ for IK Tau and R~Dor, respectively. }
   {The observations presented here provide important constraints on the properties of these two oxygen-dominated AGB stellar winds. In particular, the ALMA data prove that both the dynamical and chemical properties are vastly different for this high mass-loss rate (IK~Tau) and low mass-loss rate (R~Dor) star. }

   
\titlerunning{ALMA spectral and imaging atlas of IK~Tau and R~Dor between 335 and 362\,GHz}
  \maketitle

\section{Introduction} \label{Sec:Introduction}

Asymptotic giant branch (AGB) stars form a late stage of stellar evolution shortly before losing their outer layers and moving into the state where a planetary nebula is formed. These stars show severe mass-loss rates between $10^{-8}$\,\Msun/yr at the lower AGB, increasing to some $10^{-4}$\,\Msun/yr as the star evolves to the top of the AGB while becoming slightly cooler and considerably more luminous. Pulsations allow for higher density regions to form at temperatures low enough for dust to condense from the gas phase. 
\citet{Hoyle1962MNRAS.124..417H} were the first to propose the wind acceleration in AGB stars to be caused by radiation pressure on dust grains, which generates an outwards directed force which counteracts the gravitational attraction of the star. They later argued that grain-gas collisions transfer sufficient momentum to the gas to accelerate the material to velocities above the local escape velocity, thus driving a wind \citep{Wickramasinghe1966ApJ...146..590W}. 

Solving the equations of conservation of mass, momentum, and energy shows that a wind with given lower-wind boundaries (determined by the temperature ($T_0$) and density ($\rho_0$) values of the outer layers of the star with gravity $g$) can transition from sub-sonic to supersonic values for only one specific value of the mass-loss rate (\Mdot). Adding an additional force in the sub-sonic part of the wind (in this case due to radiation pressure on the grains) results in a smaller wind velocity gradient but higher velocities than in the absence of the force. The higher density scale height also results in an increase of the mass-loss rate. 
The rate of acceleration of the wind decreases with radius, and essentially ceases from a certain radius onwards. Attenuation of the stellar radiation field by intervening dust, a decrease in relative total cross-section of the dust as it moves outwards, and a decoupling of the dust and gas in the tenuous outer wind causes the flow to reach a terminal wind velocity, $v_\infty$.

Although the idea of radiation pressure on the dust grains as a wind-driving mechanism has existed since the sixties, we still are not able to  predict the wind acceleration, terminal wind velocity, and mass-loss rate from first principles. Detailed numerical models solving the (above-mentioned) set of equations show the dependence of the outcome of these physical quantities on the stellar, and hence also pulsation, characteristics. One of the major unknowns in current theoretical models concerns the dust nucleation process. That is,\  we still do not know which molecules will form larger gas-phase clusters that transition into the first little solid-state species in oxygen-rich (C/O$<$1, O-rich) winds. Thermodynamic condensation sequences favour alumina (Al$_2$O$_3$) or Fe-free silicates (such as Mg$_2$SiO$_4$) of which the former species might be formed at slightly higher temperatures. Grains of this type should, however, already be large enough ($\sim$200\,nm--1\,$\mu$m) close to the star for photon scattering to compensate for their low near-infrared absorption cross-section, hence allowing the onset of a stellar wind.

As usual in astronomy, observations provide critical constraints for these theoretical models. Starting in the late 1960s, infrared and sub-millimeter  telescopes have determined the observational characteristics of AGB stars. Broad spectral features in the infrared were attributed to crystalline or amorphous grains \citep{Gillett1968ApJ...154..677G, Woolf1969ApJ...155L.181W}. However, no dynamical information can be retrieved from this (broad) dust features. The only way to unlock both the key chemical and dynamical processes in the sub-sonic to supersonic transition region is by studying at high spatial resolution the molecules contributing to the dust formation since, in contrast to dust, molecules can give us information on the wind dynamics and on the gaseous left-overs after dust formation.

Targeted surveys either observing one molecule in a larger sample of stars or a predefined set of atomic or molecular transitions in one (or a few) stars have successfully increased our knowledge of the velocity profile and mass-loss rate of AGB winds \citep[e.g.][]{Loup1993A&AS...99..291L, Ziurys2002ApJ...564L..45Z, DeBeck2010A&A...523A..18D, Agundez2011A&A...533L...6A, Schoier2013A&A...550A..78S, Danilovich2016A&A...588A.119D}. Recent instrumentation now allow for unbiased spectral surveys of stellar winds letting one to set up a molecular inventory of the circumstellar envelope (CSE). Some 80 molecular species have been identified in evolved stars' stellar winds. Only some 20 of these were detected in O-rich AGB winds, while some 70 were found around carbon-rich AGB stars. One reason for this difference is the observational bias towards the nearby carbon-rich AGB star CW~Leo which has a high mass-loss rate wind \citep[\Mdot\,=\,2.1$\times10^{-5}$\,\Msun/yr, ][]{Decin2010Natur.467...64D}. During the last decade, recent advances in instrumentation have offered the possibility for O-rich winds to be surveyed, which are fainter either due to a larger distance or a lower mass-loss rate. In contrast to carbon-rich winds, the study of these O-rich environments can advance our knowledge of other galactic environments, such as the interstellar medium or young stellar objects, that have a `cosmic' abundance resembling that of O-rich winds. As such, O-rich stellar winds are excellent laboratories to unravel the intriguing coupling between chemical and dynamical processes with a broad applicability for other research domains.

Unbiased sub-millimeter spectral surveys of O-rich stellar winds are rare. \citet{Kaminski2013ApJS..209...38K} presented a SMA spectral survey of the high mass-loss rate red supergiant VY~CMa (\Mdot $\sim 2 \times 10^{-4}$\,\Msun/yr) between 279 and 355\,GHz with a synthesised beam of $\sim$0.9\arcsec. The detection of 223 spectral lines arising from 19 molecules and their isotopologues allowed for a detailed investigation of the spatio-kinematical structure of the complex nebula surrounding VY~CMa. \citet{Quintana-Lacaci2016A&A...592A..51Q} have presented a 1 and 3\,mm line survey towards the yellow hypergiant IRC\,+10420 obtained with the IRAM telescope. The IRAM telescope beam size is 21-29\arcsec\ at 3\,mm and 9--13\arcsec\ at 1\,mm. 106 molecular emission lines from 22 molecular species were identified, of which approximately half of them are N-bearing species. A spectral line survey with the IRAM-30\,m telescope ($\sim$80--345\,GHz) and Herschel/HIFI ($\sim$479--1244\,GHz) of the O-rich pre-planetary nebula OH\,231.8+4.2 revealed the presence of hundreds of lines from different species, including HCO$^+$, H$^{13}$CO$^+$, SO$^+$, N$_2$H$^+$, and H$_3$O$^+$ \citep{Sanchez2014apn6.confE..88S, Sanchez2015A&A...577A..52S}.
The two targets of this publication, IK~Tau and R~Dor, have also already been subject of spectral survey campaigns thanks to either the high mass-loss rate (in the case of IK~Tau) or the nearby distance (R~Dor, D=59\,pc). Recently, \citet{Velilla2017A&A...597A..25V} published an IRAM-30\, survey towards IK~Tau between 79 and 356\,GHz. The half power width of the main beam varies between 7.5\arcsec\ and 29\arcsec. A total of $\sim$250 lines of 34 different molecular species (including different isotopologues) were detected, including rotational lines in the ground vibrational state of HCO$^+$, NS, NO, and H$_2$CO.  \citet{DeBeck2015ASPC..497...73D} used the SMA to obtain a spectral line imaging survey at 279--355\,GHz of IK~Tau at $\sim$0.9\arcsec\ spatial resolution. The survey shows over 140 emission lines, belonging to more than 30 species. In addition, there exists an unpublished APEX spectral scan of R~Dor between 159--368.5\,GHz \citep{DeBeck2017A&A...598A..53D}. 

The high spatial and spectral resolution and sensitivity offered by ALMA now opens a new window for sub-millimeter spectral surveys allowing us to study the rich chemistry in stellar winds with unprecedented detail. The presence of various transitions of a specific molecule covering a large range in excitation temperatures allows the derivation of the wind velocity profile \citep[see, e.g.][for the case of CW~Leo and IK~Tau, respectively]{Fonfria2008ApJ...673..445F, Decin2010A&A...516A..69D}. However, spatially unresolved velocity features in (sometimes asymmetric) line profiles complicate this task.
Another way to determine the wind acceleration and terminal wind velocity is by using high spatial and spectral resolution data in such a way that one can resolve spatially and spectrally the molecular emission of various transitions. Plotting the measured wind velocities (as determined from the half-line width) versus half of the largest detectable scale or half of the spatial full width at half maximum (FWHM, representing the dominant line formation region) yields a direct way of visualising and interpreting the wind profile \citep{Decin2015A&A...574A...5D}. In addition, complex morphologies can be taken into account, if present. If a high enough spatial resolution is achieved, one can examine the chemical processes across the sub-sonic to supersonic transition region, that is,\ the region where dust condenses and grows, via the study of the corresponding depletion of gas-phase molecules. This ultimately allows one to couple the derived abundance variations in the CSE to chemical models. Additionally, observing several isotopologues allows one to determine isotopic ratios in the CSE and from that constrain the AGB nucleosynthesis, initial stellar mass, and age.

In this paper, we present the first ALMA spectral survey of two O-rich AGB stars, IK~Tau and R~Dor. Data have been obtained between 335 and 362\,GHz at a spatial resolution of $\sim$120$\times$150\,mas ($\sim$5.3$\times$10$^{13}$\,cm for R~Dor and $\sim$2.4$\times$10$^{14}$ for IK~Tau; see Table~\ref{Table:parameters}). This spatial resolution corresponds to the main locus of dust formation in both stars. We have selected  IK~Tau and R~Dor as being the best representatives for the class of high mass-loss rate and low mass-loss rate O-rich AGB stars due to their proximity (260\,pc and 59\,pc, respectively). Many dust and molecular species have been detected in their wind \citep[for an overview, see][]{Decin2017arXiv170405237D}. Solid-state aluminium-bearing species have been detected in the spectral energy distribution (SED) of the semi-regular variable R~Dor, while the SED of Mira-type star IK~Tau seems dominated by magnesium-iron silicates. This supports the hypothesis that the dust condensation sequence in R~Dor has experienced a freeze-out after the formation of the aluminium dust species, while in IK~Tau the silicates might have coated the previously formed aluminium grains \citep[see][for more details]{Decin2017arXiv170405237D}. Molecules detected in the wind of both stars include CO, HCN, SiO, SiS, SO, SO$_2$, NaCl, PO, PN, etc.  \citep{Milam2007ApJ...668L.131M, Kim2010A&A...516A..68K, Decin2010A&A...516A..69D, Decin2010A&A...521L...4D, DeBeck2013A&A...558A.132D, DeBeck2015ASPC..497...73D, Velilla2017A&A...597A..25V}. Some molecules in IK~Tau, including SiO and SiS, show depletion in the CSE suggesting depletion due to condensation onto dust grains \citep{Decin2010A&A...516A..69D}. An overview of the most important stellar and wind properties of both stars can be found in Table~\ref{Table:parameters}.

\begin{table}[htp]
\begin{center}
\caption{Overview of stellar and wind properties of R~Dor and IK~Tau \citep{Decin2017arXiv170405237D}. Listed are the distance D, the luminosity L, the stellar temperature \Tstar, the stellar radius \Rstar, the mass-loss rate \Mdot, the locus of the dust condensation in radius $r_{\rm{dust}}$ and in diameter $\theta_{\rm{dust}}$, and the terminal velocity $v_\infty$.}
\label{Table:parameters}
\begin{tabular}{lrr}
\hline
 & R Dor & IK Tau \\
 \hline
 D [pc] & 59 & 260 \\
 L [\Lsun] & 6500 & 7700 \\
 \Tstar\ [K] & 2400 & 2100 \\
 \Rstar\ [cm] & 2.5$\times10^{13}$ & 3.8$\times10^{13}$ \\
 \Rstar\ [mas] & 30 & 10 \\
 \Mdot\ [\Msun/yr] & 1.6$\times10^{-7}$ & 5$\times10^{-6}$ \\
 $r_{\rm{dust}}$ [cm] & 5.3$\times10^{13}$ & 2.38$\times10^{14}$ \\
 $\theta_{\rm{dust}}$ [mas] & 120 & 122 \\
 $v_\infty$ [km/s] & 5.5 & 17.7 \\
 \hline
\end{tabular}
\end{center}
\end{table}

The aim of this paper is to provide the community with a spectral atlas of all species detected in the ALMA data, including the relevant flux density, velocity parameters, and angular extension. Detailed studies on the morphology, molecular excitation regions and abundance structures as derived from in-depth non-local thermodynamic equilibrium radiative transfer simulations will be presented in dedicated papers; one of which on the aluminium-bearing molecules (AlO, AlOH, and AlCl) has already been published \citep{Decin2017arXiv170405237D}. In this paper, we focus on the presentation of the ALMA data and the data reduction (Sect.~\ref{Sec:ALMA_obs_red}). We discuss the dust properties as retrieved from the ALMA continuum maps in Sect.~\ref{Sec:Imag_results}. The spectral line results, including the line identification, the measurements of the flux density, line width parameters and angular sizes, and a first overview of azimuthally averaged flux densities and zeroth moment maps are given in Sect.~\ref{Sec:spectral_lines}. In Sect.~\ref{Sec:Discussion}, we weigh up the different molecular content in both stars (Sect.~\ref{Sect:disc_species}), give a first view on the molecular morphology (Sect.~\ref{Sec:spatial}) and concentrate on the kinematic structure as deduced from the ALMA data (Sect.~\ref{Sec:kinematic}). As shown in Sect.~\ref{Sec:kinematic}, the wings of the ALMA spectral lines bear witness to projected velocities larger than the canonical terminal wind velocity and, hence, showcase a surprisingly rich morpho-kinematical behaviour in the inner wind region of both stars. In Sect~\ref{Sect:disc_isotopes} we use the integrated line intensity ratios of isotopologue species to deduce the isotopic ratios of $^{29}$Si/ $^{30}$Si, $^{34}$S/$^{33}$S, and $^{35}$Cl/$^{37}$Cl.  We summarise our conclusions in Sect.~\ref{Sec:conclusions}.


\section{ALMA observations and data reduction} \label{Sec:ALMA_obs_red}

\subsection{ALMA observations} \label{Sec:ALMA_obs}

IK Tau and R Dor were observed with ALMA in Band~7 during
August-September 2015 (proposal 2013.1.00166.S, PI L.\ Decin).  A full
spectral scan between 335--362\,GHz was made using four separate
observations per star. Each observation used four 1.875\,GHz spectral
windows.  An average of 39 good antennas were
present. The observations took place over 16 days for IK~Tau and 5
days for R~Dor using standard ALMA procedures. The IK~Tau reference
source is $\sim$5$^{\circ}$ from the target, the two reference sources
used for R~Dor are $3^{\circ}-4^{\circ}$ away. All have sub-mas
position accuracy except for J0428-6438, where this is (5.1, 3.5)\,mas.
Conditions varied between good and very good, so there are small
changes in sensitivity between spectral windows  due to this as well as to
intrinsically higher atmospheric opacity at higher frequencies.  More
details are given in Table~\ref{obs}.  The range of baseline lengths
was 0.04--1.6\,km, allowing imaging of structure on angular scales up
to 2\arcsec\ at angular resolution $\sim$150\,mas. Hanning
smoothing was applied in the correlator and the data were finally averaged
every four (two) channels for IK~Tau (R~Dor) so the imaged channels are
effectively independent.

\begin{table*}
\caption{Summary of observations. Listed are the observing data, the precipitable water vapour (PWV), the time on source per tuning (ToS), the channel spacing used in standard line imaging after averaging (Chan), the flux scale reference source, the band pass reference source, the phase reference source and the check source.}
\label{obs}
\begin{tabular}{lcccccccc}
\hline
\hline
Source & Date range & PWV range & ToS & Chan & Flux    & Bandpass & Phase-ref& Check     \\
       &(yyyy-mm-dd)&  (mm)     &(min)&(MHz) & Scale   &          &          &           \\
\hline       
IK Tau & 2015-08-13 & 0.2--1.7  & 10  & 1.95 &J0423-013&J0423-0120&J0407+0742&J0409+1217 \\
       & 2015-08-28 &           &     &      &         &          &          &           \\
R Dor  & 2015-08-27 & 0.3--0.7  & 25  & 0.98 &J0519-454&J0457-2324&J0428-6438&J0506-6109 \\
       & 2015-09-01 &           &     &      &         &J0522-3627&J0506-6109&J0428-6438 \\
       &            &           &     &      &         &J0538-4405&          &           \\
\hline
\end{tabular}
\end{table*}

\subsection{Data reduction} \label{Sec:ALMA_red}

All data reduction was done using CASA\citep{McMullin2007ASPC..376..127M}\footnote{http://casa.nra.edu}, following standard scripts for
the application of calibration from instrumental measurements such as
system temperature and water vapour radiometry, and using standard
sources for bandpass calibration etc. The flux density of the primary
calibration sources is accurate to 5\%; the scatter in amplitude
solutions for calibration sources is a few percent or less, so the net
flux scale accuracy is $\sim$7\% or better.  The astrometric accuracy
is dominated by errors in the phase transfer and, possibly, antenna
position errors. These were estimated by examining the phase-reference
solutions and the positions of the check sources. The IK~Tau
phase-reference solution rms were around 30${^\circ}$ which
corresponds to a position uncertainty (30/360)$\times$ the beam size,
or 17\,mas for a 200\,mas natural beam, slightly more higher than the observed
check-source offsets.  The R~Dor averages were a little smaller (as
expected for the better conditions) but, given the greater position
uncertainty of J0428-6438, it seems reasonable to adopt 17\,mas
astrometric accuracy for both targets.  The relative accuracy
(e.g.\ aligning different lines in the same source) is limited only by
the signal-to-noise (S/N), and would be $<$1\,mas for S/N=200, or 33\,mas (about 1 pixel) for
a faint 3$\sigma$ detection of a compact source.  For each source, the
target data were then shifted to constant velocity with respect to the
local standard of rest ($v_{\mathrm{LSR}}$). Line-free continuum
channels were identified and used for self-calibration, applied to all
data. The peak positions changed negligibly during self-calibration
and the final images rms are as expected for the conditions. The
continuum images were made using a first-order spectral slope,
although the fractional bandwidth is too small for the spectral index
measurements to be very reliable.  The continuum was subtracted and
the entire cubes were imaged.  The final continuum image parameters
are summarised in Table~\ref{reduction}.  Unless otherwise stated, all
images discussed used partial uniform weighting (CASA `robust 0.5').

\begin{table}
\caption{Summary of final continuum image parameters. Width is the continuum bandwidth (spread over total spectral span), beam and rms are the continuum restoring beam and image noise, and `imsize' denotes the area imaged.}
\label{reduction}
\setlength{\tabcolsep}{1mm}
\begin{tabular}{lccccc}
\hline
\hline
Star   & Width &Beam                        & rms   & Imsize &Pixel     \\
       & (MHz) &(mas$\times$mas, P.A.)      & (mJy) & (arcsec) & (arcsec)  \\
       \hline
IK Tau &15.2e3 &180$\times$160, 27$^{\circ}$ &0.047 & 10   &0.04 \\
R Dor  &11.6e3 &150$\times$140, $-$5$^{\circ}$ &0.047 &   7.6 &0.03 \\
\hline
\end{tabular}
\end{table}

The exact spectral restoring beam parameters depend on elevation and
frequency, and are in the range 120--180\,mas for IK~Tau and 130--180\,mas for R~Dor.  The channel $\sigma_{\mathrm{rms}}$ noise varies between spectral
windows due to elevation, weather and the intrinsically higher
atmospheric opacity at higher frequencies.  The range in IK~Tau was
3--9\,mJy and in R~Dor the range was 2.7--5.7\,mJy, except in the
brightest channels where the dynamic range limit of $\sim$1000 gives
$\sigma_{\mathrm{rms}}\le$50\,mJy/beam. The $\sigma_{\mathrm{rms}}$ for the continuum image is 0.047\,mJy.
The velocity resolution, also
depending on frequency, was 1.6--1.7\,km s$^{-1}$ for IK~Tau and
0.8--0.9\,km s$^{-1}$ for R~Dor.

The full width half maximum (FWHM) of the ALMA primary beam over our
frequency range is 14\farcs4--15\farcs6.  Detectable continuum emission is
well within a 1\arcsec\ radius in both sources. Most lines have a maximum
angular extent $<5$\arcsec\ but the largest are 10\farcs3 in IK~Tau and 7\farcs4 in R~Dor\footnote{We note that the data are not sensitive to emission which is smooth on scales $\ga$2\arcsec. \citet{Castro-Carrizo2010A&A...523A..59C} detected $^{12}$CO $J$\,=\,1--0 and $J$\,=\,2--1 in regions up to $\sim$40\arcsec\ and 25\arcsec\ in diameter, respectively. The $^{12}$CO $J$\,=\,3--2 is expected to have a smaller extent, but it is also possible that we have resolved-out extended emission.}.  
We did not apply the primary beam correction initially, since it
is more straightforward to measure the largest angular size (LAS) and
azimuthally-averaged profiles assuming flat noise per plane.  The S/N,
and thus the LAS, is unaffected.  For the profiles, the flux densities
of the most extended emission should be increased by 60\% at 5\arcsec\ radius
or 30\% at 3\farcs5 radius.



\section{Continuum image} \label{Sec:Imag_results}

\begin{figure*}[htbp!]
\includegraphics[width=0.48\textwidth]{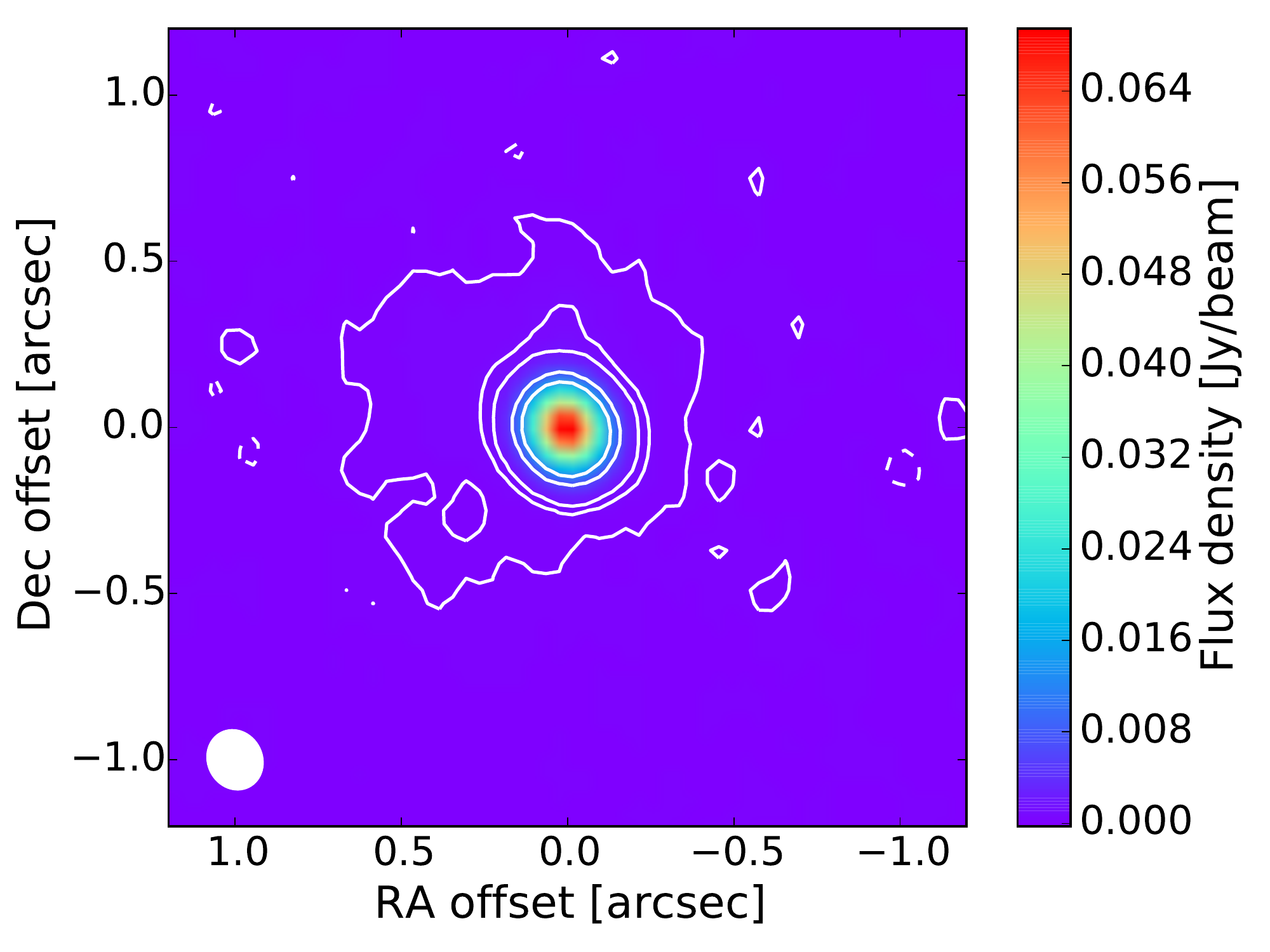}
\includegraphics[width=0.48\textwidth]{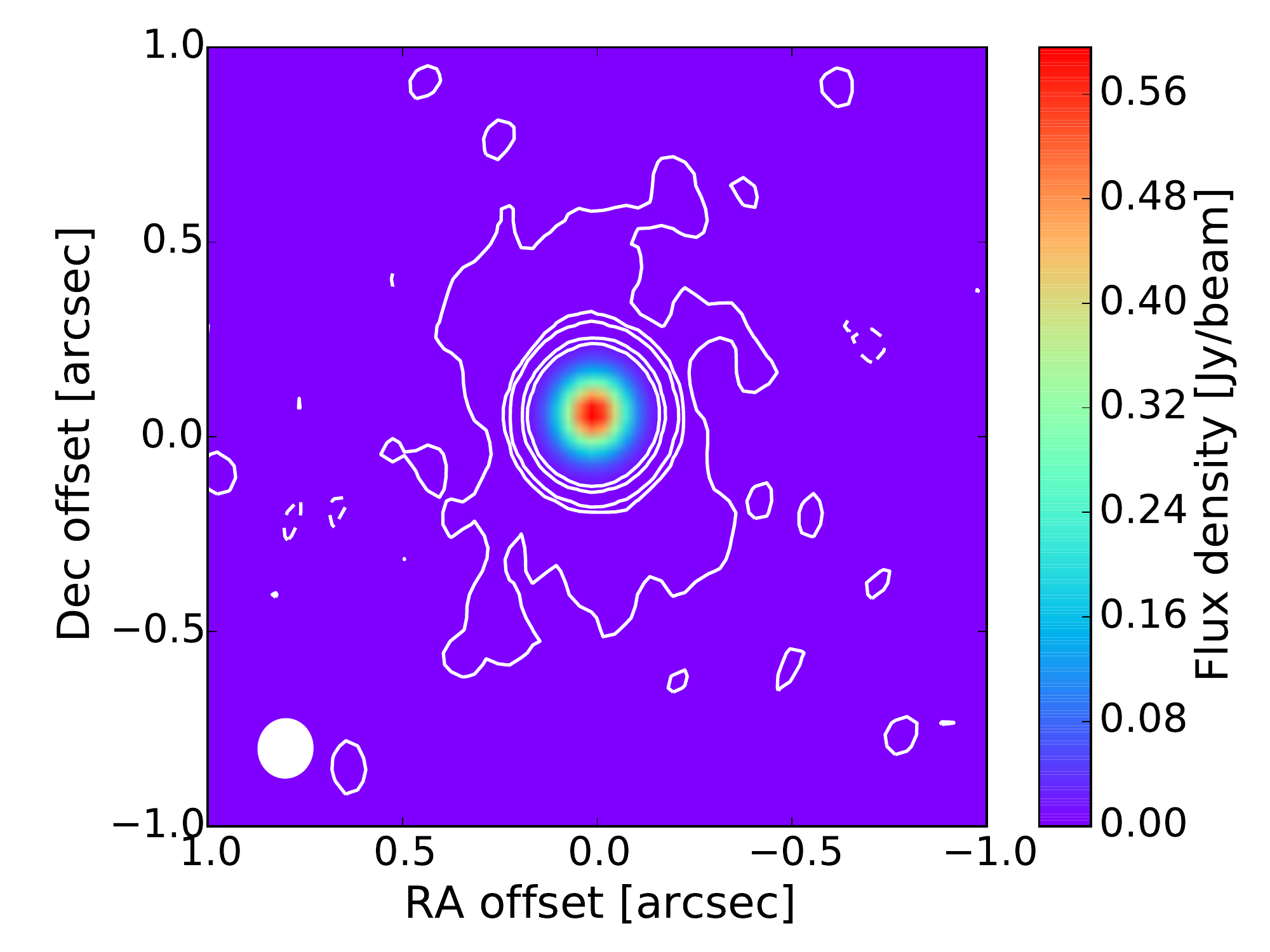}
\caption{Continuum image of IK~Tau for a beam of $0\farcs18\times0\farcs16$ (left) and of R~Dor for a beam of $0\farcs15\times0\farcs14$ (right)  (beam shown as a filled ellipse in the bottom left corner). The contour levels are at [$-$1, 1, 5, 10, 50, 100] $\times$ 0.141\,mJy beam$^{-1}$ (3\,${\sigma_{\mathrm{rms}}}$).}
\label{Fig:continuum}
\end{figure*}

The continuum emission for IK~Tau and R~Dor is shown in Fig.~\ref{Fig:continuum}. The continuum peak is 0.069\,Jy/beam for IK~Tau and 0.596\,Jy/beam for R~Dor, with a total flux density within the 3$\sigma$ contour being 0.088\,Jy and 0.648\,Jy, respectively. The radius for the 3$\sigma$ contour is $\sim$520\,mas (or 52\,\Rstar) for IK~Tau and 450\,mas (or 15\,\Rstar) for R~Dor.
The stellar photospheric emission contributes part of the total flux density. Using the stellar parameters listed in Table~\ref{Table:parameters}, the stellar flux contribution at 345\,GHz is $\sim$0.056\,Jy for IK~Tau and $\sim$0.581\,Jy for R~Dor. With the contribution from free-free emission at 345\,GHz being negligible \citep{Groenewegen1997A&A...317..503G}, we can estimate the 345\,GHz emission from dust by subtracting a (stellar) point source, convolved with the natural synthesised beam, from the continuum leaving a residual of 0.027\,Jy for IK~Tau and 0.067\,Jy for R~Dor. The onset of the dust nucleation is around 6 and 2\,\Rstar, respectively (see Table~\ref{Table:parameters}) or around $\sim$120\,mas which is comparable to the spatial resolution of the ALMA data.

Assuming that the dust is optically thin and for an opacity of $\sim$7\,cm$^2$/g \citep{Demyk2017arXiv170609801D} at a dust temperature of varying between 200--650\,K for IK~Tau and 400--1530\,K for R~Dor \citep{Decin2017arXiv170405237D}, we obtain a dust mass within the 3$\sigma$ contour of $3.7 \times 10^{-7}$\,\Msun\ and $2 \times 10^{-8}$\,\Msun\ for IK~Tau and R~Dor, respectively. Solving the momentum equation \citep{Decin2006A&A...456..549D}, we calculate the gas, drift, and dust velocity profile; the example for IK~Tau being shown in \citet{Decin2010A&A...516A..69D}. We force the drift velocity to be lower than 20\,km/s, at which value sputtering of the dust grains starts to become important \citep{Kwok1975ApJ...198..583K}. This gives us the possibility to calculate the total amount of time for the dust species to reach the observed ALMA 3$\sigma$ contour, being $\sim$36\,yr for IK~Tau and $\sim$4.5\,yr for R~Dor. This translates into a dust mass loss rate  of $\sim$1$\times10^{-8}$\,\Msun/yr for IK~Tau and of $\sim$5$\times10^{-9}$\,\Msun/yr for R~Dor. 
Since the gas mass-loss rates differ largely, the resulting dust-to-gas mass ratio are fairly different, being $\sim$1/800 for IK~Tau and $\sim$1/250 for R~Dor. 
The estimated dust mass-loss rates and dust-to-gas mass ratios are subject to large uncertainties (being around a factor of 5), mainly due to the uncertain stellar contribution, the unknown composition of the dust, the large range in dust temperatures involved and the fact that cooler dust might not be detected in the current ALMA data. 

For both targets, the dust mass-loss rate has been determined from the SED and/or from SPHERE polarimetry data. The derived values hinge on a number of assumptions, the most important ones being the dust velocity, the grain size, the dust optical constants, and/or the polarisation efficiency. \citet{KhouriPhD} and \citet{VandeSande2017} have derived both the dust and gas mass-loss rate for R~Dor, yielding a dust-to-gas ratio of $\sim$1/560 and $\sim$1/780 respectively. Analysing the SPHERE data, \citet{Khouri2016A&A...591A..70K} derived a very low value for the dust-to-gas ratio in R~Dor being $\le$1/1200 at 1.5\,\Rstar\ and $\le$1/5000 at 5\,\Rstar. Due to a lack of infrared spectroscopic data of IK~Tau, its dust composition and dust mass-loss rate are less well constrained compared to R~Dor. \citet{LeSidaner1996A&A...314..896L} derived a value for the dust mass-loss rate of 2.65$\times$10$^{-8}$\,\Msun/yr for a dust velocity assumed to be twice the gas-velocity. Combined with the results on the gas mass-loss rate and gas velocity of \citet{Decin2010A&A...516A..69D}, the dust-to-gas ratio is $\sim$1/640. 

The derived dust-to-gas mass ratio for IK~Tau around 1/800 seems low. However, the ALMA data suggest that the wind speed reaches higher velocities than predicted by the standard momentum equation (see Sect.~\ref{Sec:kinematic}), which would reduce the time taken to reach the 3$\sigma$ dust radius and increase the dust-to-gas ratio. Moreover, following the suggestion by \citet{Khouri2016A&A...591A..70K}, it is well possible that the dust-to-gas ratio varies through the inner wind region. In other words, the ISO-SWS data and other photometric data of R~Dor were analysed by \citet{VandeSande2017}. They used a structure of the dust envelope consisting of a gravitationally bound dust shell (GBDS\footnote{The region in the wind where the ratio of the radiation pressure on the grains to the gravitational attraction (called the $\Gamma$-factor) is still smaller than one so that the dust particles can reside close to the star without being pushed outwards}) located close to the star ($\sim$1.6\,\Rstar) and dust flowing out in the stellar wind as of 60\,\Rstar\ onwards, similar to the structure of the dust envelope of W~Hya developed by \citet{Khouri2015A&A...577A.114K}. They derived a dust mass-loss rate of $4.1 \times 10^{-10}$\,\Msun/yr for R~Dor for the wind beyond 60\,\Rstar. The ALMA data seems to suggest a higher dust density  in the region between 2--15\,\Rstar, the potential cause of which might be the presence of the GBDS and/or an equatorial density enhancement (a.k.a.\ disk; see Sect.~\ref{Sec:kinematic}) in the inner wind of R~Dor.

\section{Spectral line results} \label{Sec:spectral_lines}

The ALMA observations reveal a rich molecular spectrum of the inner wind material for both stars. The high spatial resolution of the ALMA data allows us to resolve the molecular excitation regions and to study the complex spatial distribution in function of the projected velocity for many molecular transitions. The data were explored using different extraction apertures, ranging from 75\,mas to 1000\,mas. Most of the spectral analysis results reported here are based on the circular extraction aperture with radius of 320\,mas for IK~Tau and 300\,mas for R~Dor. The apertures of 320 and 300\,mas radius optimise the sensitivity to lines dominated by central, compact  emission, with negligible artefacts due to dynamic range limits or resolved-out flux.  These radii are also similar to the most extended continuum detected. For the most extended lines, the extraction radii of 800\,mas (IK~Tau) and 1000\,mas (R~Dor) maximised the flux measured in the channels with the most extended emission; more extended flux may exist but is too poorly sampled to be measured accurately. The smallest apertures of 80\,mas (IK~Tau) and 75\,mas (R~Dor) were chosen as being comparable to the restoring beam size and in particular best reveal any absorption against the star, although they do not detect all emission in most cases.


\subsection{Line identification} \label{Sect:line_identification}
\begin{figure*}[htp]
\includegraphics[width=\textwidth]{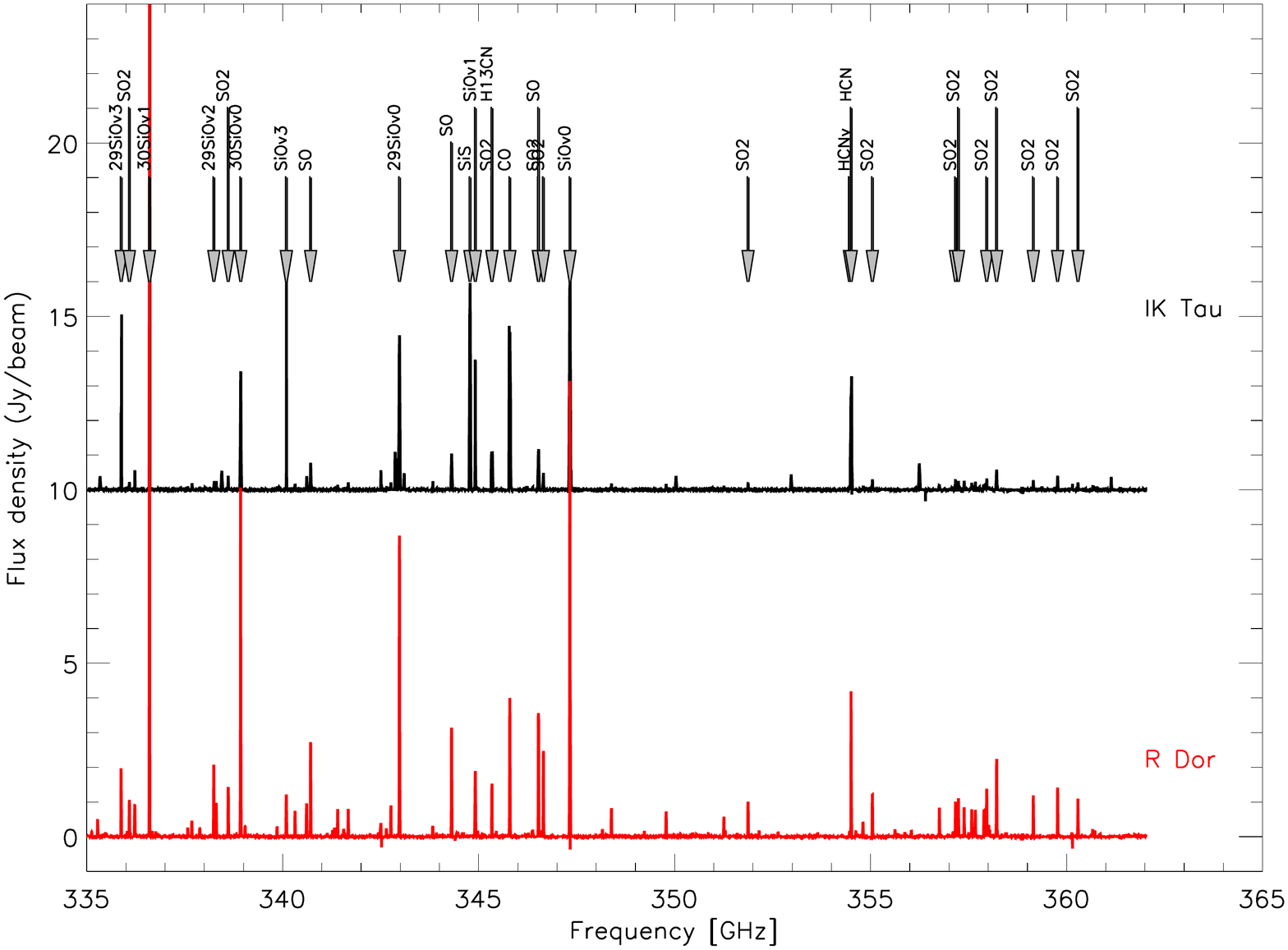}
\caption{Full spectrum of IK~Tau (in black, shifted upwards) and R~Dor (in red) obtained with ALMA band~7 between 335--362\,GHz for an extraction aperture with radius of 300\,mas. The brightest lines are identified. A zoom of the spectra (both in frequency and flux density scale) can be found in Fig.~\ref{Fig:atlas_IKTau} for IK~Tau and  in Fig.~\ref{Fig:atlas_RDor} for R~Dor.}
\label{Fig:both_spectra}
\end{figure*}

We extracted spectra in circular apertures, centred  on the continuum peak, of 80, 320 and 800\,mas in radius (IK~Tau) or 75, 300 and 1000\,mas (R~Dor).  These were used to identify the spectral lines (see below) and thus the rest frequencies. Where multiple hyperfine components overlap, subsequent measurements are based on the rest frequency of the brightest component, and in all cases the measurements use the rest frequencies shifted to the observed frequency using a local standard of rest velocity, $v_{\rm{LSR}}$, of 33\,km/s for IK~Tau and of 7\,km/s for R~Dor\footnote{Values for the local standard of rest velocity, $v_{\rm{LSR}}$, are difficult to derive with high accuracy. Methods often used include taking the mid-point between peaks of CO v=0 lines or from the OH 1612\,MHz maser emission are reported. The CO line profiles are, however, often asymmetric due to the effect of blue wing absorption and/or non-homogeneous winds. Values for IK~Tau vary between 33-35\,km/s and for R~Dor around 6-7\,km/s \citep{Boboltz2005ApJ...625..978B, Decin2010A&A...516A..69D, KhouriPhD, VandeSande2017}. The ALMA high spectral resolution data of (optically thin) high-excitation lines offer another way to estimate $v_{\rm{LSR}}$. Doing so, we derive $v_{\rm{LSR}}$\,=\,33$\pm$1\,km/s for IK~Tau and $v_{\rm{LSR}}$\,=\,7$\pm$0.5\,km/s for R~Dor.}.  For each spectrum, a noise spectrum was also extracted off-source, giving a per-channel $\sigma_{\mathrm{chn\_rms}}$. The full spectrum for the $\sim$300\,mas extraction beam is shown in Fig.~\ref{Fig:both_spectra}; a magnification  of the spectra (both in frequency and flux density scale) and the line identifications are displayed in Fig.~\ref{Fig:atlas_IKTau} for IK~Tau and  in Fig.~\ref{Fig:atlas_RDor} for R~Dor.

By default, the 320 or 300\,mas aperture spectra were used for line identification; lines were required to have continuous emission above
3$\times$ the local channel-map off-source rms. In a few cases, such as lines affected by absorption, this was relaxed after inspection.

The identification of the spectral features combined different procedures. 
A first identification was made using the SMA spectral atlas of VY~CMa as published by \citet{Kaminski2013ApJS..209...38K} and spectral line catalogues of the Jet Propulsion Laboratory \citep[JPL,][]{Pickett1998JQSRT..60..883P} and the Cologne Database for Molecular Spectroscopy \citep[CDMS,][]{Muller2001A&A...370L..49M, Muller2005JMoSt.742..215M, Endres2016JMoSp.327...95E}. 
For all molecules identified in this first step, a non-LTE spectrum was calculated for a fixed excitation temperature using the {\sc{RADEX}}-code \citep{vanderTak2007A&A...468..627V}\footnote{http://home.strw.leidenuniv.nl/~moldata/radex.html}. 
Most tools for line identification such as {\sc MADCUBAIJ}\footnote{http://cab.inta-csic.es/madcuba/MADCUBA\_IMAGEJ/\\ImageJMadcuba.html} and {\sc XCLASS}\footnote{https://www.astro.uni-koeln.de/projects/schilke/XCLASSInterface} are using spectroscopic data from standard databases such as JPL or CDMS. In contrast, we have expanded the spectroscopic data in some cases to include excited vibrational states and the infrared transitions that can pump the excited states. 
These calculations did not aim to fit the data or to retrieve molecular abundances, but served the goal of correctly identifying the molecular transitions based on a correlation with the predicted intensity. For most molecules, more than one transition was detected. 
In a last step, the spectral features at the identified (rest) frequencies of the molecular transitions were fitted using a so-called soft-parabola-function \citep{DeBeck2010A&A...523A..18D}
\begin{equation}
I(v) = I_c\left[1-\left(\frac{v-v_{\rm{LSR}}}{v_{\rm{max}}}\right)^2\right]^{\alpha/2}\,,
\end{equation}
where $\alpha$ is a measure for the shape of the profile function, $I_c$ the maximum flux density, and $v_{\rm{max}}$ the maximum width of the line (which is often larger than the canonical value for the terminal velocity, $v_\infty$; see Sect.~\ref{Sec:kinematic}).
This fit was not aimed to retrieve the best-fit parameters since many lines display complex profiles. The purpose of the fitting was to identify in a last step by manual inspection blended lines  and to determine the main component in the case of blends.

\begin{table}[!htbp]
\caption{Summary of molecules detected in the ALMA band 7 survey of IK~Tau and R~Dor. For each molecule, the total number of detected lines and the diameter of the maximum angular size (in mas) is listed, if detected. The hypercomponents for AlO, AlOH, and NS are counted separately, albeit often they can not be spectrally resolved. }
\label{Tab:summary}
\begin{tabular}{l|rr|rr}
\hline
\hline
                & \multicolumn{2}{c}{IK Tau} & \multicolumn{2}{c}{R Dor} \\
\hline
Molecule & Total & Max ang size & Total & Max ang size \\
\hline          
AlCl & 1 & 434 & 1 & 623\\
AlO & 8 & 1760 & 14 & 529\\
AlOH & 12 & 333 & 9 & 247\\
CO & 2 & 9760 & 3 & 5294\\
CS & 1 & 3360 & 0 & $-$\\
C$^{34}$S & 1 & 1188 & 0 & $-$\\
H$_{2}$O & 3 & 434 & 3 & 505\\
HCN & 3 & 2318 & 3 & 5920\\
H$^{13}$CN & 1 & 6880 & 1 & 4240\\
NS & 13 & 652 & 0 & $-$\\
NaCl & 2 & 797 & 0 & $-$\\
Na$^{37}$Cl & 2 & 333 & 0 & $-$\\
SO & 8 & 3680 & 8 & 6640\\
$^{33}$SO & 2 & 362 & 11 & 1120\\
$^{34}$SO & 2 & 724 & 2 & 4960\\
SO$_2$ & 54 & 6880 & 75 & 4720\\
SO$^{17}$O & 0 & $-$ & 4 & 447\\
SO$^{18}$O & 1 & 434 & 2 & 517\\
$^{34}$SO$_{2}$ & 5 & 3360 & 19 & 1023\\
SiO & 6 & 2028 & 6 & 7360\\
$^{29}$SiO & 4 & 3360 & 4 & 5200\\
$^{30}$SiO & 2 & 1231 & 2 & 6880\\
Si$^{18}$O & 2 & 507 & 3 & 705\\
$^{29}$Si$^{18}$O & 0 & $-$ & 3 & 364\\
$^{30}$Si$^{18}$O & 1 & 405 & 1 & 658\\
SiS & 6 & 3040 & 0 & $-$\\
$^{29}$SiS & 4 & 1913 & 0 & $-$\\
$^{30}$SiS & 2 & 884 & 0 & $-$\\
Si$^{33}$S & 2 & 811 & 0 & $-$\\
Si$^{34}$S & 3 & 1028 & 0 & $-$\\
$^{29}$Si$^{34}$S & 1 & 405 & 0 & $-$\\
$^{30}$Si$^{34}$S & 2 & 507 & 0 & $-$\\
TiO & 2 & 289 & 3 & 400\\
TiO$_{2}$ & 11 & 550 & 20 & 880\\
 \hline
total &          168 & &          197 & \\
 \hline
\end{tabular}
\end{table}

\begin{table}[!htbp]
\caption{Frequencies (in GHz) of the unidentified lines.  Frequencies written in boldface refer to lines detected in both IK~Tau and R~Dor, in italics to spectral lines only detected in IK~Tau, and in normal font only detected in R~Dor.}
\label{Tab:unidentified}
\begin{tabular}{ccccc}
\hline
\hline
335.993          & 336.307          & 336.471          & 336.815          &
336.973          \\
\textbf{337.421} & 338.170          & 338.447          & 338.520          &
338.677          \\
338.814          & 339.103          & 339.398          & 339.417          &
340.787        \\
\textit{341.141} & 341.233          & 344.223          & 344.719          &
345.020          \\
345.086          & 345.420          & 346.120          & 346.196          &
\textit{346.954} \\
347.221          & \textit{347.956} & 348.117          & \textbf{348.256} &
348.497          \\
348.525          & 348.634          & 348.878          & 348.998          &
349.530          \\
349.890          & \textbf{349.970} & 350.145          & \textit{350.475} &
350.356          \\
350.559          & 350.810          & 351.049          & 351.512          &
351.581          \\
351.790          & 351.920          & 352.034          & \textbf{352.074} &
352.502          \\
353.570          & 353.796          & 355.284          & 355.875          &
355.967          \\
356.461          & 356.462          & 356.933          & 356.974          &
357.767          \\
358.468          & 358.680          & 359.660          & 361.596          &
361.734          \\
361.896          &   &   &   &   \\
\hline
\end{tabular}
\end{table}


The spectral identifications of each feature are given for the $\sim$300 and $\sim$800\,mas extraction apertures in Table~\ref{Tab:IKTau} and Table~\ref{Tab:RDor} and are shown in the spectral atlas in Fig.~\ref{Fig:atlas_IKTau} and Fig.~\ref{Fig:atlas_RDor} for IK~Tau and R~Dor, respectively.
A summary of the number of molecular lines detected is given in Table~\ref{Tab:summary}. A total of 168 lines is detected in IK~Tau and of 197 lines in R~Dor arising from 15 molecules (not counting isotopic species of some of them). 66 lines remain unidentified (see Table~\ref{Tab:unidentified}). Out of these, the line at 345.42\,GHz is suggested to potentially arise from $^{17}$OH, the line at 339.103\,GHz from OH, the line at 348.634\,GHz from H$_2$O, and the lines at 336.377\,GHz and 352.074\,GHz might be $^{34}$SO$_2$. Both lines at 348.497 and 348.525\,GHz might be blends with NS. The 339.103\,GHz line from OH is a high-J rotational transition in the $v=1$ state. Although OH lines of high J are almost impossible to excite by collisions at low temperatures or by IR-pumping, high-J rotational lines are a signature of energetic dissociation of H$_2$O \citep{Tappe2008ApJ...680L.117T}, which could occur deep in the envelope of an AGB star where pulsation shocks are strong enough to generate some H-Lyman $\alpha$ emission.

We note that for various spectral lines in Tables~\ref{Tab:IKTau} and~\ref{Tab:RDor} the peak flux, the integrated flux, and velocity extent in the $\sim$300\,mas region are larger than in the $\sim$800\,mas aperture. The reason for this (maybe counterintuitive) result is linked to the noise-value which is higher in the largest extraction apertures.  This has two effects.  Most importantly, even if the flux measured per channel in the faint line wings are similar, the 3-times local off-source rms is higher, and hence the estimated line extent above this cut-off is smaller.  The effects of resolved out flux and dynamic range limitations may also contribute negative artefacts. The apparent total line extent is in some cases smaller for the largest aperture compared to the 300 or 320\,mas radii, affecting the integrated flux in a second way.  These effects occur more often for R~Dor, which has the brighter continuum and is more likely to have large-scale smooth continuum and line flux, as it is closer. The narrower channels also slightly worsen the uv-coverage and thus the dynamic range problems. As a consequence, the large-aperture measurements are less reliable for compact emission.

\begin{figure*}[!htp]
\includegraphics[angle=0,width=\textwidth,height=7.4truecm]{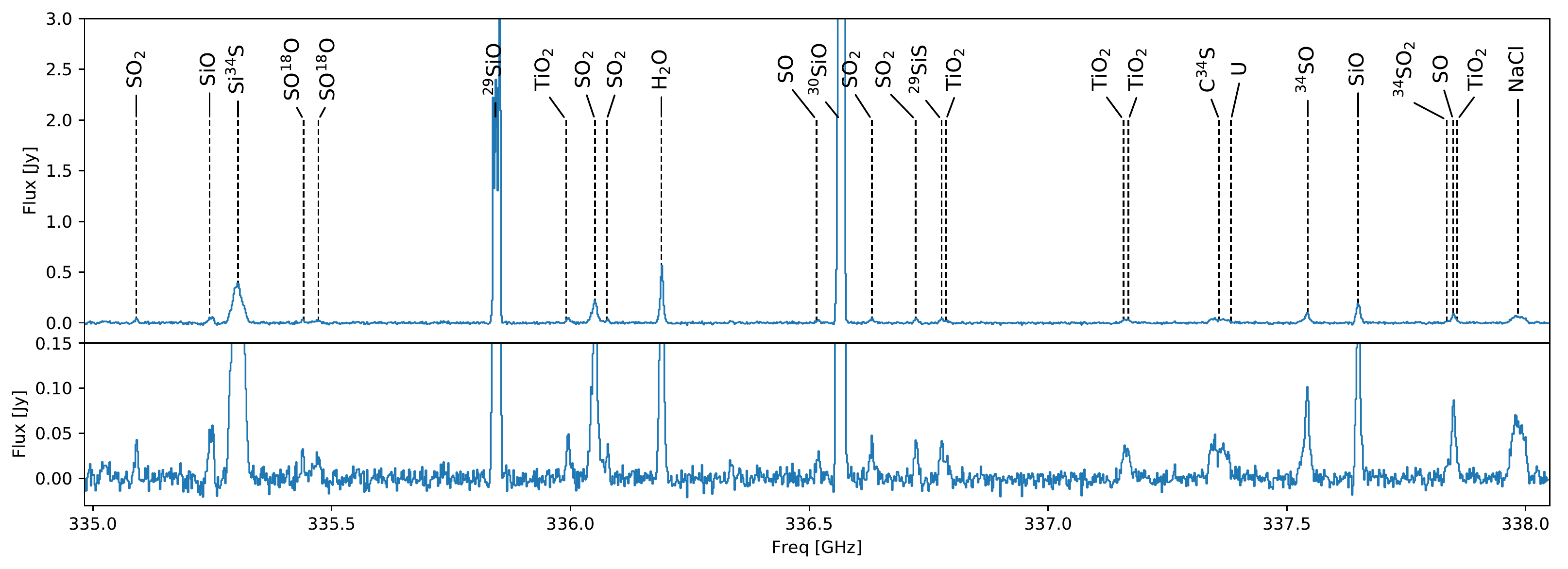}
\medskip
\includegraphics[angle=0,width=\textwidth,height=7.4truecm]{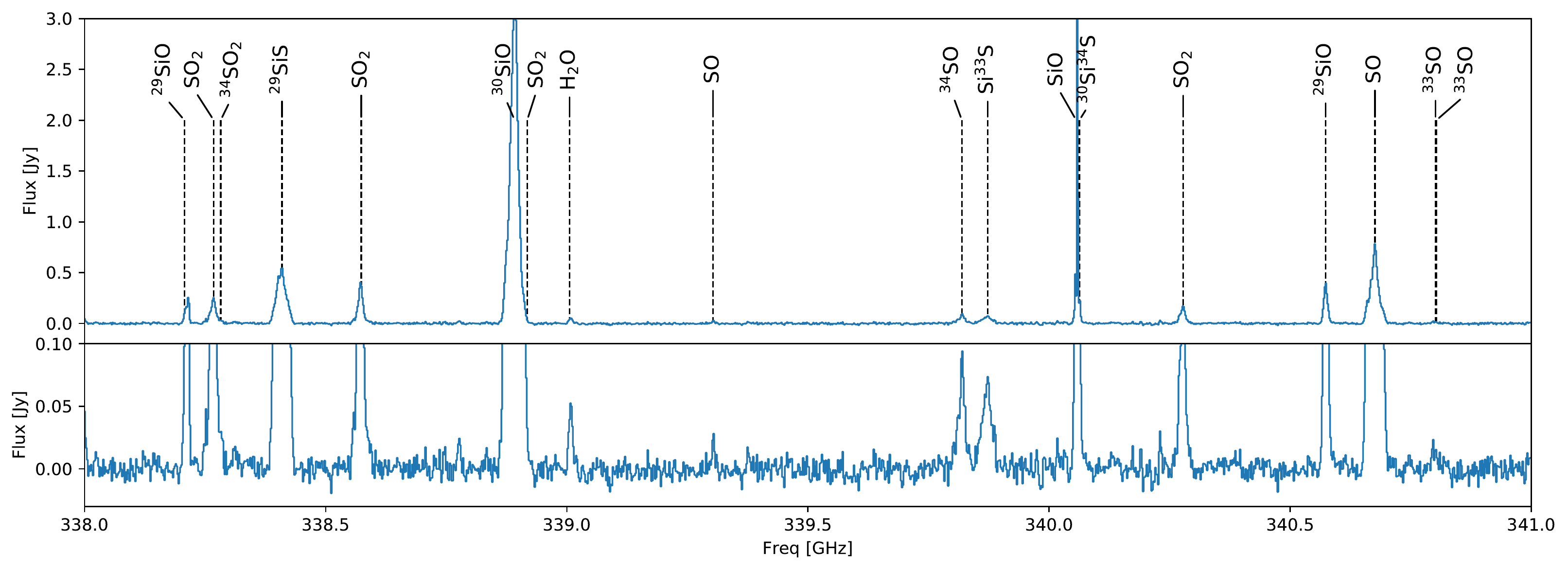}
\medskip
\includegraphics[angle=0,width=\textwidth,height=7.4truecm]{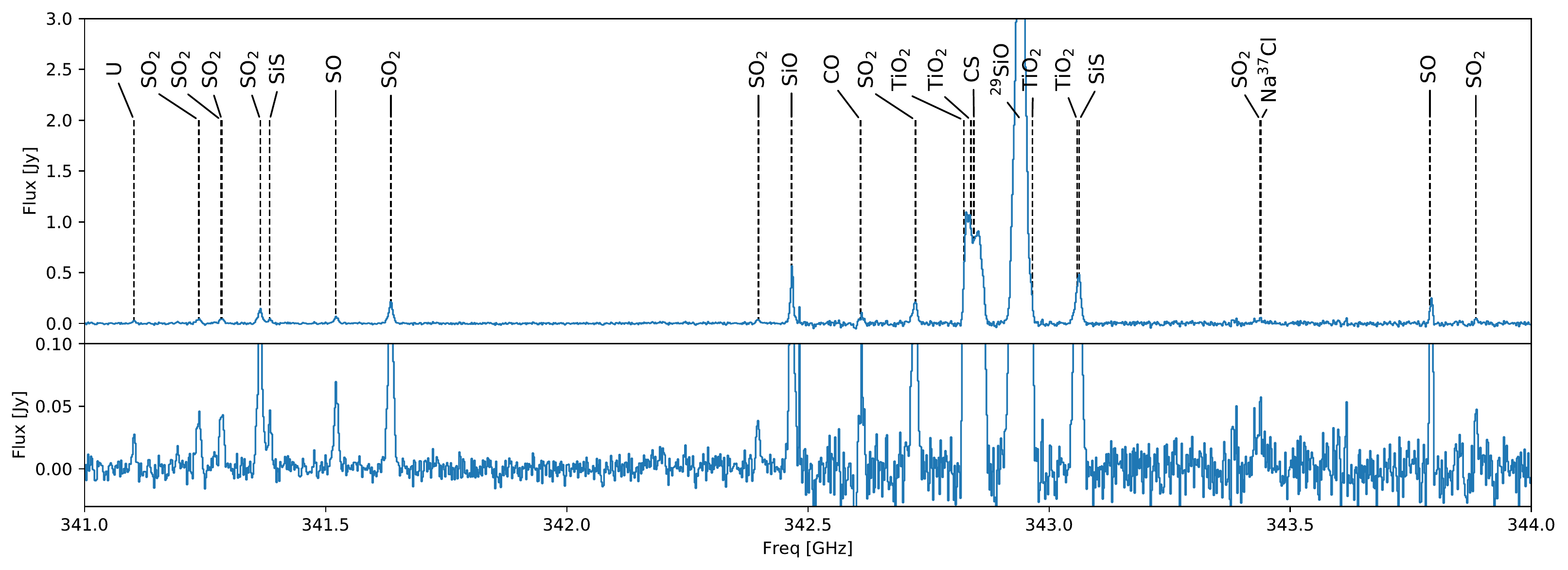}
\caption{ALMA spectral atlas of IK~Tau between 335 and 362\,GHz extracted for a circular beam with an aperture of 320\,mas. Data are displayed per 3\,GHz frequency band; a zoom to better visualise the weaker lines is shown in the bottom panels. }
\label{Fig:atlas_IKTau}
\end{figure*} 

\begin{figure*}[!htp]
\includegraphics[angle=0,width=\textwidth,height=7.4truecm]{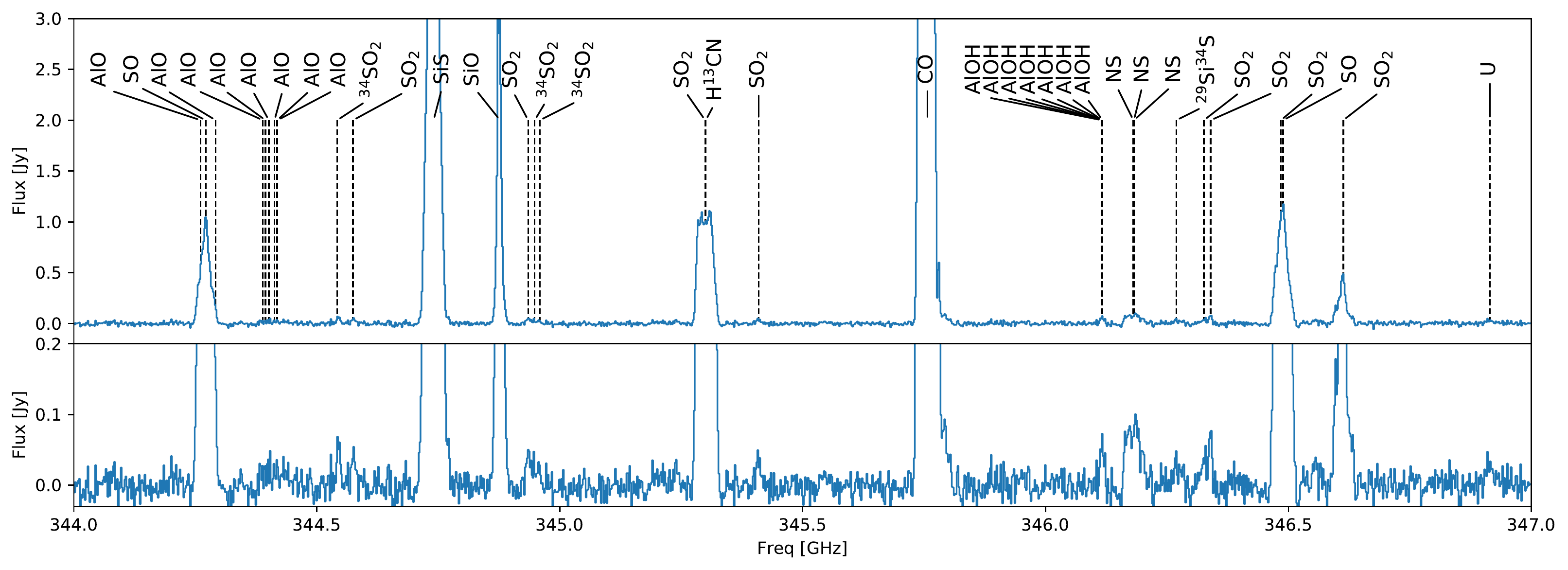}
\medskip
\includegraphics[angle=0,width=\textwidth,height=7.4truecm]{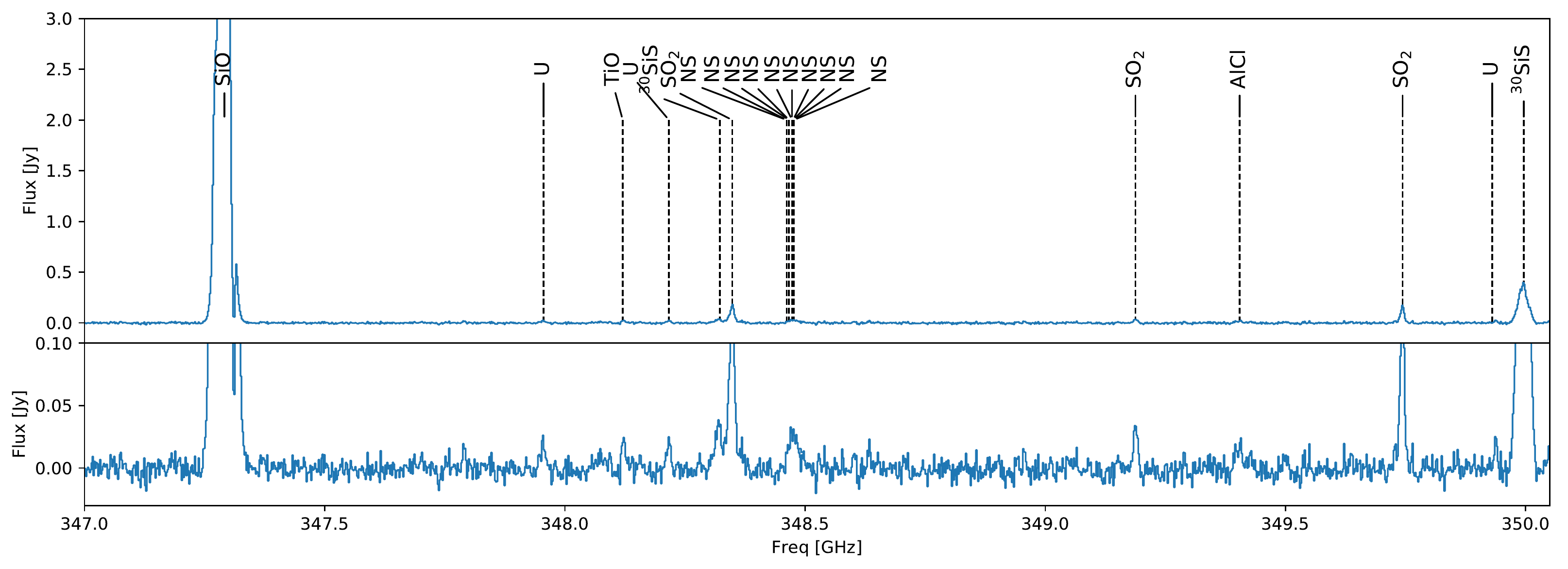}
\medskip
\includegraphics[angle=0,width=\textwidth,height=7.4truecm]{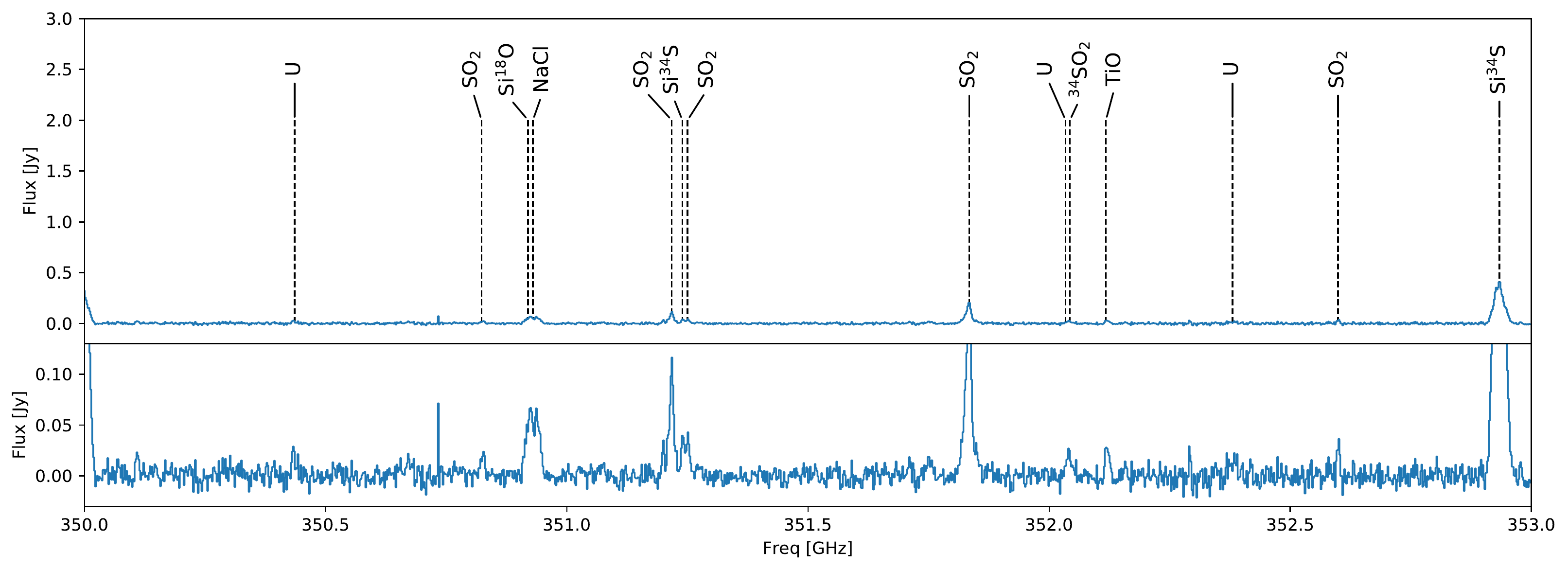}
\addtocounter{figure}{-1}
\caption{\it (Continued)}
\end{figure*}  

\begin{figure*}[!htbp]
\includegraphics[angle=0,width=\textwidth,height=7.4truecm]{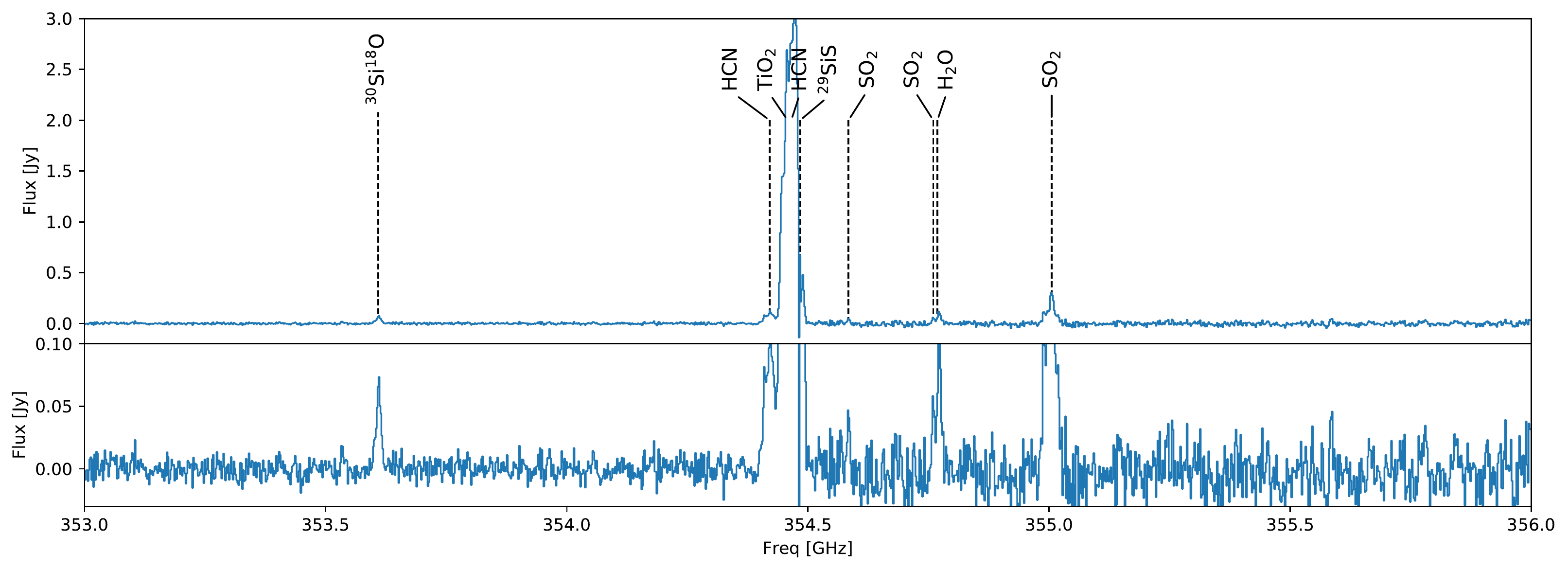}
\medskip
\includegraphics[angle=0,width=\textwidth,height=7.4truecm]{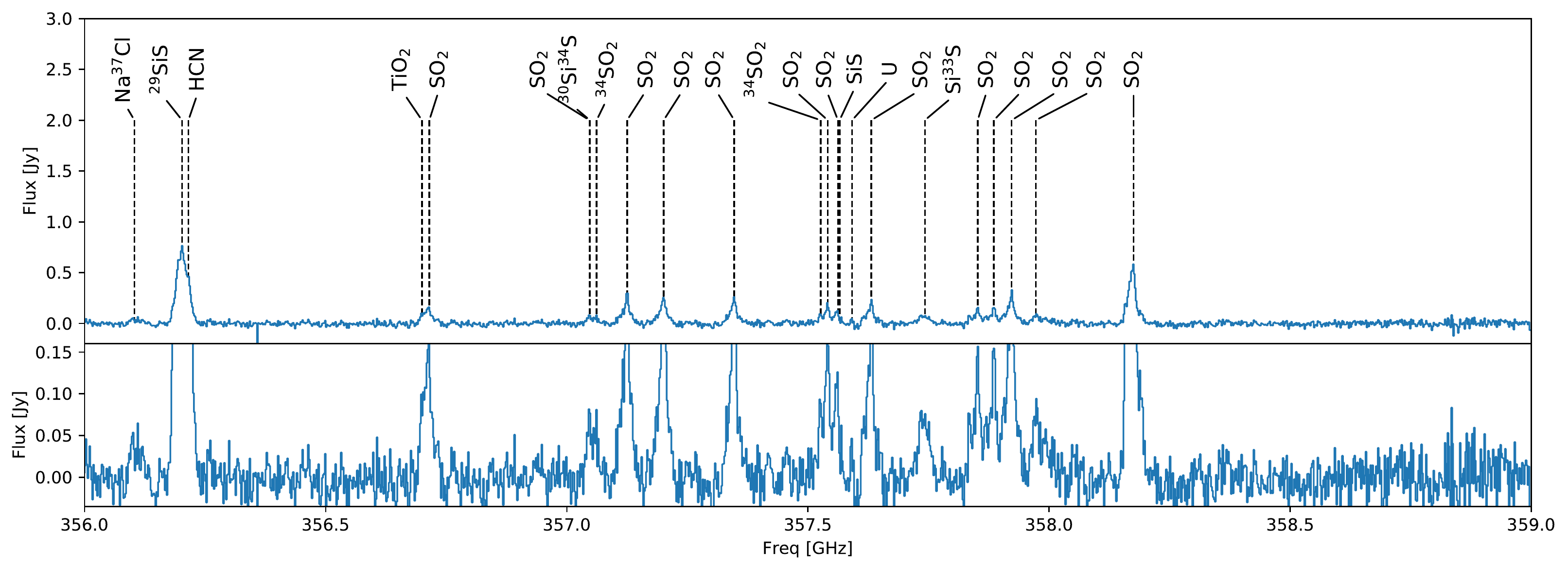}
\medskip
\includegraphics[angle=0,width=\textwidth,height=7.4truecm]{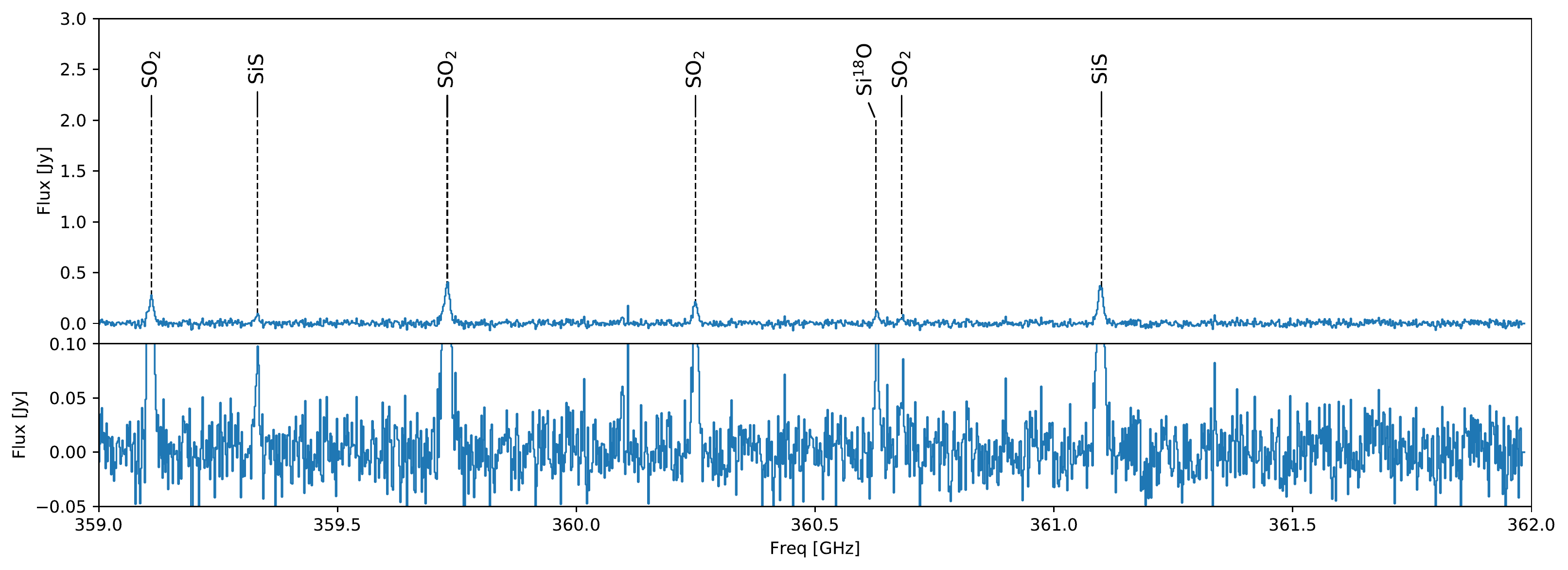}
\addtocounter{figure}{-1}
\caption{\it (Continued)}
\end{figure*} 

\begin{figure*}[!htbp]
\includegraphics[angle=0,width=\textwidth,height=7.4truecm]{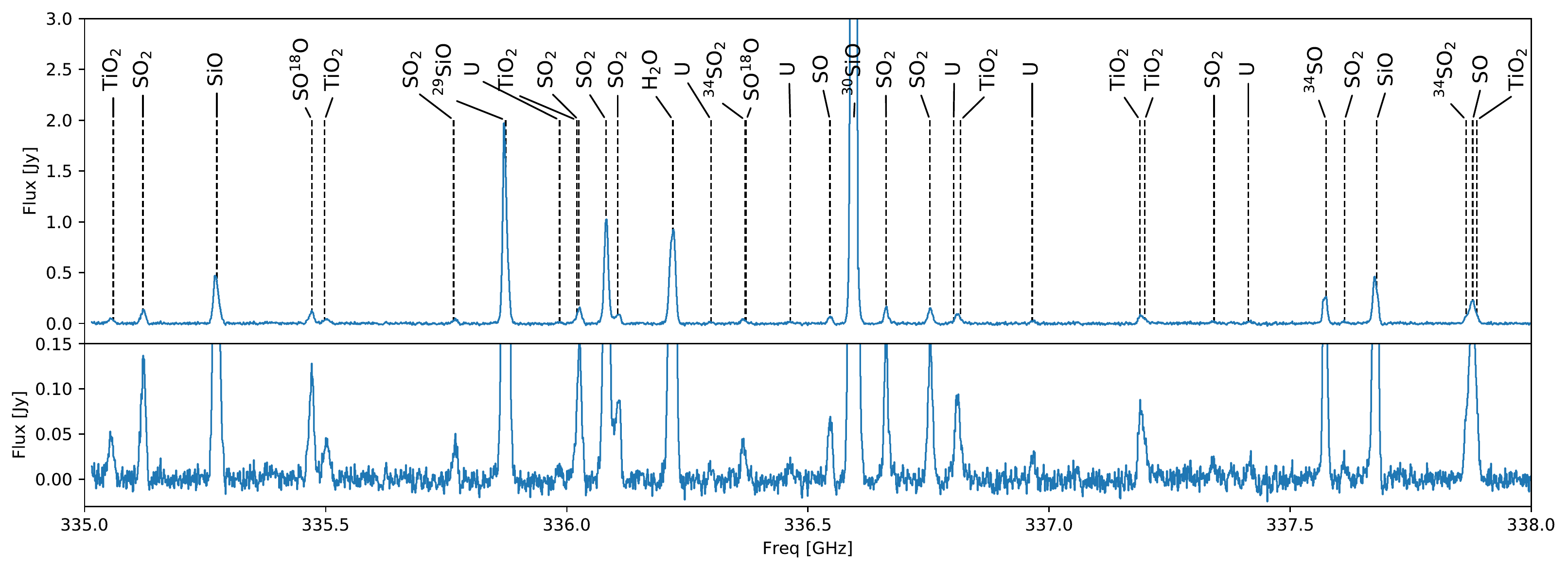}
\medskip
\includegraphics[angle=0,width=\textwidth,height=7.4truecm]{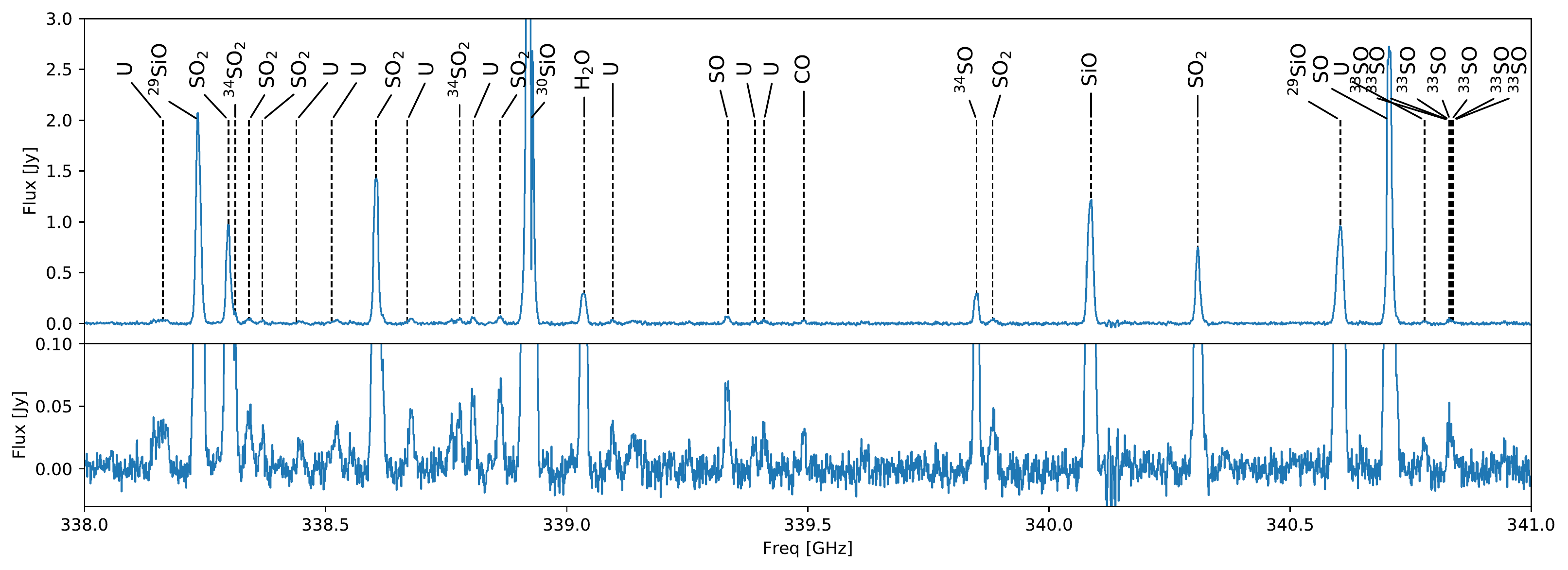}
\medskip
\includegraphics[angle=0,width=\textwidth,height=7.4truecm]{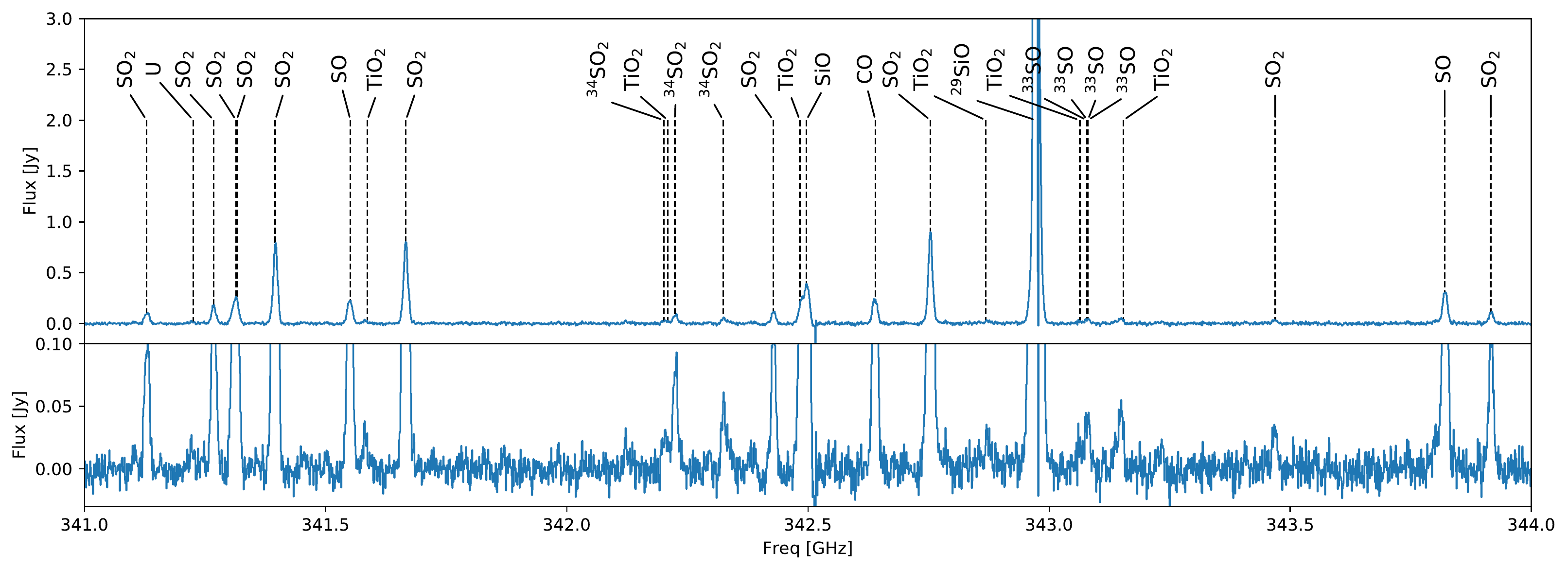}
\caption{ALMA spectral atlas of R~Dor between 335 and 362\,GHz extracted for a circular beam with an aperture of 300\,mas. Data are displayed per 3\,GHz frequency band; a zoom to better visualise the weaker lines is shown in the bottom panels. }
\label{Fig:atlas_RDor}
\end{figure*} 

\begin{figure*}[!htbp]
\includegraphics[angle=0,width=\textwidth,height=7.4truecm]{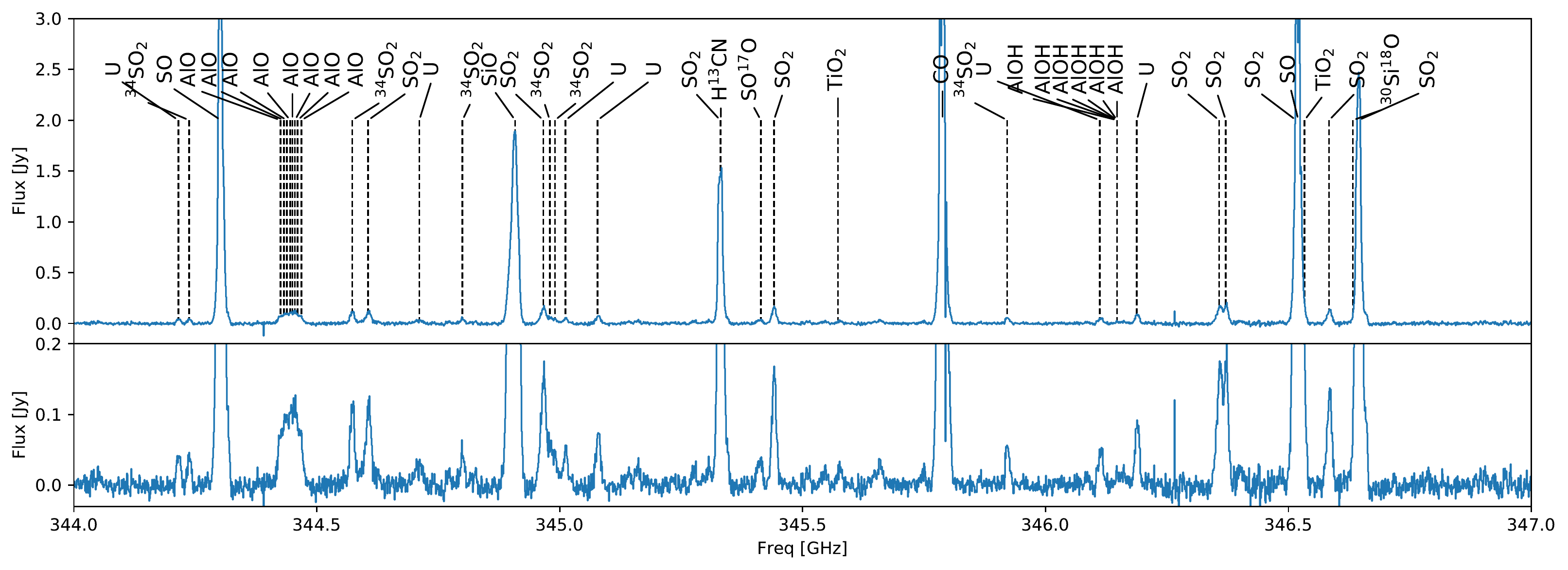}
\medskip
\includegraphics[angle=0,width=\textwidth,height=7.4truecm]{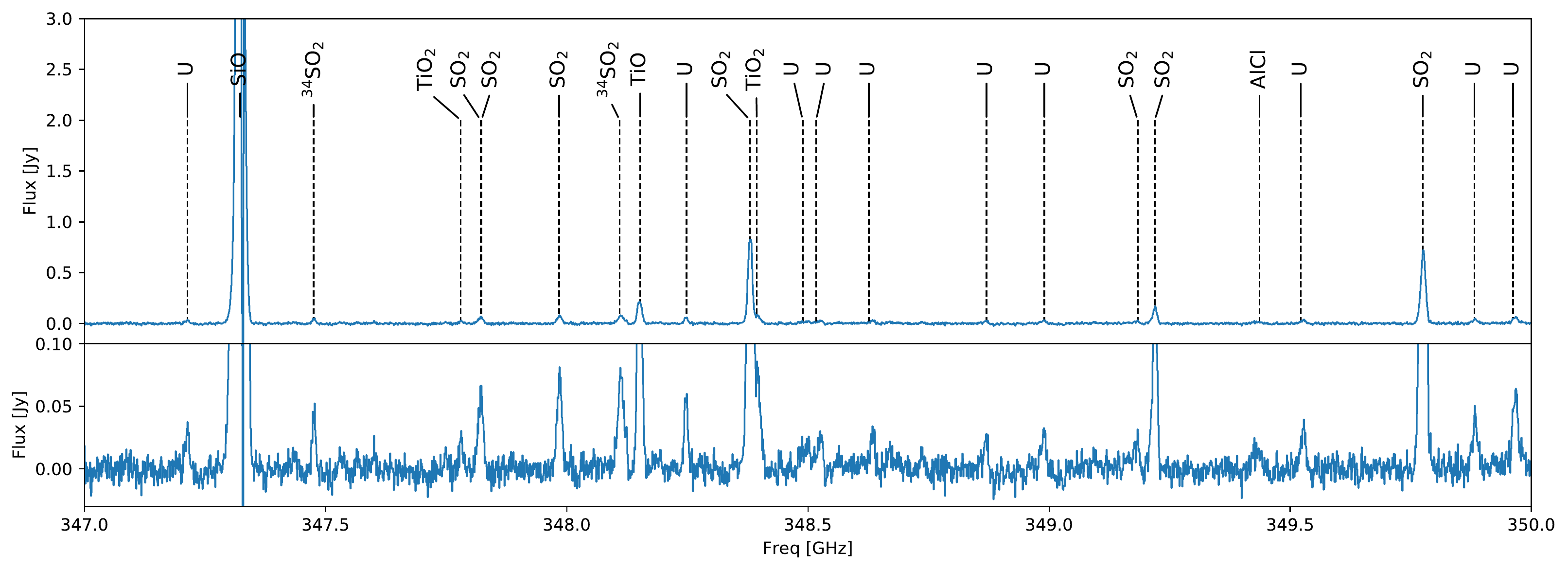}
\medskip
\includegraphics[angle=0,width=\textwidth,height=7.4truecm]{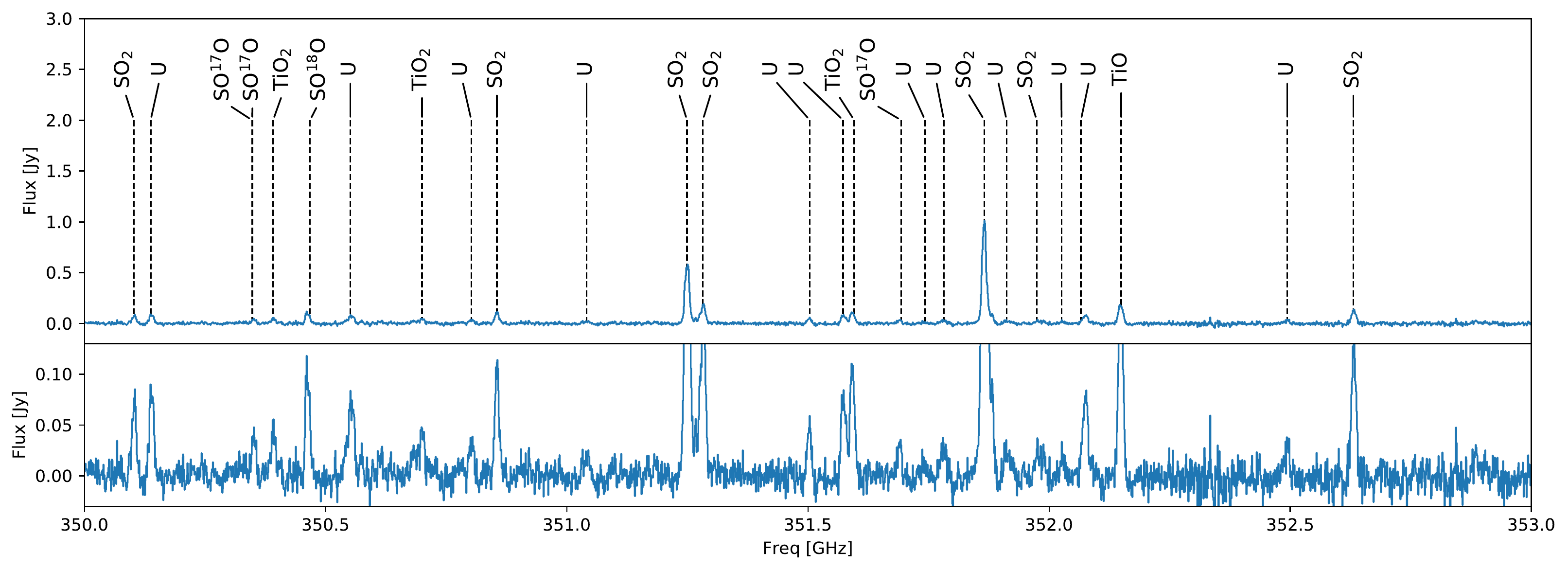}
\addtocounter{figure}{-1}
\caption{\it (Continued)}
\end{figure*}  

\begin{figure*}[!htbp]
\includegraphics[angle=0,width=\textwidth,height=7.4truecm]{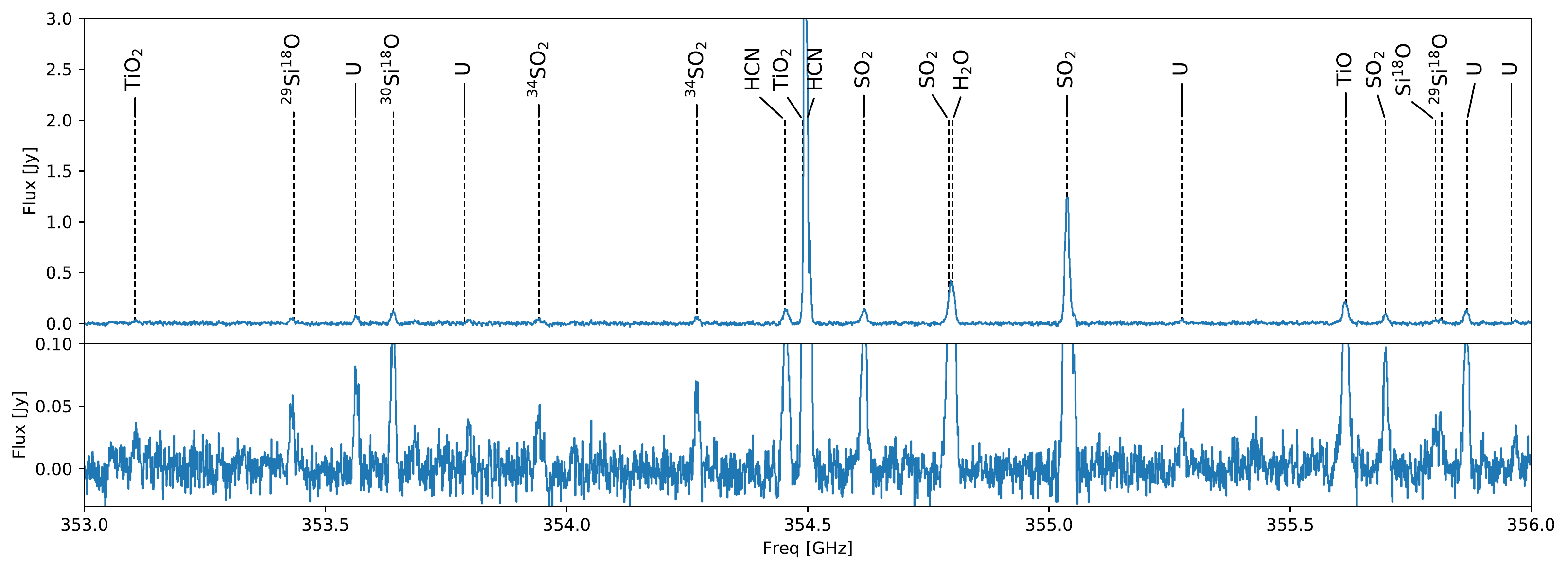}
\medskip
\includegraphics[angle=0,width=\textwidth,height=7.4truecm]{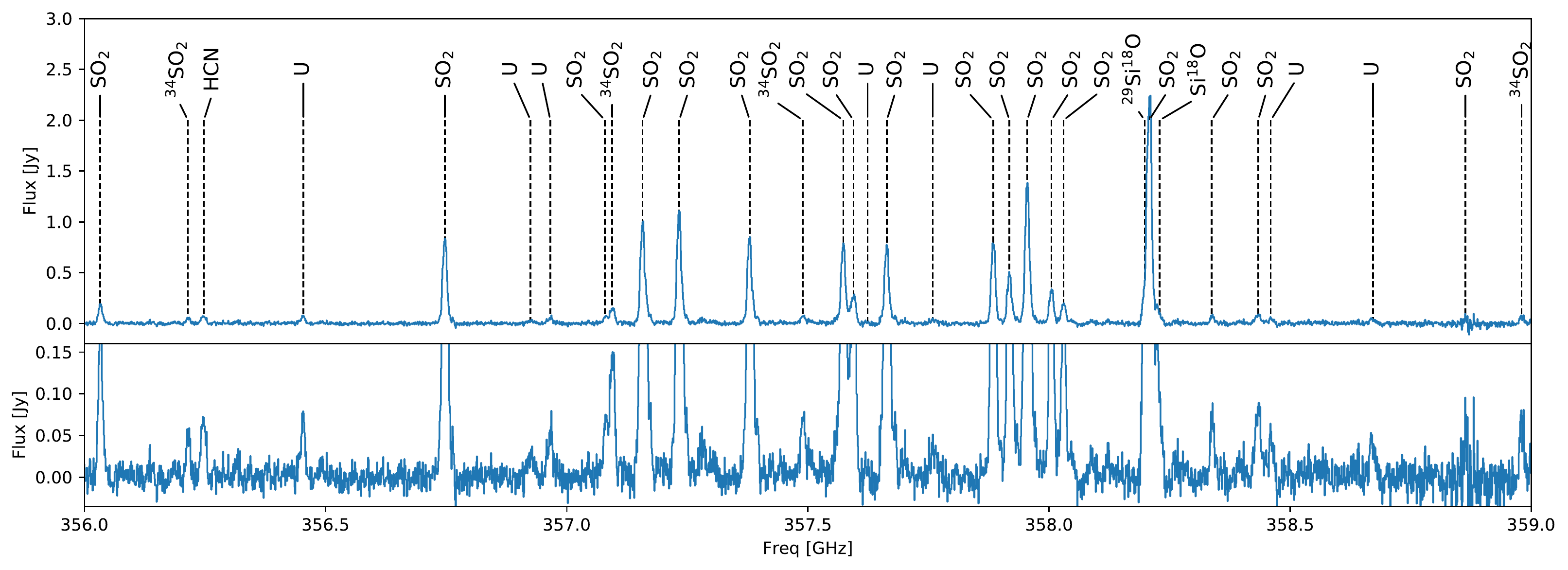}
\medskip
\includegraphics[angle=0,width=\textwidth,height=7.4truecm]{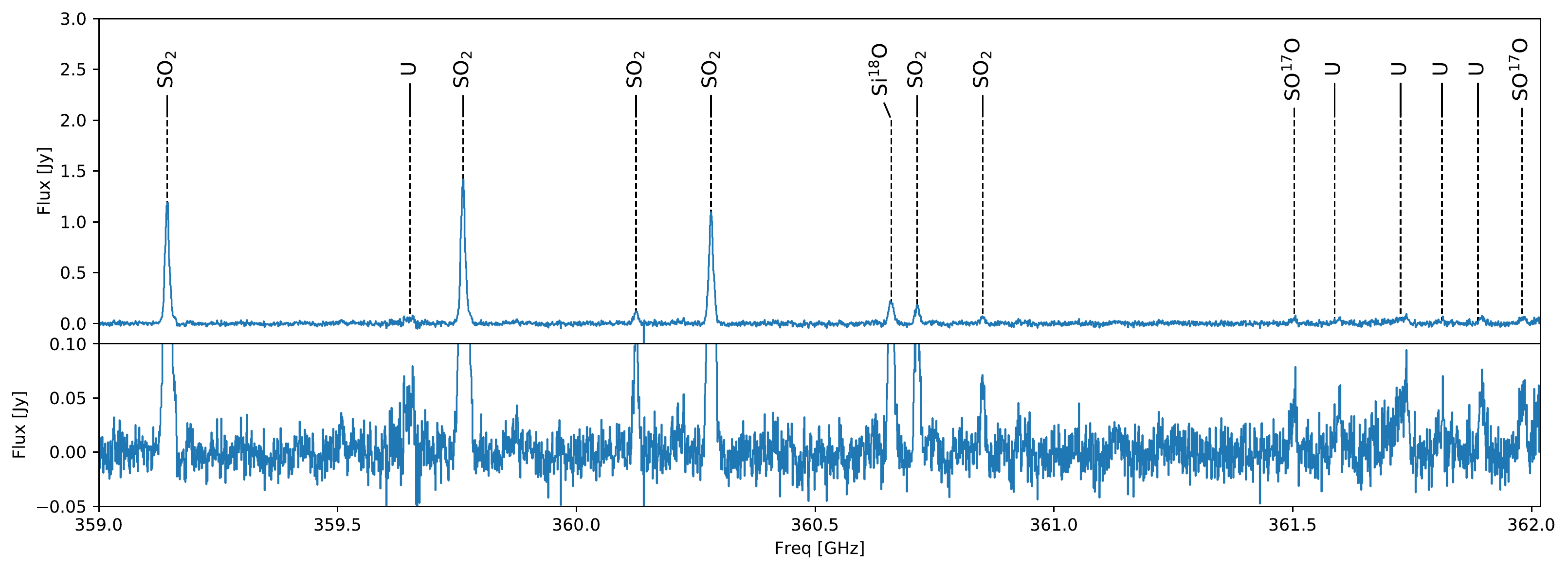}
\addtocounter{figure}{-1}
\caption{\it (Continued)}
\end{figure*}

\subsection{Analysis of the spectral cubes} \label{Sec:spectral_analysis}

For each spectral feature, different properties were measured including the velocity width, the shift of the centroid with respect to\ the LSR velocity of the star, the angular width, and the peak flux and integrated flux within the specified velocity interval.

The velocity span $\Delta v$ of single lines (or hyperfine blends of a single species) is taken to be the continuous extent above 3.5$\sigma_{\mathrm{chn\_rms}}$. Asymmetry is defined as the excess velocity on the side extending further from $v_{\rm{LSR}}$ (positive on the red-shifted side).  For lines blended on one side only, twice the width on the unblended side is used (so no asymmetry can be measured).  If a line has a blend on both sides (a so-called mid blend), a measurement is only made if one blend is very faint, in which case $\Delta v$ is an upper limit.

To derive the angular extent of the molecular emission, we have created zero-moment (total intensity) maps (see some examples in Sect.~\ref{Sec:spatial}). For that, the total velocity extent of unblended lines was used or only the unblended side in the case of blends. The angular extent of the emission  was estimated  from the moment maps by measuring  the flux density in one-pixel annuli until the flux density fell below 1.5 times the off-source rms of the moment map, and then taking the diameter of this annulus. However, in order to investigate asymmetries and distinguish these from beam sampling, we also measured the angular extent of the continuous 1.5$\sigma_{\rm{rms}}$ contour enclosing the stellar position. This was used for the measurements and morphology characterisation given in Tables~\ref{Tab:IKTau}-\ref{Tab:RDor}. In general the measurements were in agreement, the difference providing the uncertainty in angular size due to source asymmetry.

\begin{figure*}[!htbp]
\begin{minipage}[t]{.33\textwidth}
        \centerline{\resizebox{\textwidth}{!}{\includegraphics[angle=0]{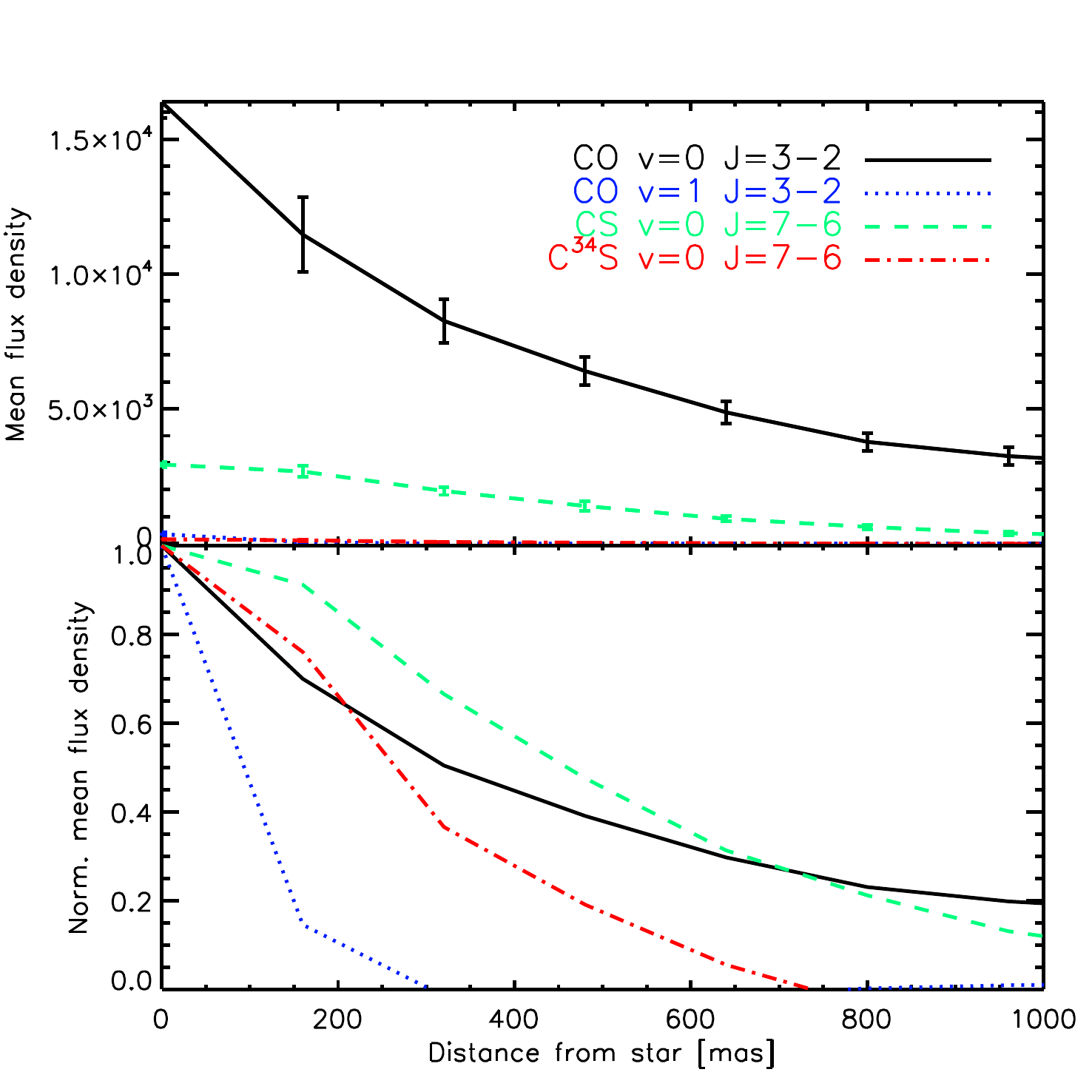}}}
    \end{minipage}
    \hfill
\begin{minipage}[t]{.33\textwidth}
        \centerline{\resizebox{\textwidth}{!}{\includegraphics[angle=0]{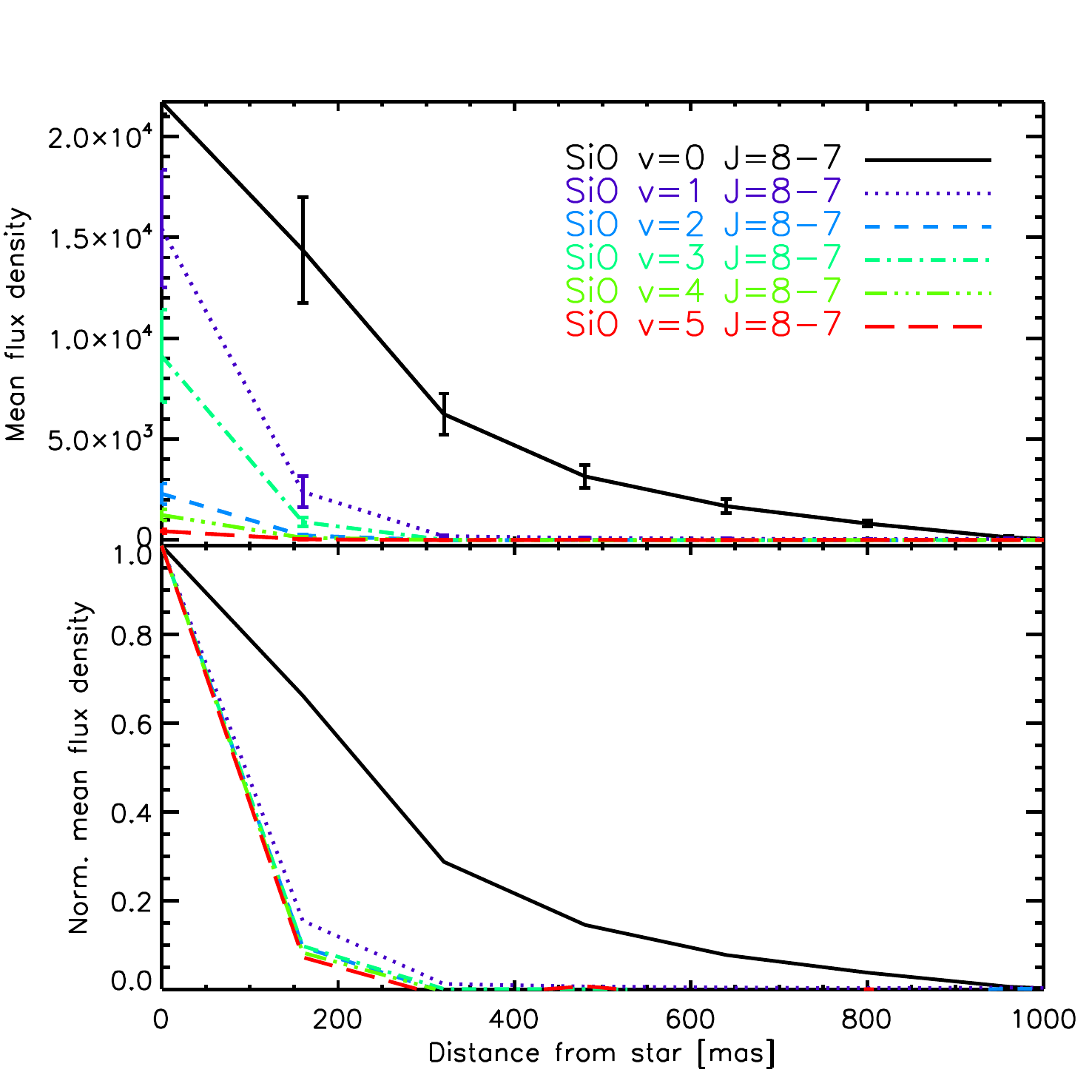}}}
    \end{minipage}
    \hfill
\begin{minipage}[t]{.33\textwidth}
        \centerline{\resizebox{\textwidth}{!}{\includegraphics[angle=0]{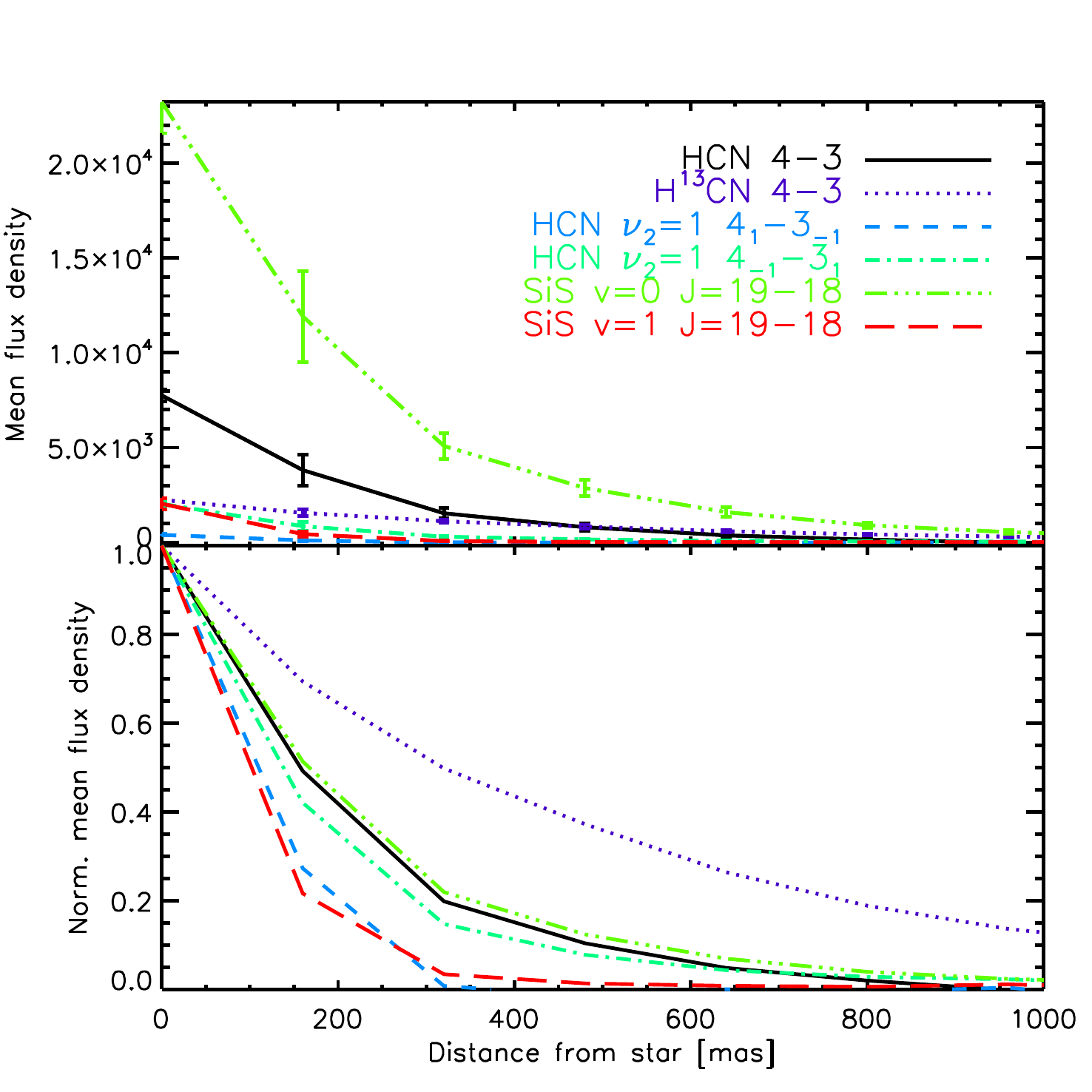}}}
    \end{minipage}
\begin{minipage}[t]{.33\textwidth}
        \centerline{\resizebox{\textwidth}{!}{\includegraphics[angle=0]{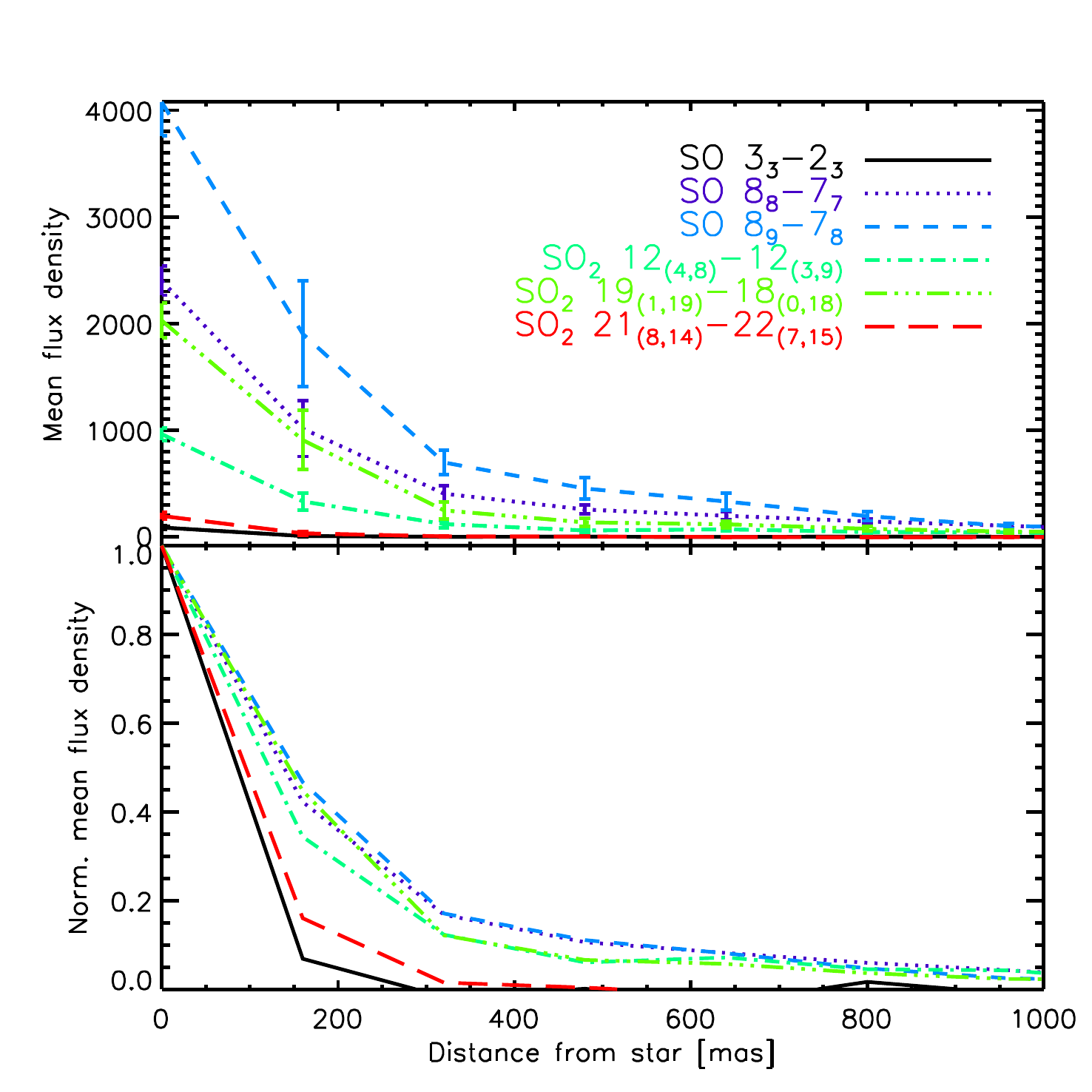}}}
    \end{minipage}
    \hfill
\begin{minipage}[t]{.33\textwidth}
        \centerline{\resizebox{\textwidth}{!}{\includegraphics[angle=0]{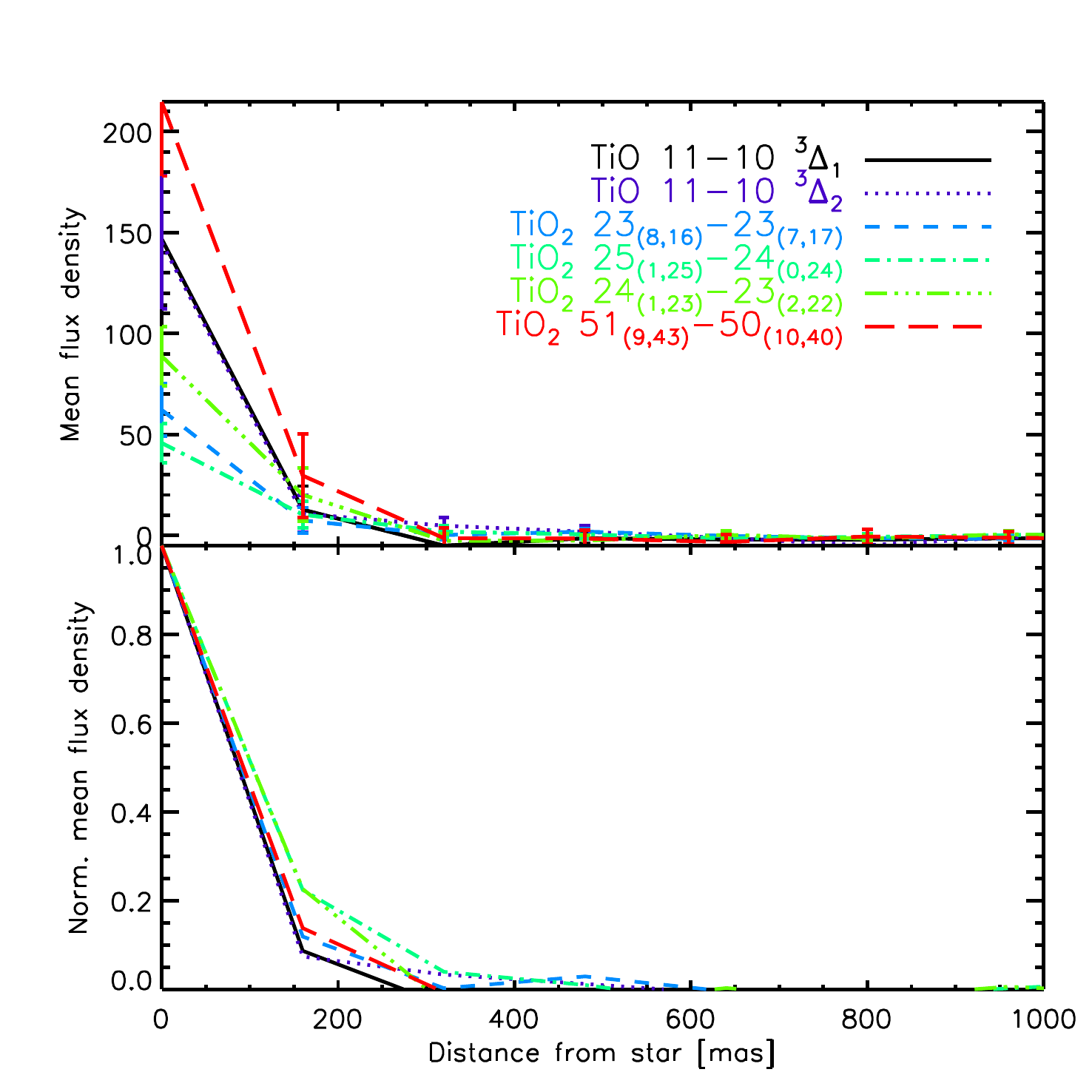}}}
    \end{minipage}
    \hfill
\begin{minipage}[t]{.33\textwidth}
        \centerline{\resizebox{\textwidth}{!}{\includegraphics[angle=0]{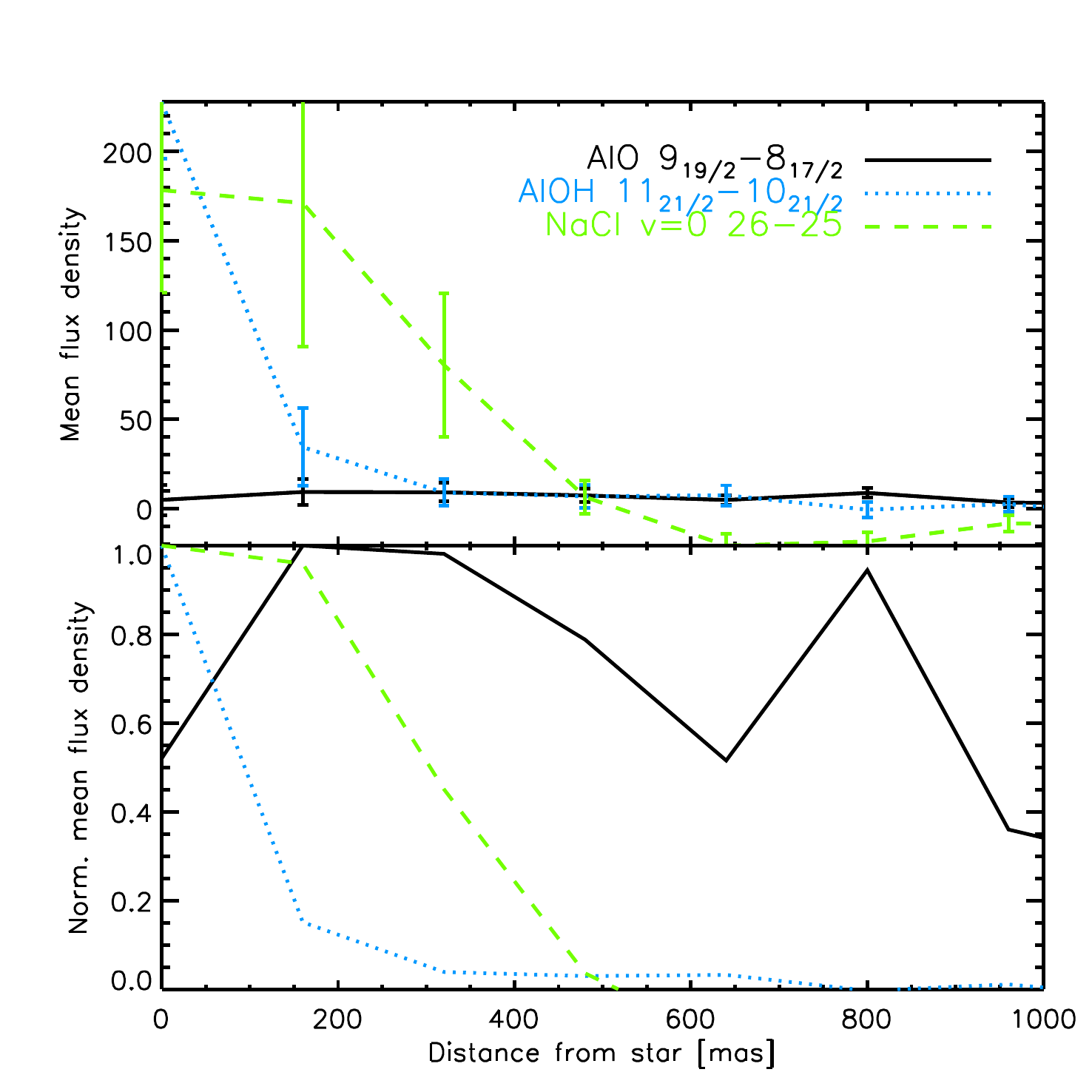}}}
    \end{minipage}
  \caption{Azimuthally averaged flux density (in units of [mJy/beam km/s]) as a function of the angular distance from the central star of IK~Tau for a selected sample of line transitions. For some transitions, a blend with another line impacts the mean flux density; notes on that can be found in Table~\ref{Tab:IKTau}. In the upper panel, the error bars designate the scatter w.r.t.\ the mean flux density (see text for further details); 
  the lower panel displays the normalised mean flux densities, omitting the standard deviations for clarity.}
  \label{Fig:IKTau_RadProf}
\end{figure*}

\begin{figure*}[!htbp]
           \includegraphics[angle=0, width=58mm]{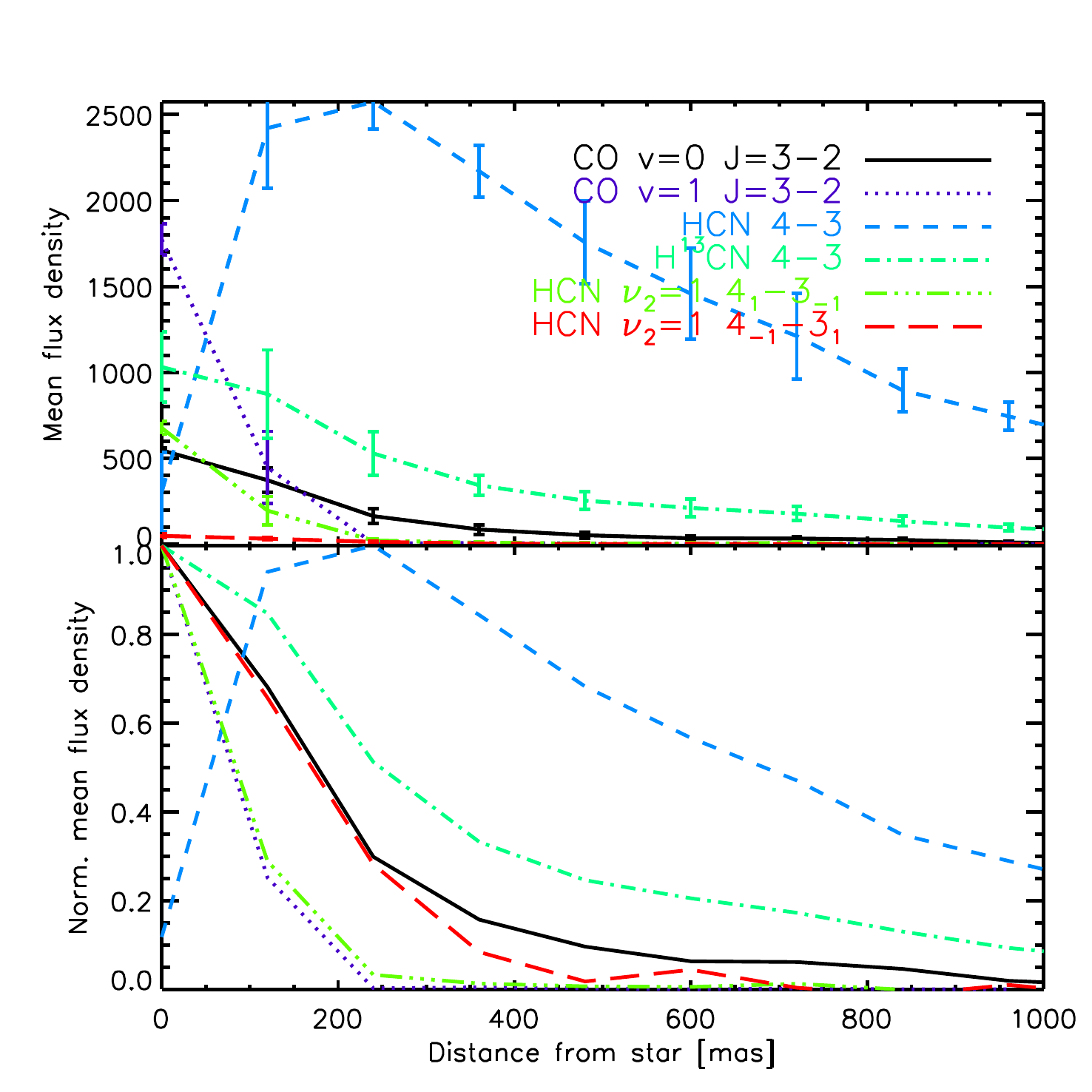}
      \includegraphics[angle=0,width=58mm]{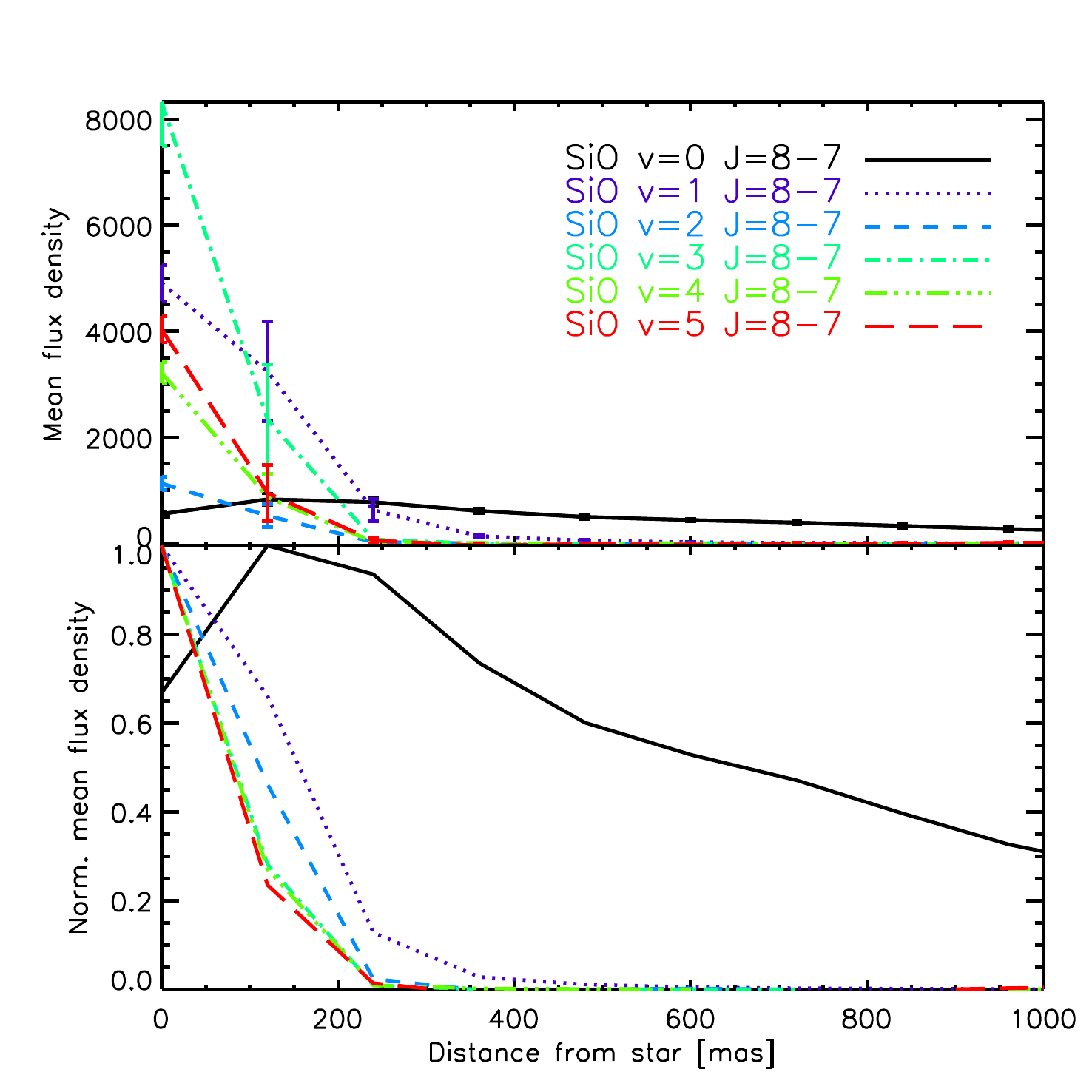}

      \includegraphics[angle=0, width=58mm]{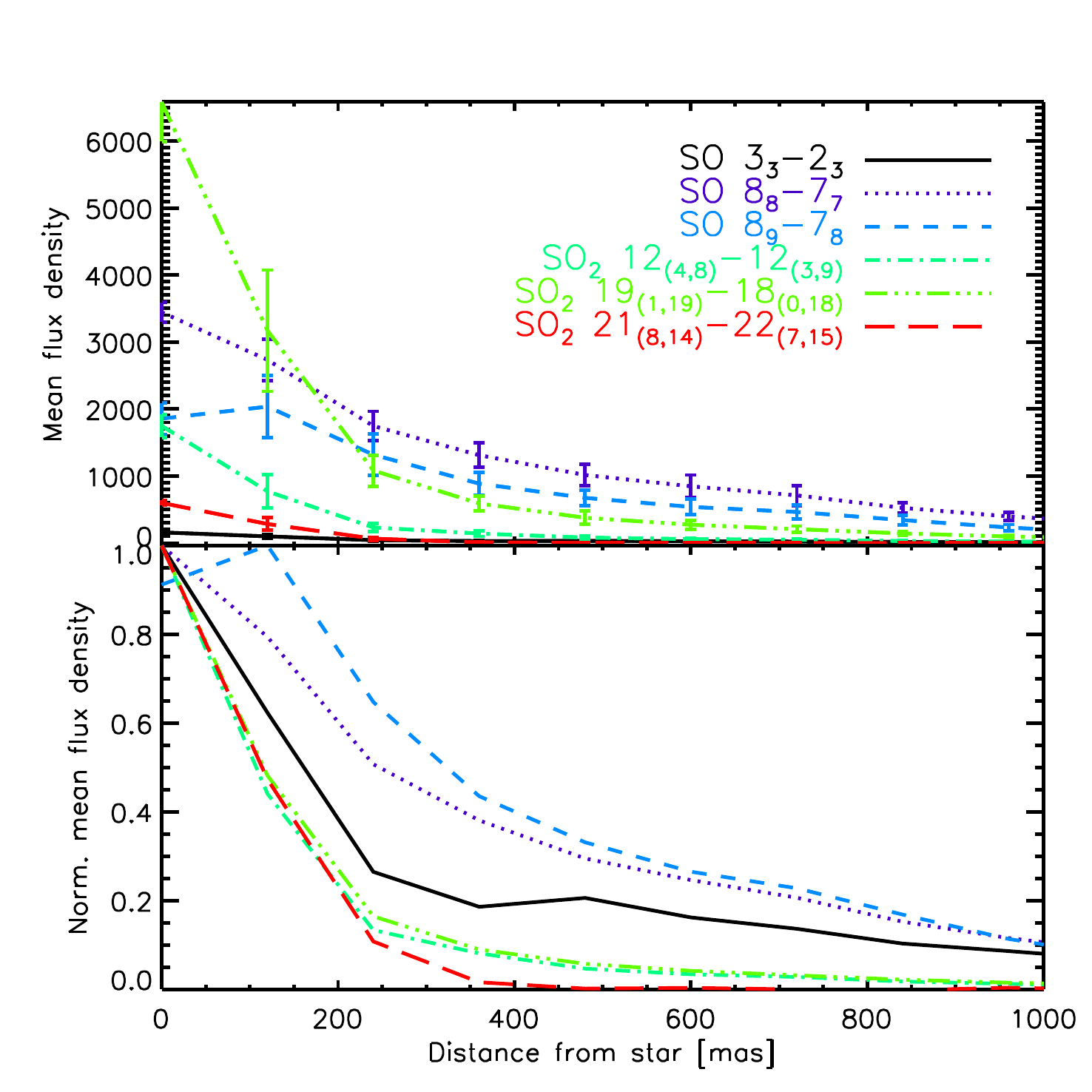}
        \includegraphics[angle=0, width=58mm]{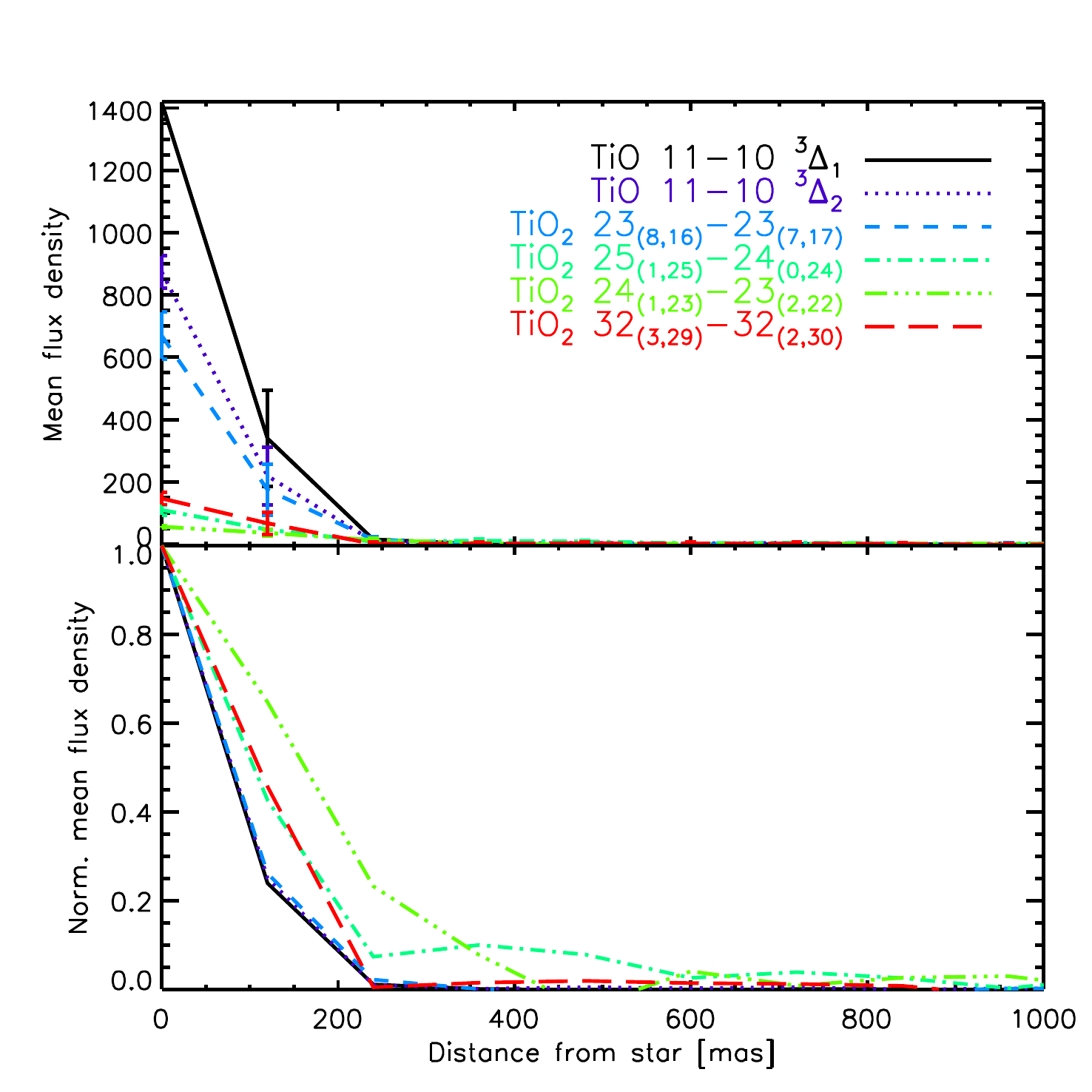}
   \caption{Azimuthally averaged flux density (in units of [mJy/beam km/s]) as a function of the angular distance from the central star of R~Dor for a selected sample of line transitions. For some transitions, a blend with another line impacts the mean flux density; notes on that can be found in Table~\ref{Tab:RDor}. In the upper panel, the error bars designate the scatter w.r.t.\ the mean flux density (see text for further details); 
  the lower panel displays the normalised mean flux densities, omitting the standard deviations for clarity.}
  \label{Fig:RDor_RadProf}
\end{figure*}

The zeroth moment maps were also used to compute azimuthally averaged flux densities; see some examples in Fig.~\ref{Fig:IKTau_RadProf}  for IK~Tau and  Fig.~\ref{Fig:RDor_RadProf} for R~Dor.  These azimuthal averages are used to get a more accurate estimate of the angular extent of molecular emission in the case of genuinely fragmented or resolved-out extended emission. These plots can also serve to retrieve the mean molecular density assuming  (1D) spherically symmetric radiative transfer models, as done for example,\ by \citet{Decin2017arXiv170405237D}. There are two main sources of error in the profiles measured from azimuthally averaged annuli. Where the emission is strong, the fluctuations within an annulus are caused by clumpiness or asymmetries in the distribution of the emission, represented by the `absolute value of the median deviation from the mean flux density'.  Where the emission is weak, the uncertainty in the mean flux density is noise dominated, represented by the `rms normalised by the square root of the number of beam areas' in the annulus.  We took the error bars as the lesser of these two metrics for the mean flux density in each annulus.
The emission of NaCl (see Fig.~\ref{Fig:IKTau_RadProf}) in IK~Tau is a clear example of large error bars that arise due to a very inhomogeneous morphology (see also Sect.~\ref{Sec:spatial}).

The retrieved values for each of the described quantities can be found in Table~\ref{Tab:IKTau} for IK~Tau and in Table~\ref{Tab:RDor} for R~Dor. The velocity widths and spatial extents are used in Sect.~\ref{Sec:kinematic} to discuss the kinematic structure of the stellar winds.

\section{Discussion} \label{Sec:Discussion}

\subsection{Detected molecular species} \label{Sect:disc_species}

For both targets, the brightest line is the vibrationally excited $^{30}$SiO v=1 J=8-7 line at 336.603\,GHz which has a flux density of 55\,Jy (IK~Tau) and 182\,Jy (R~Dor) for the spectrum with extraction beam of $\sim$300\,mas radius. The detected $^{28}$SiO, $^{29}$SiO, and $^{30}$SiO  lines might be prone to masering; the vibrationally excited lines trace the regions close to the star.

Table~\ref{Tab:summary} shows the large difference in chemical composition between both stars: while only SO and SO$_2$ are detected as sulphur-bearing species in R~Dor, SiS and CS are also prominently present in the high mass-loss rate star IK~Tau. This is in accord with the conclusions drawn by \citet{Danilovich2016A&A...588A.119D}, who suggested that for low mass-loss rate O-rich AGB stars almost all sulphur seems to be locked up in SO and SO$_2$, while these two molecules can not account for the total sulphur budget in high mass-loss rate O-rich AGB stars. Current theoretical models solving the (gas-phase) chemical network \citep[e.g.][]{Li2016A&A...588A...4L} are not able to explain this dichotomy. Different gas-grain chemical interactions, the effect of density clumps allowing for a varying penetration of the interstellar UV photons affecting the photochemistry \citep{Agundez2010ApJ...724L.133A}, or pulsation-induced non-equilibrium chemistry \citep{Cherchneff2006A&A...456.1001C} might be the reason for this distinction. We are currently investigating each of these effects; the results of which will be presented in forthcoming papers \citep[][Boulangier et al.\ \textit{(in prep.)}]{VandeSande2018}. We note that both CS and SiS are expected not to be present in high amounts in O-rich winds, a conclusion drawn from chemical equilibrium calculations. The presence of these two molecules, as well as the detection of HCN in both targets, fortifies earlier claims that chemical abundances in stellar winds can not be reliably predicted from thermodynamic equilibrium considerations, but that the complex interaction between morphology, dynamics and (gas-grain) chemistry should be considered.

The species AlO, AlOH, TiO, TiO$_2$, SiO, and H$_2$O (can) play a key role in the nucleation of dust grains; the latter two molecules as reactants in a heteromolecular formation process for silicates \citep{Gouamns2012MNRAS.420.3344G}, the first ones might form via homogeneous nucleation. \citet{Decin2017arXiv170405237D} has shown that AlO, AlOH, and AlCl only consume $\sim$2\% of the total aluminium content in both targets, hence leaving ample of room for large gas-phase (Al$_2$O$_3$)$_n$ clusters and Al-bearing dust grains to form. The ALMA data reveal that TiO gas is traceable up to 2.4$\times r_{\rm{dust}}$ (IK~Tau) and 3.3$\times r_{\rm{dust}}$ (R~Dor) and TiO$_2$ even up to 4.5$\times r_{\rm{dust}}$ (IK~Tau) and 7.3$\times r_{\rm{dust}}$ (R~Dor), that is,\ well beyond the onset of the dust condensation. A similar result is found in Mira by \citet{Kaminski2017A&A...599A..59K}. We note here that only the blue-shifted emission of the two TiO lines in IK~Tau is detected. The channel map of both lines shows a clear unresolved blob of emission at those frequencies. A search in various spectroscopic database did not bring forwards another potential identification of these two lines. Moreover, since these two lines are also detected in R~Dor (at the correct central frequency), we are convinced that both lines belong to TiO. This compact blue-shifted emission suggests outflow (towards us), the red-shifted emission being obscured by the star. This points towards a fast radial decrease in TiO abundance. TiO can get oxidised via the reaction TiO+OH$\rightarrow$TiO$_2$+H, which is exothermic by $-$141\,kJ/mol \citep{Plane2013RSPTA.37120335P}. The forwards reaction is fast, but the reverse reaction is extremely slow, so that the TiO and TiO$_2$ concentrations become nearly equal at 1000\,K because the equilibrium constant increases at lower temperatures. In addition, both the concentrations of TiO and TiO$_2$ decay because of the formation of metal titanates, which might be important as predecessors of condensation nuclei \citep{Plane2013RSPTA.37120335P}.

Three water vapour lines are detected in each source, including the vibrational excited H$_2$O 5$_{2,3}$-6$_{1,6}$ ($\nu_2$\,=\,1) line. The two other lines are in the vibrational ground-state with upper $J$-values of 16 or 17. The three lines are predicted to be a maser \citep{Amano1991CPL...182...93A, Feldman1993LNP...412...65F, Gray2016MNRAS.456..374G}. All these three lines appear compact with the diameter of the maximum angular extent being 434\,mas for IK~Tau and 505\,mas for R~Dor, or $\sim$22\,\Rstar\ and $\sim$8.5\,\Rstar in radius, respectively. They are, however, not bright enough to be definitely identified as masers; higher spectral and spatial resolution would be needed to verify these predictions.

Sodium chloride, NaCl, is only detected in IK~Tau  both in the main stable isotope Na$^{35}$Cl and in the less abundant Na$^{37}$Cl gas specimen. Salt has already been detected and spatially resolved with ALMA in the high mass-loss rate supergiant VY~CMa \citep[\Mdot\,$\sim$2$\times10^{-4}$\,\Msun/yr,][]{Decin2016A&A...592A..76D}. It is well know that this metal halide is highly refractory and its detection in gaseous form points towards a chemical process preventing all NaCl from condensing onto dust grains such as photodesorption, thermal desorption or shock-induced sputtering. Its high dipole moment \citep[$\mu$\,=\,9\,D,][]{Muller2005JMoSt.742..215M} enables efficient radiative excitation and facilitates its detection. 
The fact that NaCl, as is the case with SiS and CS, remains undetected in R~Dor proves again the vastly distinct chemistry in both objects where (subtle) differences in pulsation-induced shocks, optical thickness, and gas-grain chemistry can dictate the chemical structure in the wind. As pointed out by \citet{Quintana2016ApJ...818..192Q} in their study of the NaCl emission in the ALMA data of the carbon-rich AGB star CW~Leo, NaCl can be thermalised in high-density regions, and as such serve as a good density tracer in collisional dominated regions. This fact is used in Sect.~\ref{Sec:spatial} when giving a first view on the morphological structure of IK~Tau.

We cannot confirm earlier claims by \citet{Velilla2017A&A...597A..25V} on the detection with the IRAM-30\,m telescope of HCO$^+$ and NO in IK~Tau. There is a potential hint of a double peaked line profile at the rest-frequency of the H$_2$CO ($5_{1,5}-4_{1,4}$) line at 351.768\,GHz, but both for the red and blue shifted wing only one spectral point is around the 3--4$\times\sigma_{\mathrm{chn\_rms}}$ level.
A potential reason might be a pulsation-phase dependent formation of those molecules, but a follow-up study is needed to confirm this suggestion.
 Several hyperfine components of NS are detected around 348.51\,GHz.

The minor isotopes of silicon ($^{29}$Si and $^{30}$Si),  oxygen ($^{18}$O), and sulphur ($^{33}$S and $^{34}$S) are traced in both targets. A first-order analysis of the isotope ratios is presented in Sect.~\ref{Sect:disc_isotopes}.

\subsection{First view on the molecular morphology} \label{Sec:spatial}

\begin{figure*}[!htbp]
\begin{minipage}[t]{.33\textwidth}
       \centerline{\resizebox{\textwidth}{!}{\includegraphics[angle=0]{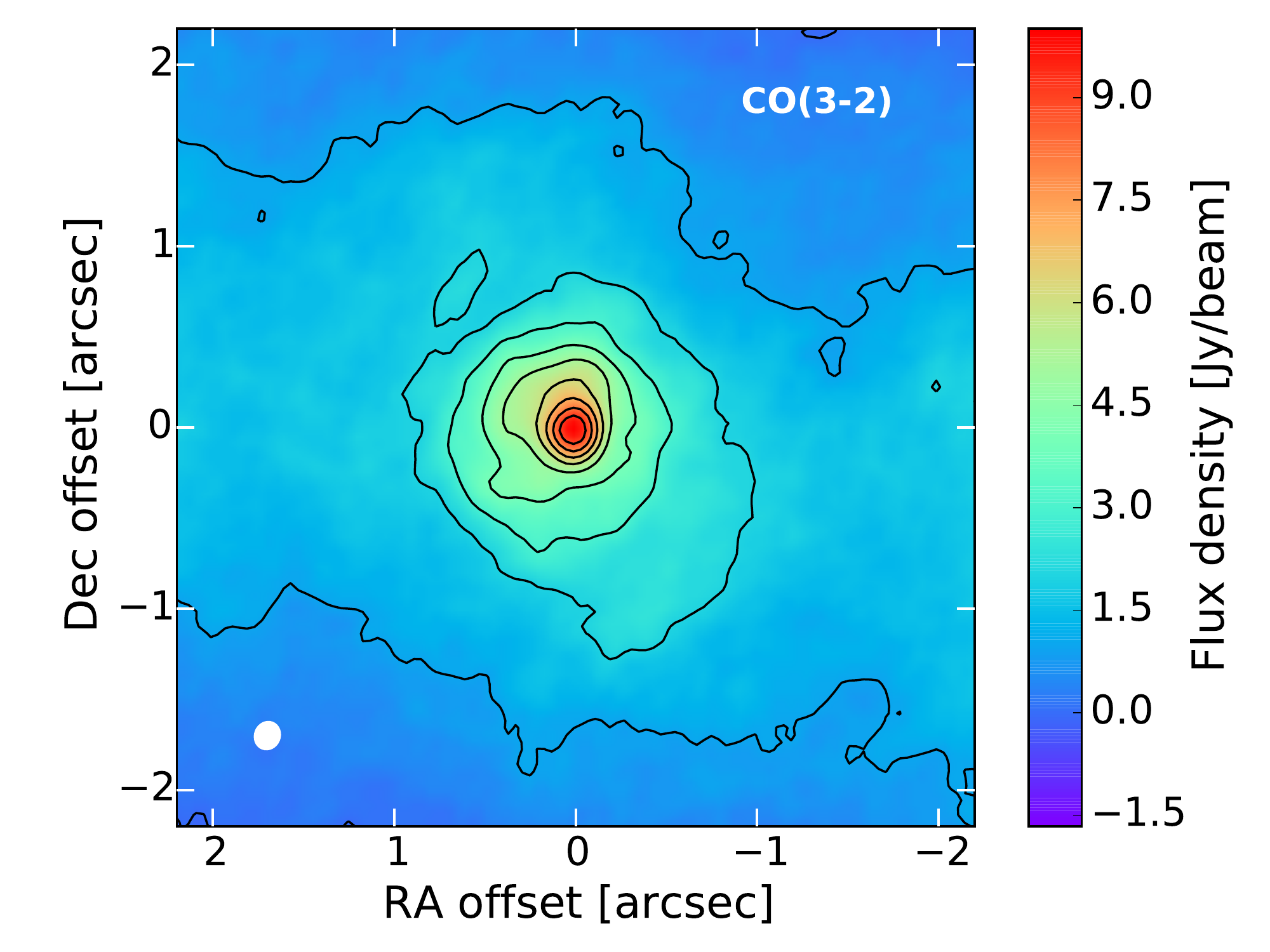}}}
    \end{minipage}
    \hfill
\begin{minipage}[t]{.33\textwidth}
       \centerline{\resizebox{\textwidth}{!}{\includegraphics[angle=0]{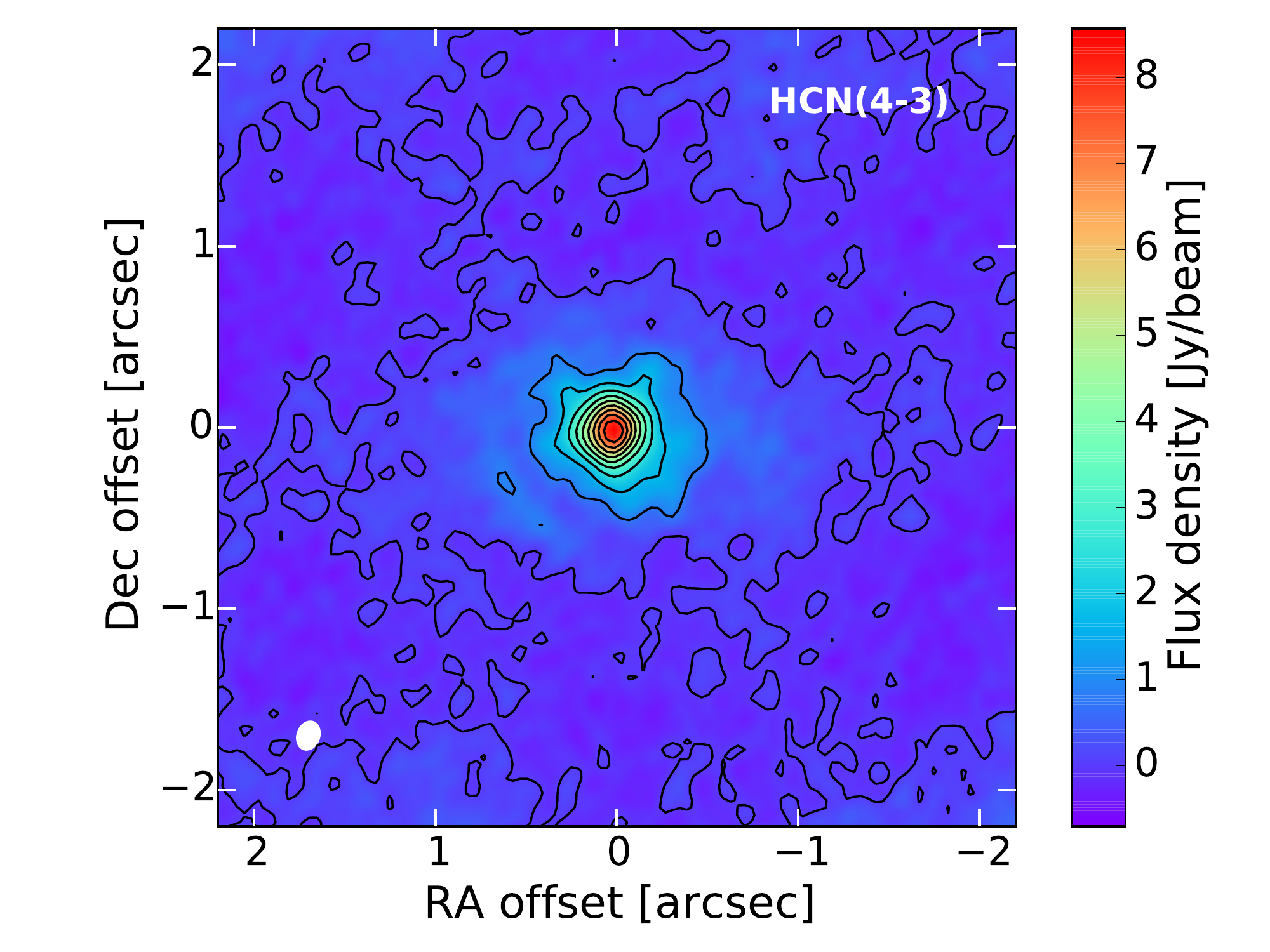}}}
    \end{minipage}
    \hfill
    \begin{minipage}[t]{.33\textwidth}
       \centerline{\resizebox{\textwidth}{!}{\includegraphics[angle=0]{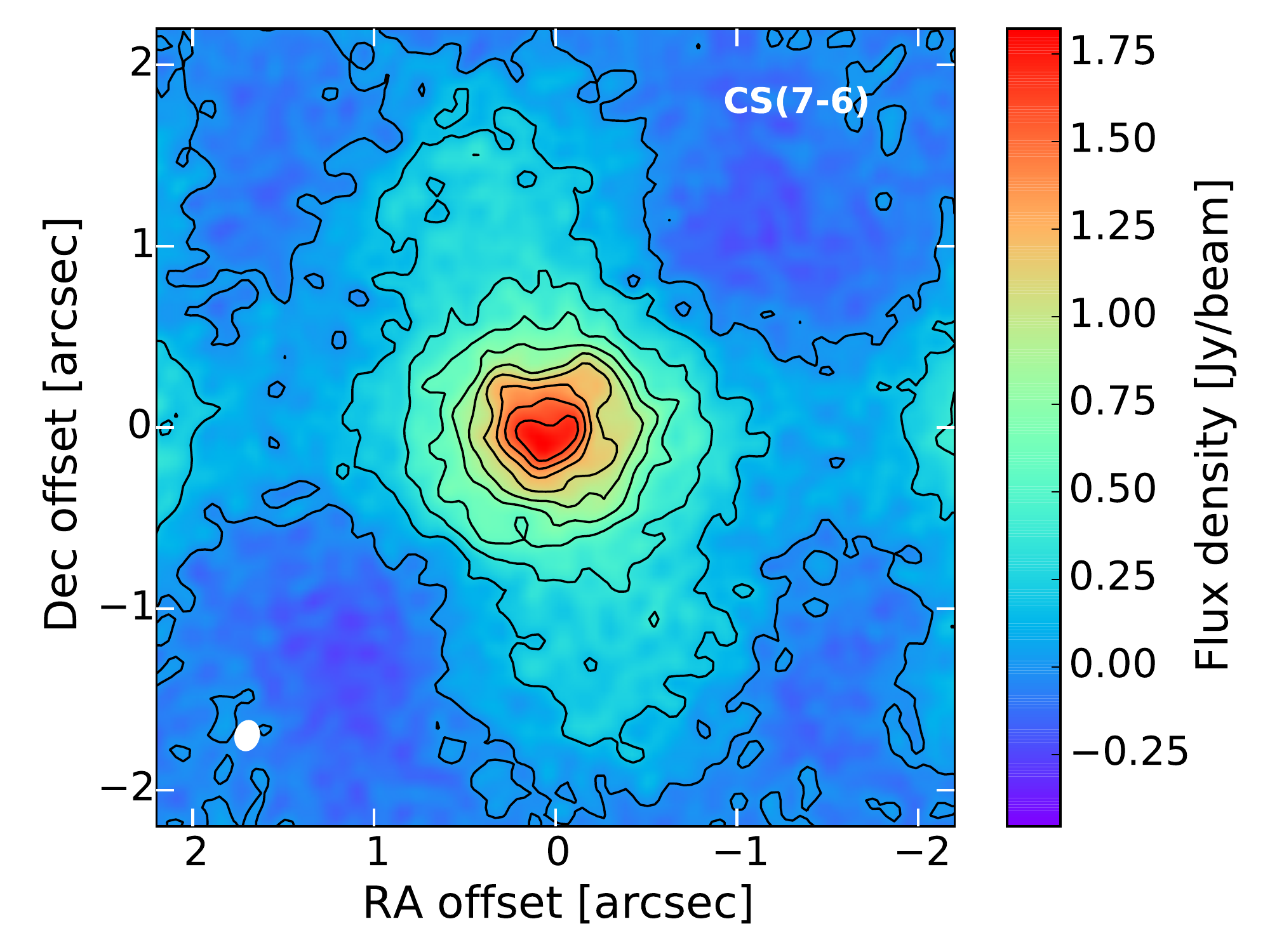}}}
    \end{minipage}

\begin{minipage}[t]{.33\textwidth}
       \centerline{\resizebox{\textwidth}{!}{\includegraphics[angle=0]{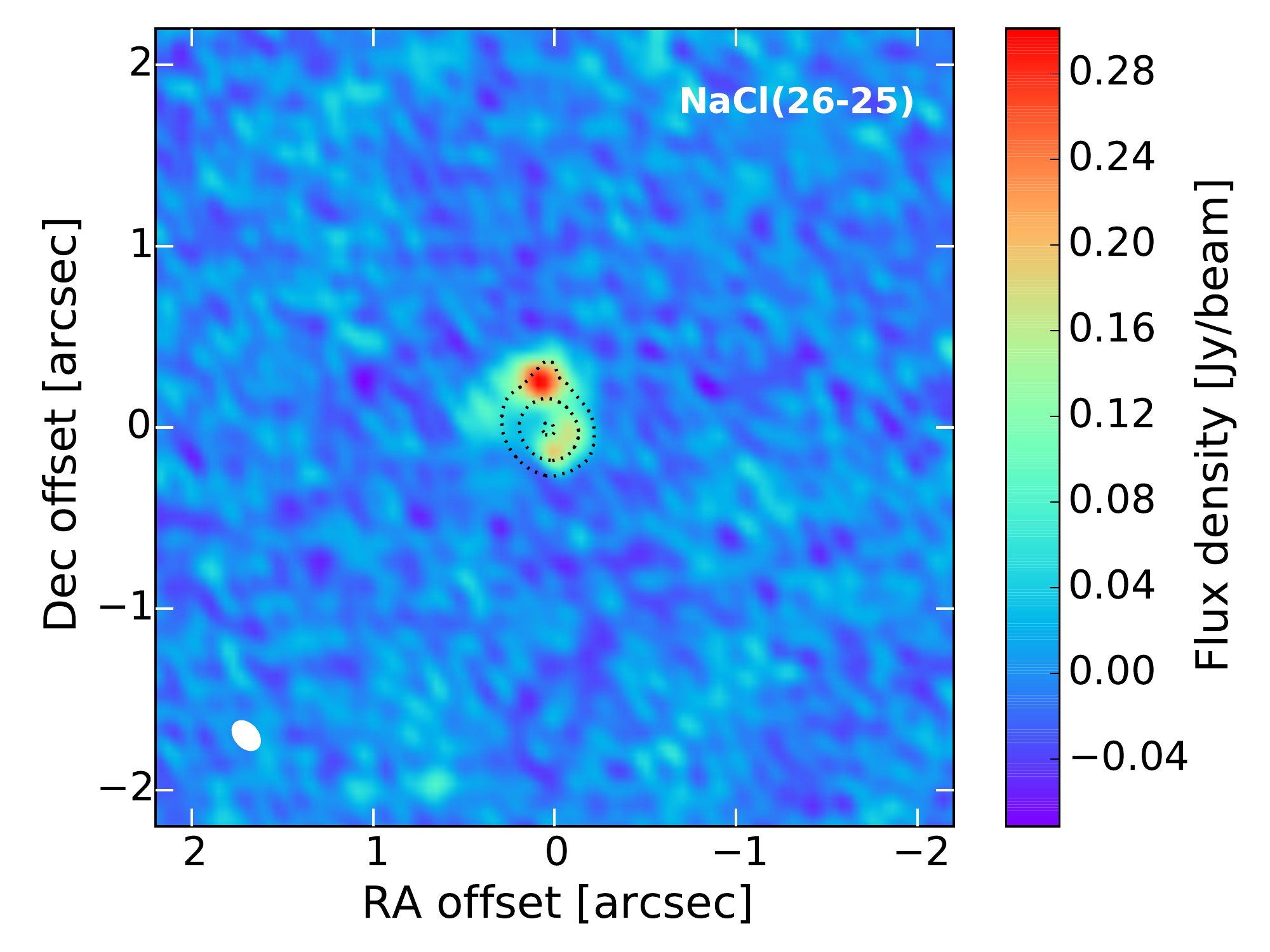}}}
    \end{minipage}
        \hfill
\begin{minipage}[t]{.33\textwidth}
       \centerline{\resizebox{\textwidth}{!}{\includegraphics[angle=0]{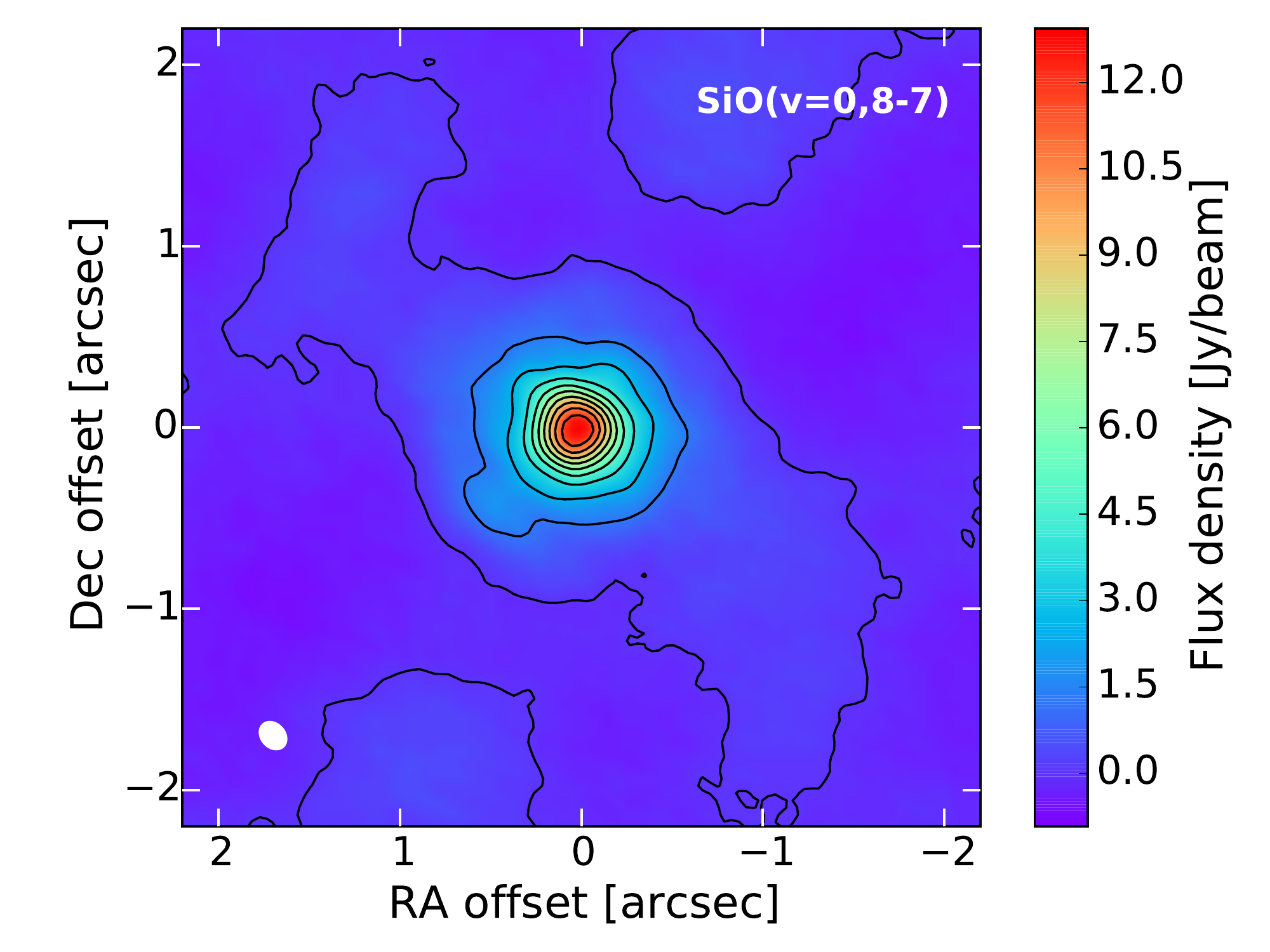}}}
    \end{minipage}
    \hfill
\begin{minipage}[t]{.33\textwidth}
       \centerline{\resizebox{\textwidth}{!}{\includegraphics[angle=0]{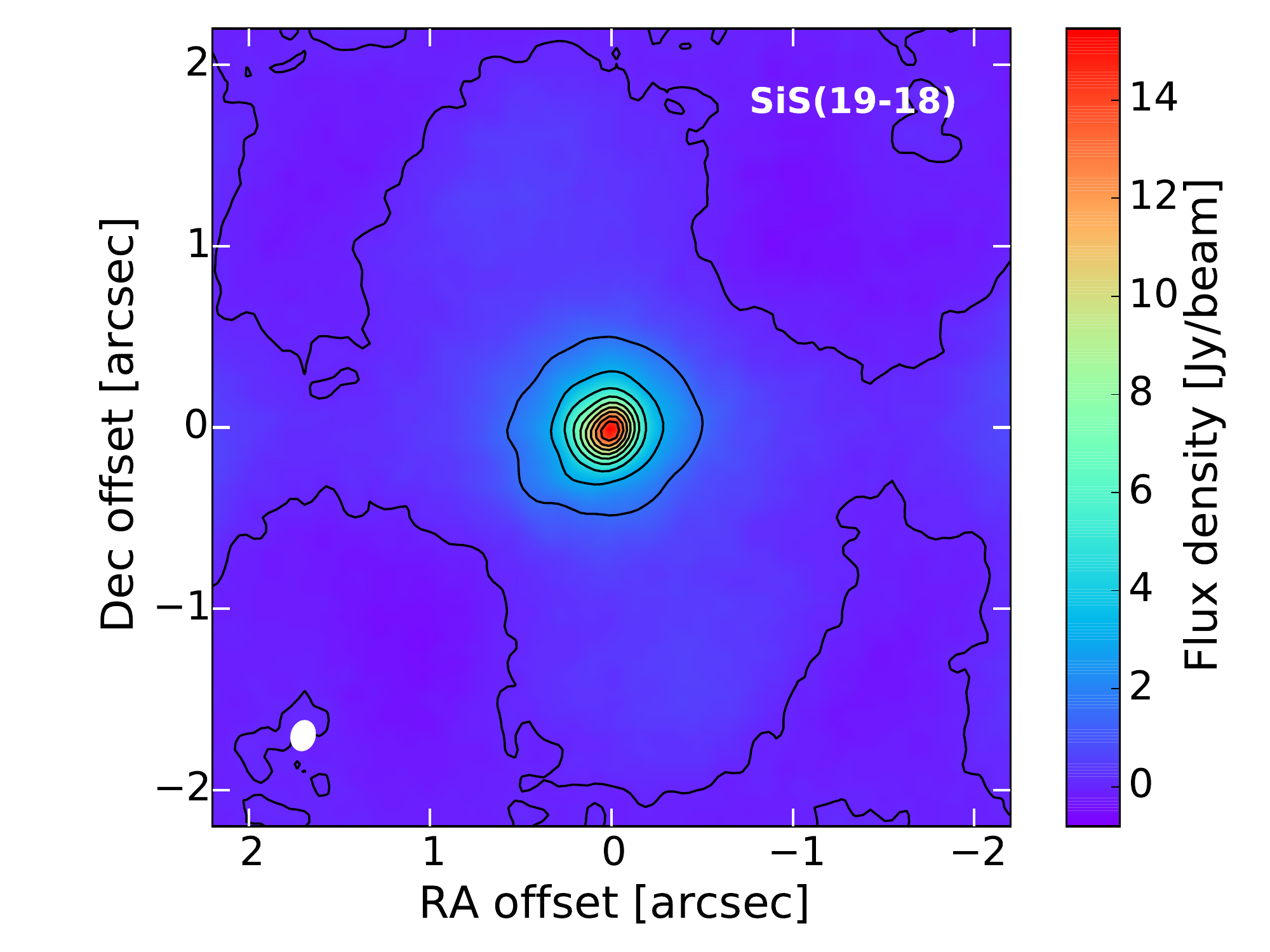}}}
    \end{minipage}
  \caption{Total intensity (zeroth moment) maps for some bright lines in IK~Tau. 
For all lines, except NaCl(26-25), the line contours are given, each step being 10\% of the peak flux density. For NaCl(26-25), we demonstrate the arcs offset from the stellar position by showing in dotted black line the dust continuum contours at 1, 10, and 90\% of the peak continuum emission. The contrast in the figure is best visible on screen.}
  \label{Fig:IKTau_Mom0}
\end{figure*}

\begin{figure*}[!htbp]
\begin{minipage}[t]{.33\textwidth}
       \centerline{\resizebox{\textwidth}{!}{\includegraphics[angle=0]{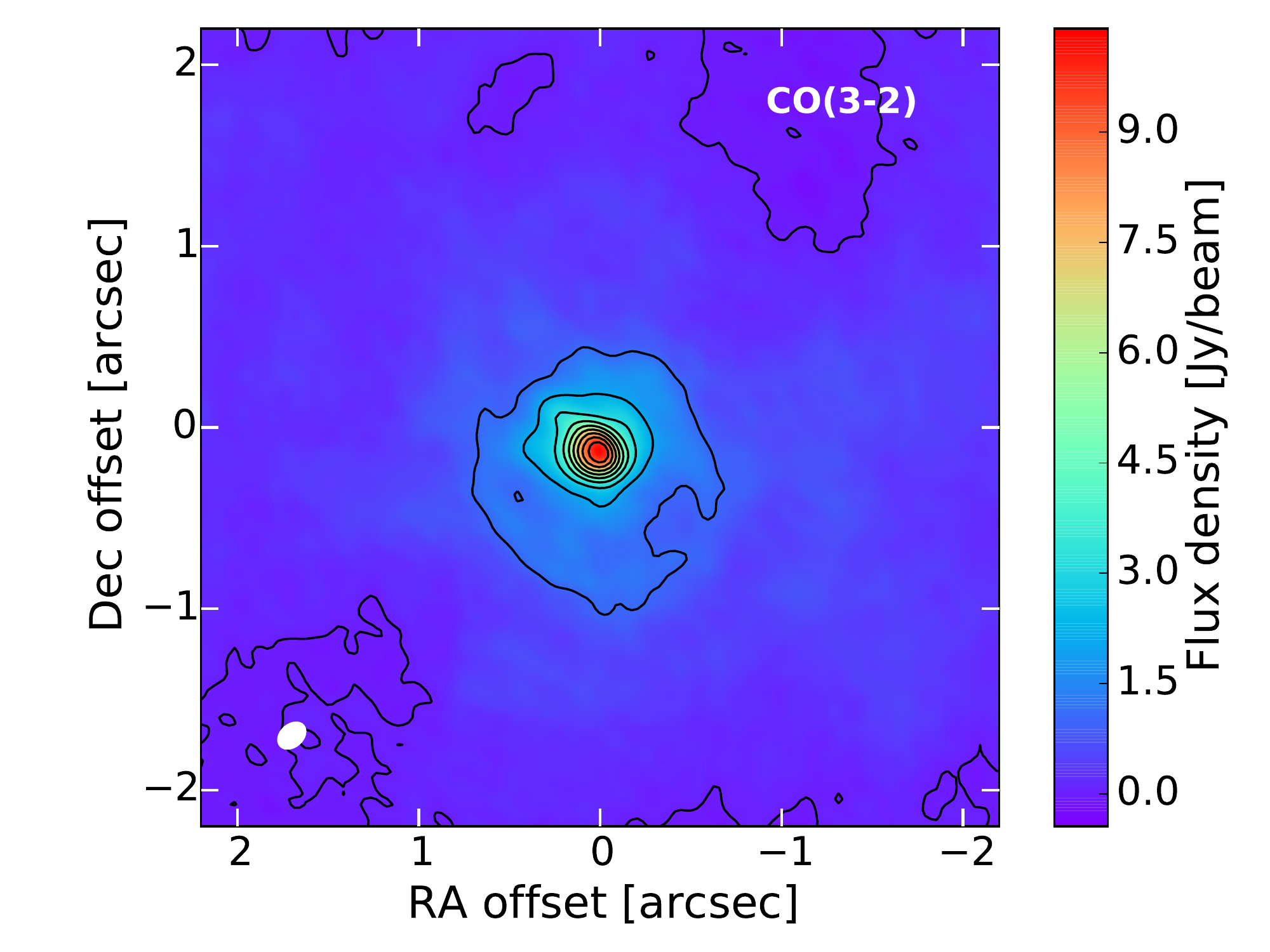}}}
    \end{minipage}
    \hfill
\begin{minipage}[t]{.33\textwidth}
       \centerline{\resizebox{\textwidth}{!}{\includegraphics[angle=0]{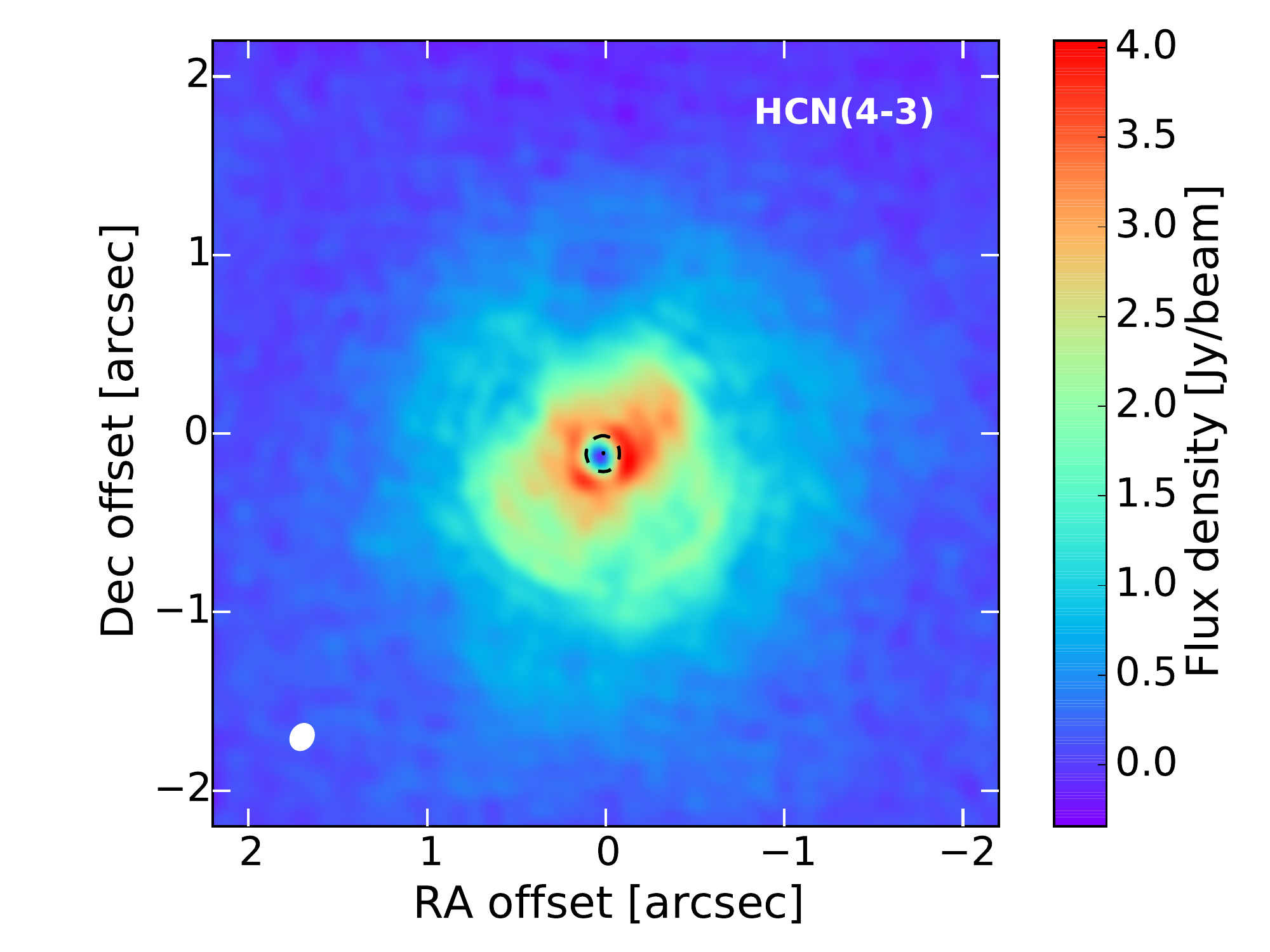}}}
    \end{minipage}
    \hfill
\begin{minipage}[t]{.33\textwidth}
       \centerline{\resizebox{\textwidth}{!}{\includegraphics[angle=0]{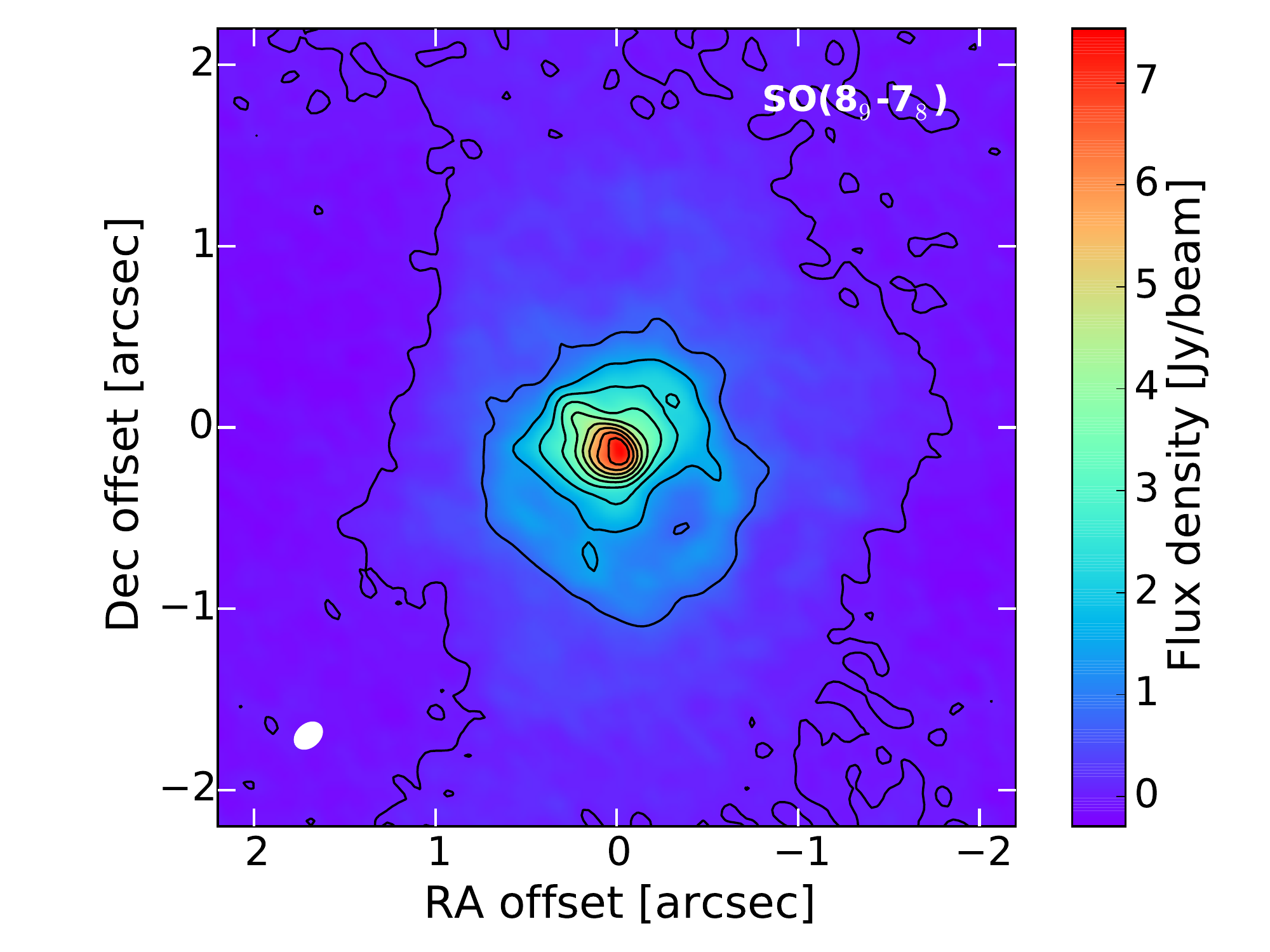}}}
    \end{minipage}
    
\begin{minipage}[t]{.33\textwidth}
       \centerline{\resizebox{\textwidth}{!}{\includegraphics[angle=0]{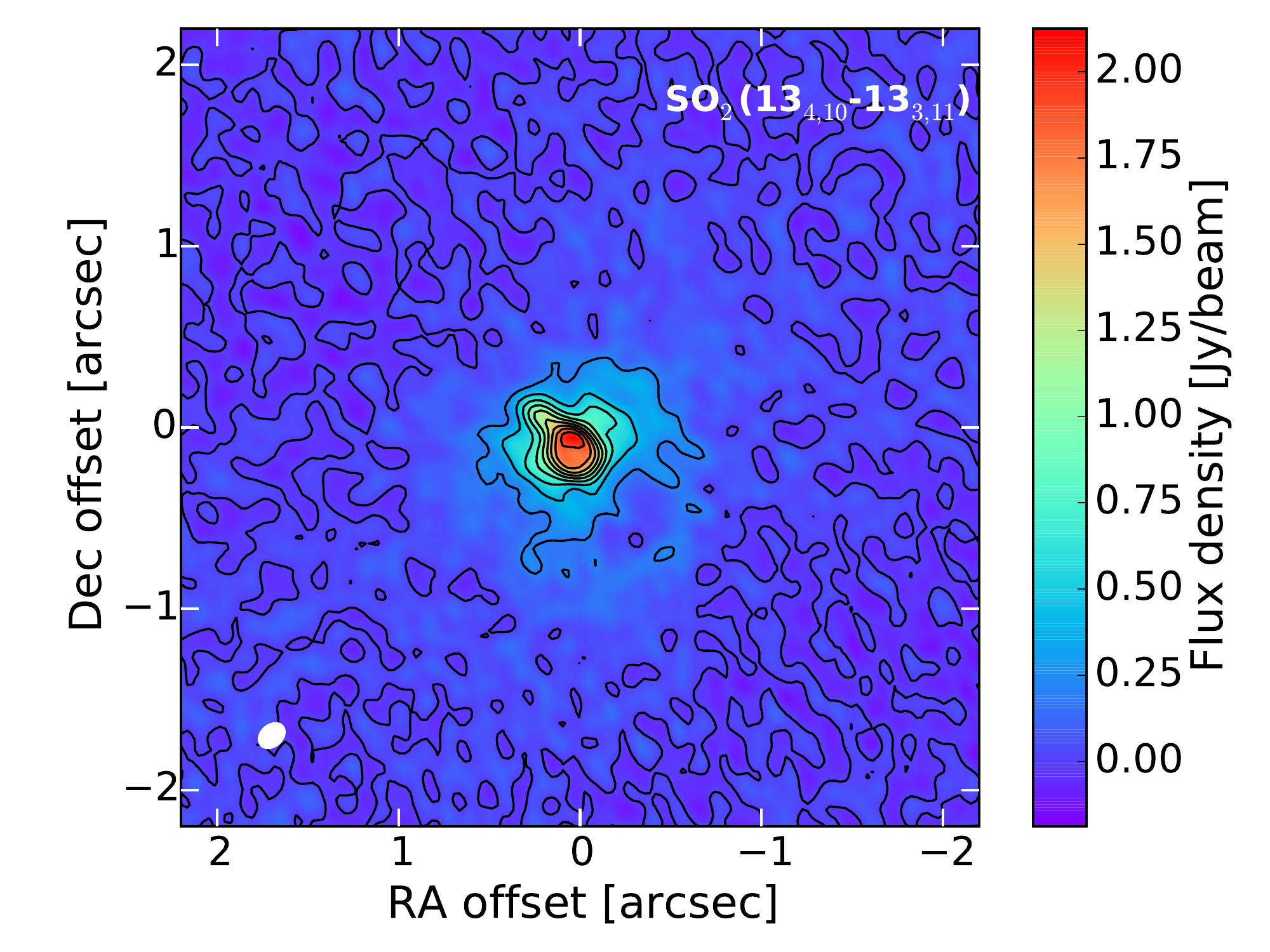}}}
    \end{minipage}
    \hfill
\begin{minipage}[t]{.33\textwidth}
       \centerline{\resizebox{\textwidth}{!}{\includegraphics[angle=0]{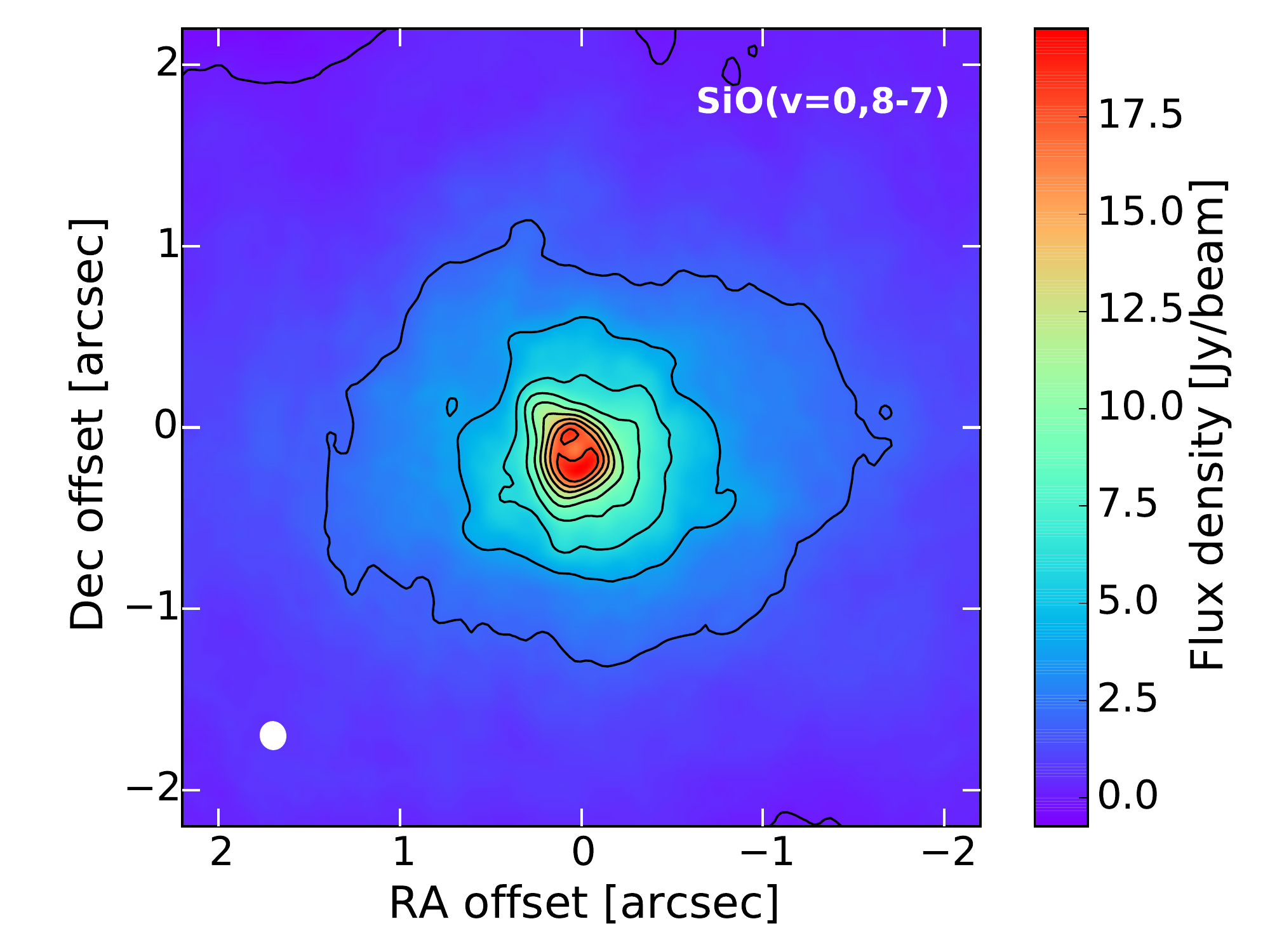}}}
    \end{minipage}
    \hfill
\begin{minipage}[t]{.33\textwidth}
       \centerline{\resizebox{\textwidth}{!}{\includegraphics[angle=0]{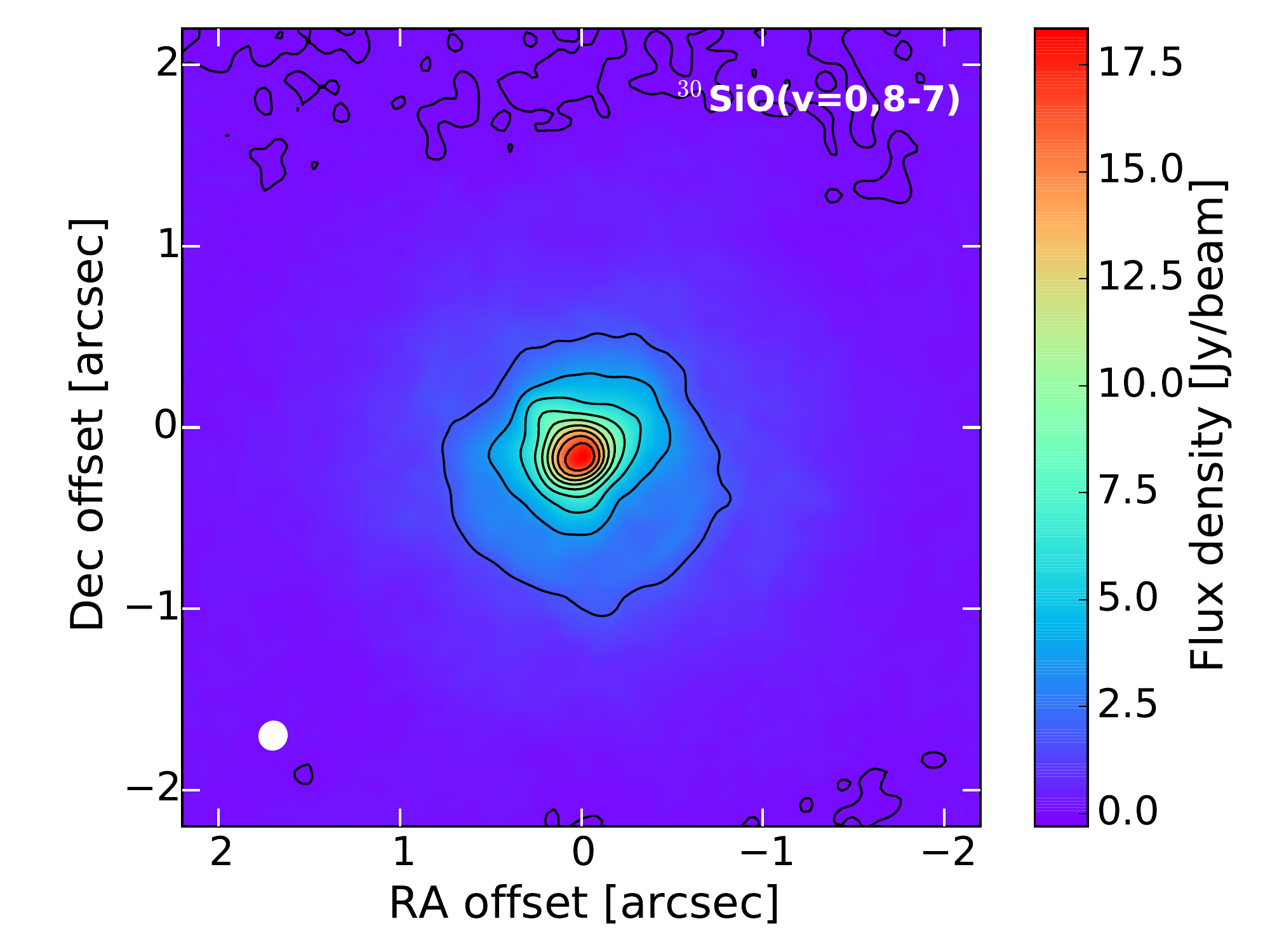}}}
    \end{minipage}
  \caption{Total intensity (zeroth moment) maps for some bright lines in R~Dor. 
For all lines, except HCN(4-3), the line contours are given, each step being 10\% of the peak flux density. For HCN(4-3), we demonstrate the position of the `blue hole' by showing  by showing in dotted black line the dust continuum contours at 1, 10, and 90\% of the peak continuum emission. The contrast in the figure is best visible on screen.}
  \label{Fig:RDor_Mom0}
\end{figure*}

The channel maps and zero-moment maps testify that the excitation region of most molecular transitions is probing a complex spatio-kinematic density distribution (see Fig.~\ref{Fig:IKTau_Mom0} and Fig.~\ref{Fig:RDor_Mom0}). These specific emission patterns will be addressed in some forthcoming papers \citep[or have already been published in the case of the Al-bearing molecules,][]{Decin2017arXiv170405237D}. This section serves to give a first snapshot of some of the interesting features seen in both stars.

\begin{figure*}[!htbp]
\begin{minipage}[t]{.44\textwidth}
       \centerline{\resizebox{\textwidth}{!}{\includegraphics[angle=0]{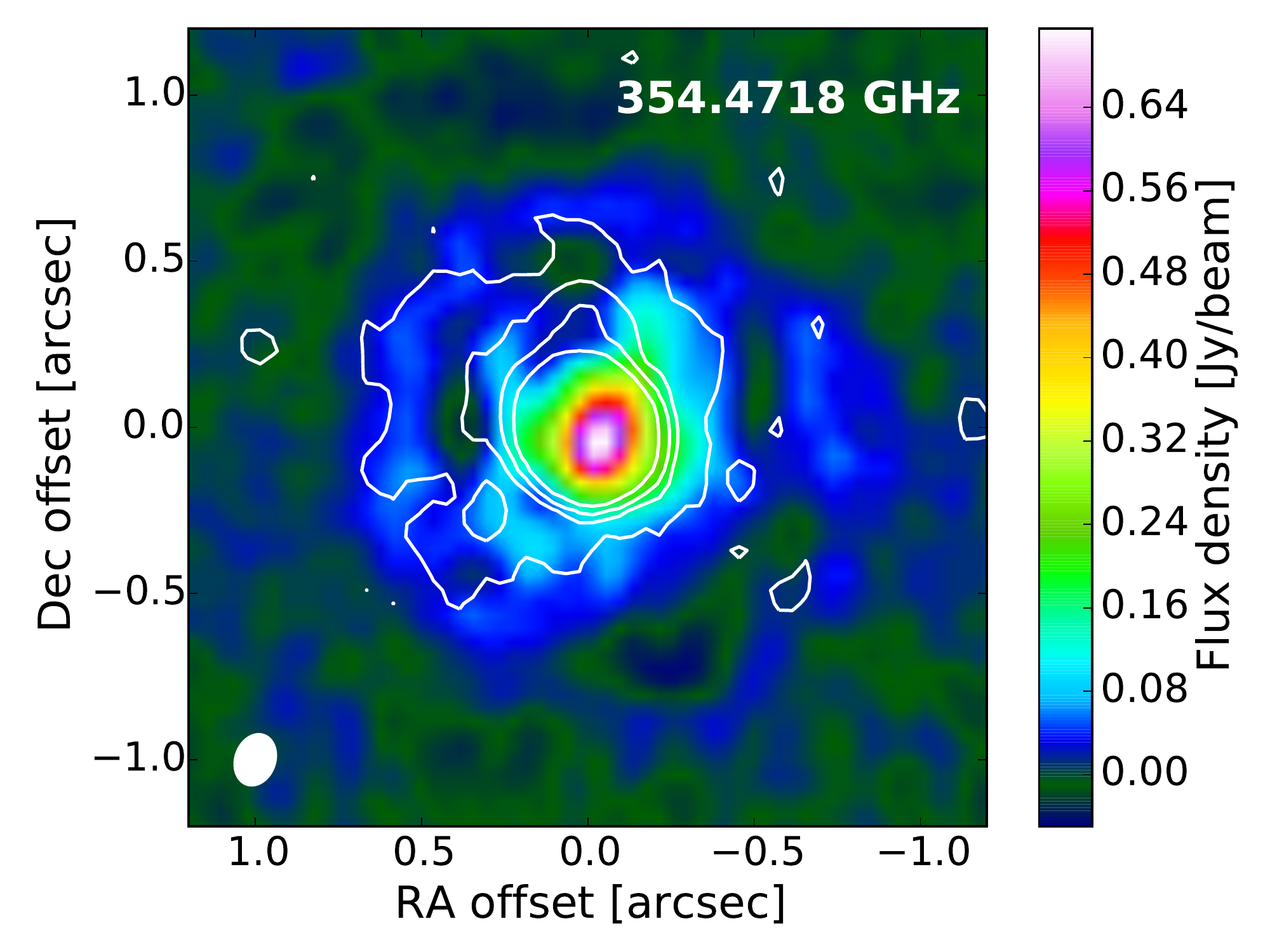}}}
    \end{minipage}
    \hfill
\begin{minipage}[t]{.44\textwidth}
       \centerline{\resizebox{\textwidth}{!}{\includegraphics[angle=0]{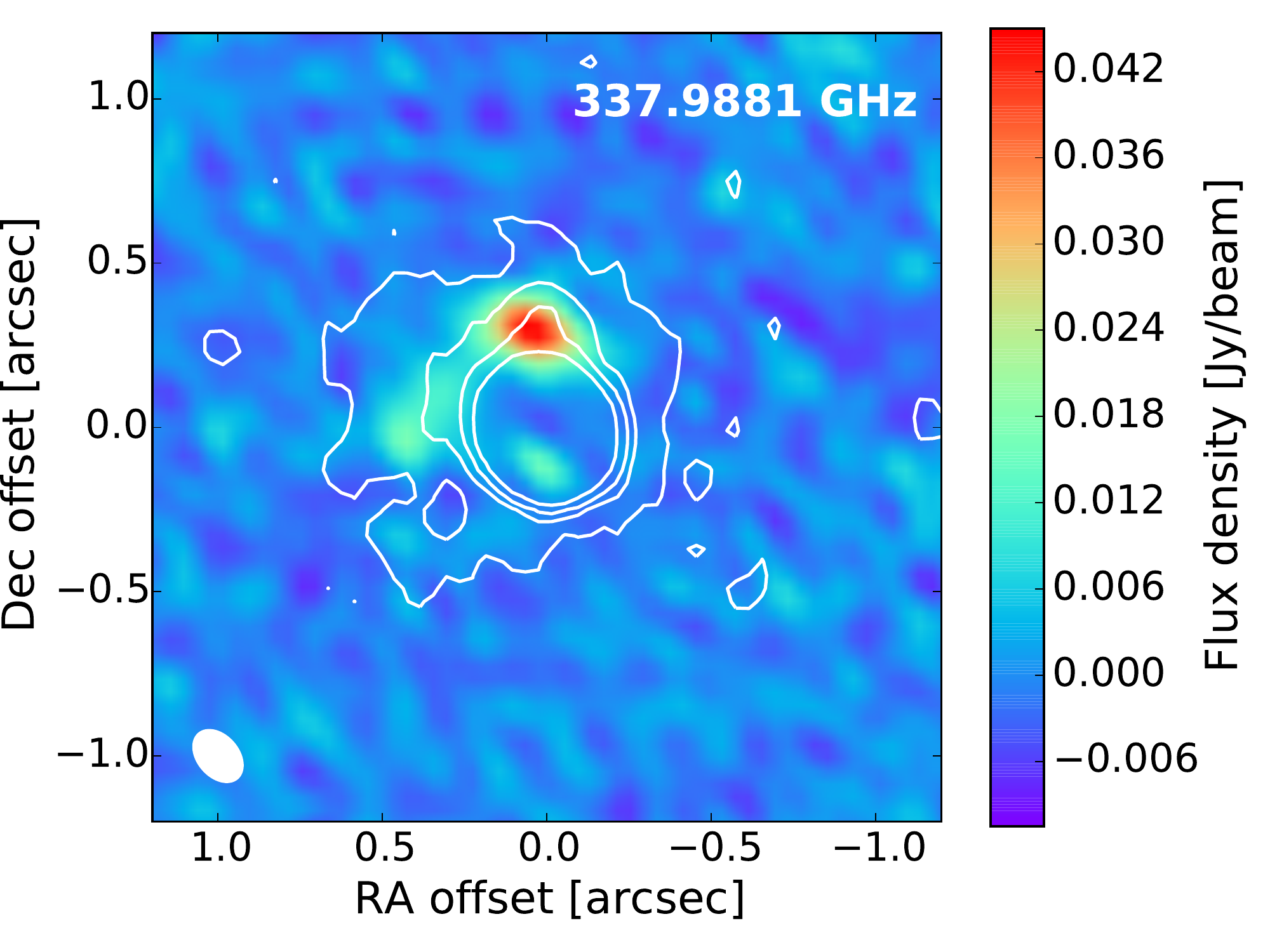}}}
    \end{minipage}
  \caption{Comparison between the continuum image of IK~Tau (white contours at (1,3,5,10)$\times$0.141\,mJy beam$^{-1}$ (3$\sigma_{\rm{rms}}$)) and
  the molecular emission of HCN(4-3) (left panel) and NaCl(26-25) (right panel) showing the co-location of the dust continuum emission and the molecular emission. The molecular emission is shown for a specific frequency not corrected for the $v_{\rm{LSR}}$ as given in the upper right corner of each panel.}
  \label{Fig:IKTau-mol-dust}
\end{figure*}

\paragraph{IK Tau:} With the exception of NaCl, the intensity of most other molecules is centrally peaked at the position of the continuum peak (see Fig.~\ref{Fig:IKTau_Mom0}). As has been discussed by \citet{Decin2016A&A...592A..76D}, NaCl is a favourable molecule to study the morphological structure in AGB stars. The fractional abundance of NaCl (w.r.t.\ total hydrogen content) is very low for temperatures above 1500\,K \citep{Milam2007ApJ...668L.131M, Decin2016A&A...592A..76D} explaining its non-detection at the stellar position.
The total intensity map of NaCl in Fig.~\ref{Fig:IKTau_Mom0} and the channel map in Fig.~\ref{Fig:IKTau_NaCl_channel} show the first indication of an arc-like structure at a distance of $\sim$0.4\arcsec\ from the central star. Other arcs are visible in the channel and total intensity maps of other molecules, often at further distances from the central star (see, e.g. HCN in Fig.~\ref{Fig:IKTau_Mom0} and its channel map in Fig.~\ref{Fig:IKTau_HCN_channel}). As discussed by Decin et al.\ (\textit{in prep.}) these arcs might be part of a spiral structure embedded within the smooth stellar outflow. The realisation that these arcs are present in the molecular data, also helps in interpreting the continuum data. As shown in Fig.~\ref{Fig:IKTau-mol-dust}, the gas and dust are co-located and the extended structure in the continuum map in the north-east direction can be explained by density enhancements related to this spiral structure.

\begin{figure*}[!htbp]
\begin{minipage}[t]{.48\textwidth}
        \centerline{\resizebox{\textwidth}{!}{\includegraphics[angle=180]{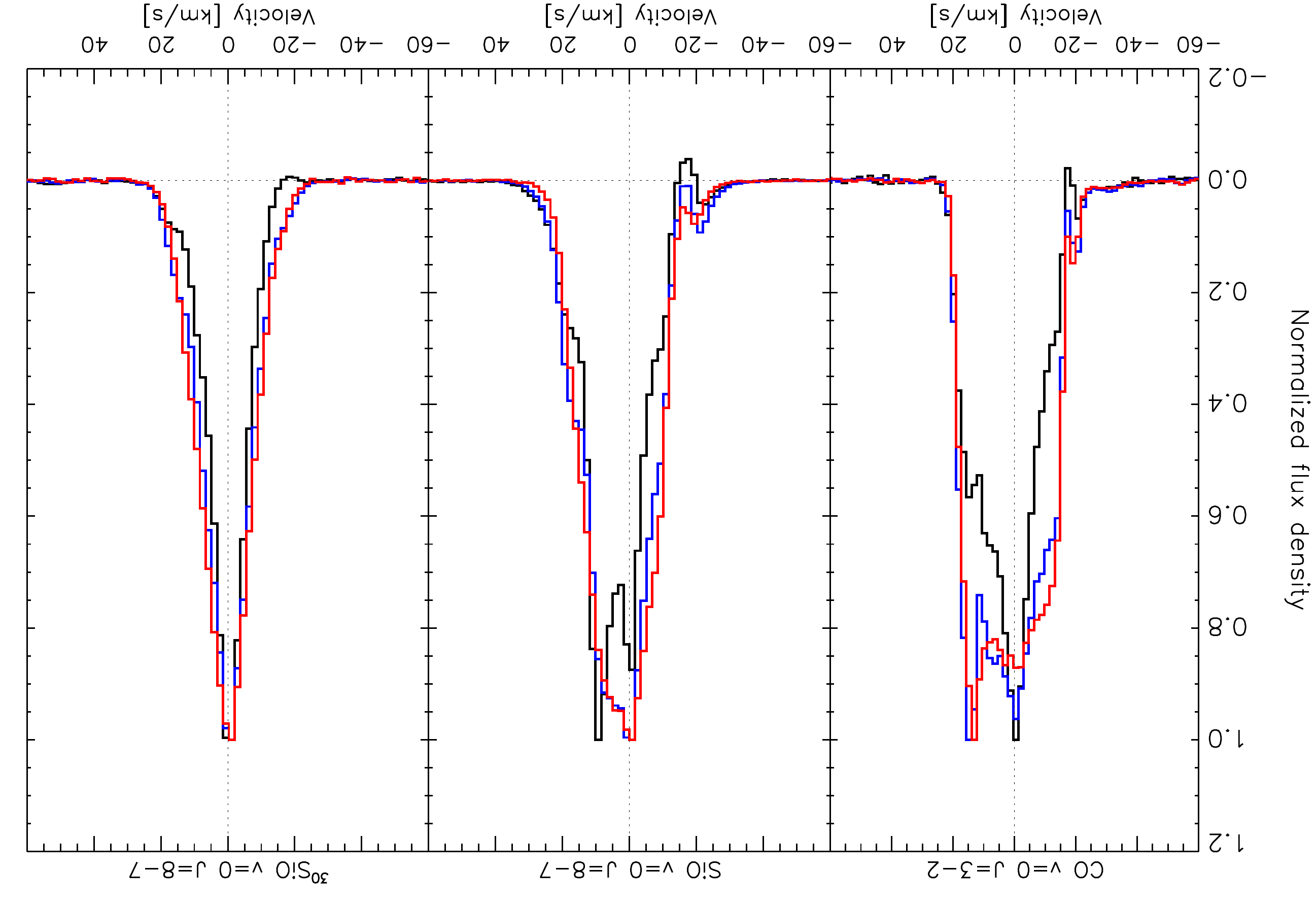}}}
        \centerline{\resizebox{\textwidth}{!}{\includegraphics[angle=0]{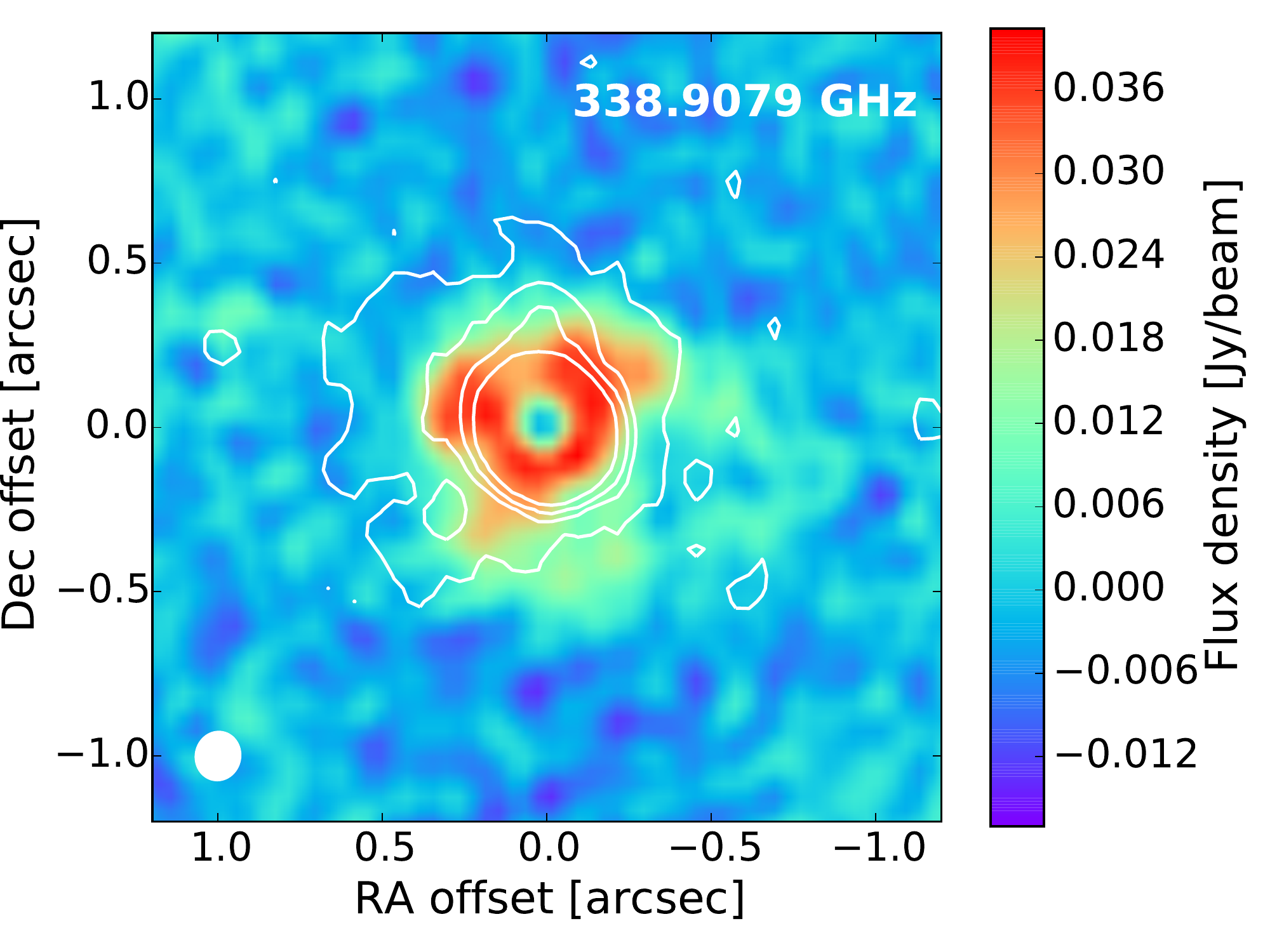}}}
    \end{minipage}
    \hfill
\begin{minipage}[t]{.48\textwidth}
        \centerline{\resizebox{\textwidth}{!}{\includegraphics[angle=180]{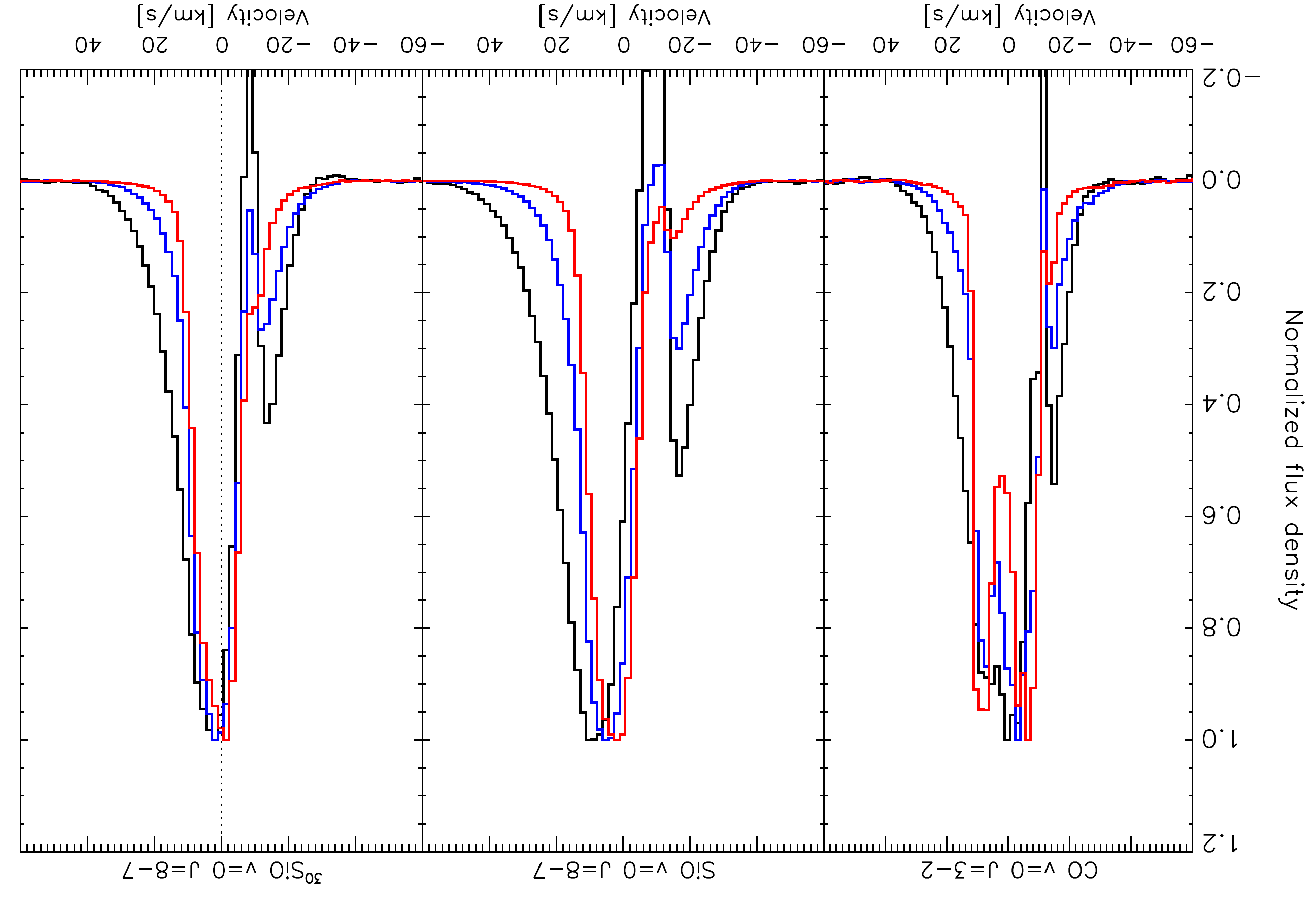}}}
        \centerline{\resizebox{\textwidth}{!}{\includegraphics[angle=0]{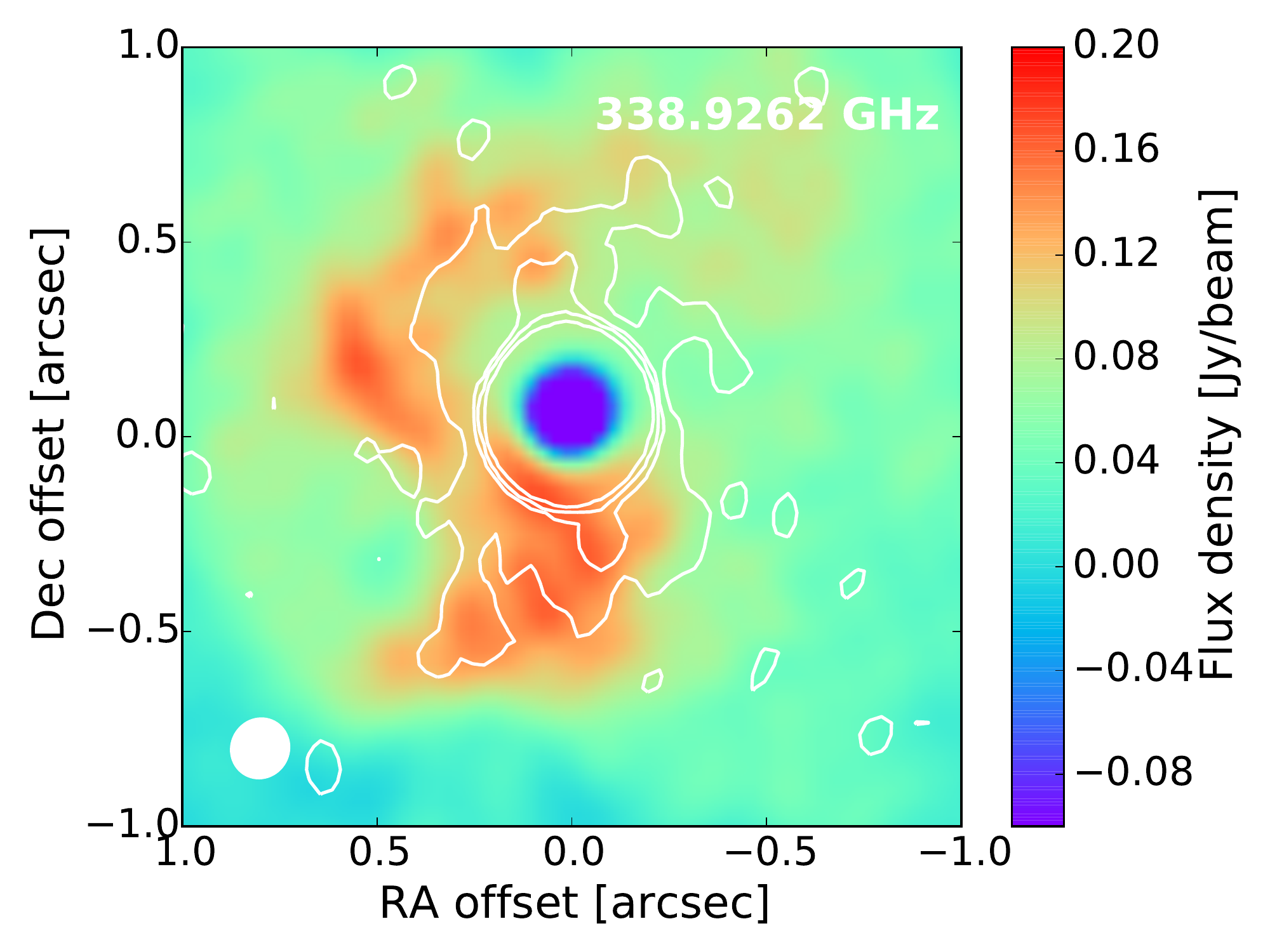}}}
    \end{minipage}
\caption{\textit{Upper panels:} Demonstration of the absorption feature in the blue wing of the CO v=0 J=3-2, $^{28}$SiO v=0 J=8-7, and $^{30}$SiO v=0 J=8-7 line in IK~Tau (left) and R~Dor (right) for an extraction aperture with radius $\sim$80\,mas (black), $\sim$300\,mas (blue), and 800-1000\,mas (red).  The smaller the extraction aperture, the better visible is the blue absorption feature. \textit{Bottom panels:} The channel map demonstrates the `blue hole' on the stellar position for  $^{30}$SiO v=0 J=8-7 (at a frequency not corrected for the v$_{\rm{LSR}}$) for IK~Tau (left) and R~Dor (right). For reference, the continuum contours are shown in white at (1,3,5,10)$\times$0.141mJy/beam.}
\label{Fig:Blue_hole}
\end{figure*}

\paragraph{R Dor:} As can be seen in the HCN zeroth moment map of R~Dor, a the `horseshoe'-like structure is visible with a bright bended feature pointing north-west out to a distance of $\sim$1.12\arcsec (see also the HCN(4-3) channel maps in Fig.~\ref{Fig:RDor_HCN_channel}). Other features and blobs are visible at smaller distances from the star.  As discussed by \citet{Homan2018}, the stereograms of R~Dor argue for the presence of a compact disk with total extent of $\sim$30\,AU (or 0.5\arcsec). It is noteworthy that we see a `hole' at the location of the star for some of the molecules (see, e.g. HCN(4-3) in Fig.~\ref{Fig:RDor_Mom0}).  The ALMA data allow us to detect an absorption feature in the blue wing for the first time (see also Fig.~\ref{Fig:Blue_hole} in Sect.~\ref{Sec:kinematic}). This feature arises from impact parameters that cross the central star (see bottom panels in Fig.~\ref{Fig:Blue_hole}) yielding a typical blue-shifted absorption profile, in analogy with the well-known P-Cygni profiles that trace outflowing material in a medium for which the  source function decreases with distance\footnote{Infalling material is recognised by an inverse P-Cygni profile, see for example\ \citet{Wong2016A&A...590A.127W}.}. A large ratio between the stellar angular diameter and angular beam size facilitates its detection, and is the reason for the `blue hole' to be more pronounced in R~Dor than in IK~Tau. For larger beams, too much emission from impact parameters not crossing the central star and with the same projected velocities (along the line-of-sight) is picked up and fills in the absorption hole. This blue hole is also the reason for, for example, the radial profile of HCN 4--3 and SiO v=0 J=8--7 not being centrally peaked in Fig.~\ref{Fig:RDor_RadProf}.

\subsection{Wind kinematics} \label{Sec:kinematic}


 \begin{figure*}[!htbp]
\sidecaption
\includegraphics[angle=0, width=120mm]{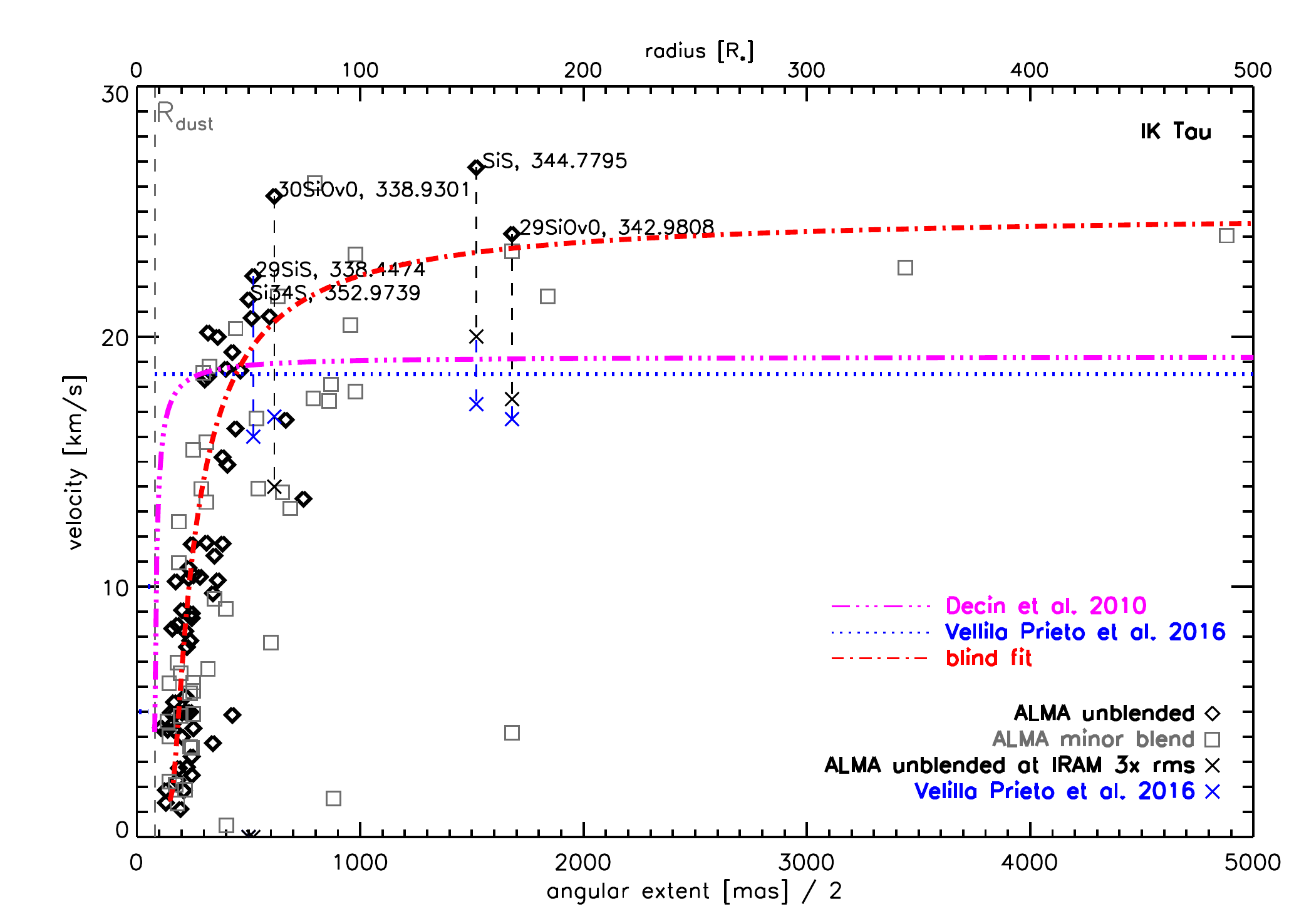}
\caption{Measured wind velocities for IK~Tau (as determined from the half-line width at zero intensity) versus half of the spatial FWHM $s_{\rm{FWHM}}$ (representing the dominant line formation region). Unblended lines are indicated with a diamond, (minor) blended lines with a grey squared box. The velocity profile deduced by \citet{Decin2010A&A...516A..69D} is shown in magenta (dashed-triple dotted line), by \citet{Velilla2017A&A...597A..25V} in blue (dotted line), and a blind fit to the unblended data for a fixed stochastic velocity of 1.5\,km/s \citep{Decin2010A&A...516A..69D} in red (dashed-dotted line).}
\label{Fig:velocities_angwid_IKTau}
\end{figure*}

 \begin{figure*}[!htbp]
\sidecaption
\includegraphics[angle=0, width=120mm]{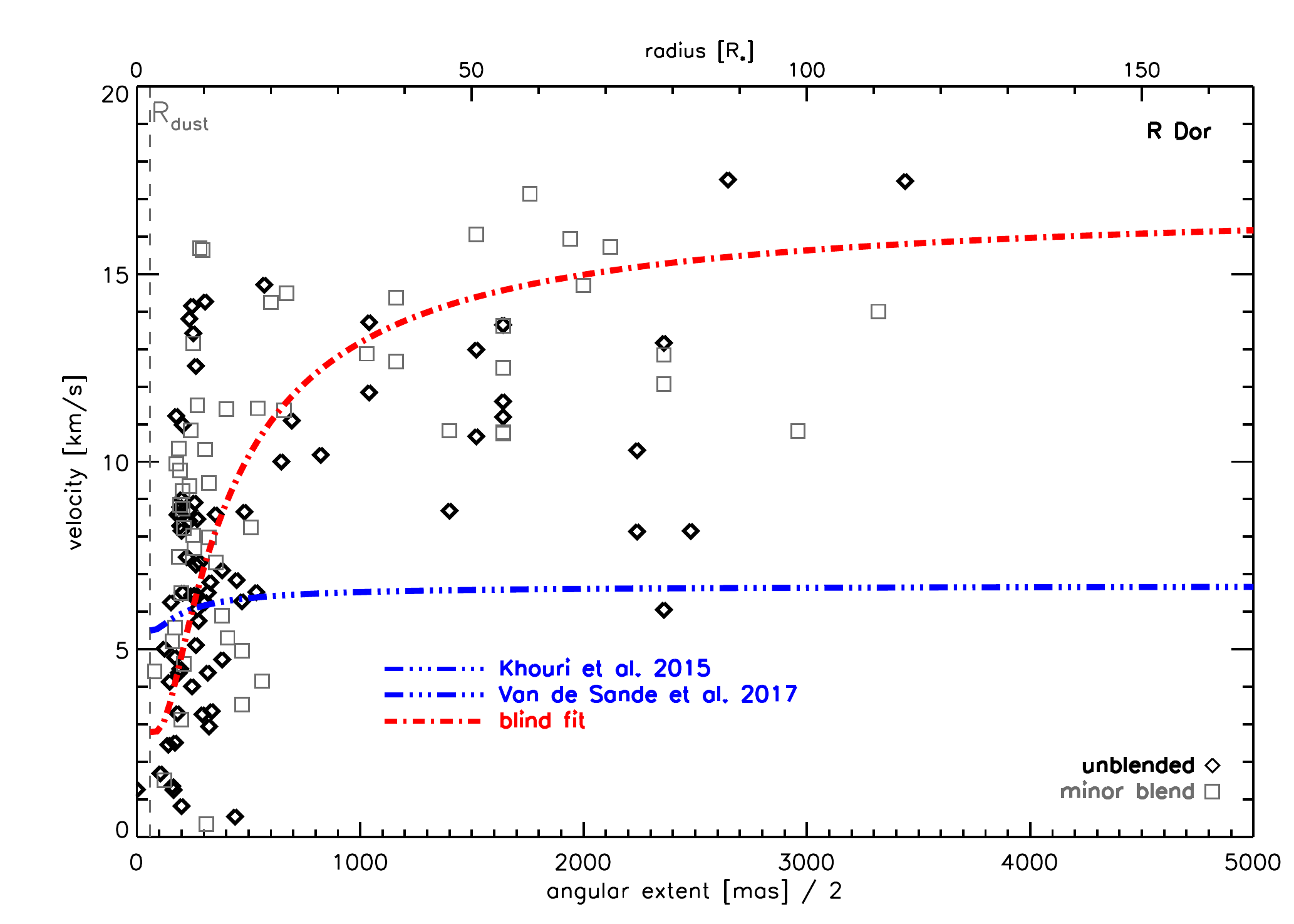}
\caption{Measured wind velocities for R~Dor (as determined from the half-line width at zero intensity) versus half of the spatial FWHM $s_{\rm{FWHM}}$ (representing the dominant line formation region). Unblended lines are indicated with a diamond, (minor) blended lines with a grey squared box. The velocity profile deduced by \citet{KhouriPhD} and \citet{VandeSande2017} is shown in blue (dashed-triple dotted line) and a blind fit to the unblended data for a fixed stochastic velocity of 1\,km/s \citep{VandeSande2017} in red (dashed-dotted line).}
\label{Fig:velocities_angwid_RDor}
\end{figure*}

The ALMA data offer an excellent tool to trace the wind acceleration in the regions where dust condensation is important and the wind velocity might transit from sub-sonic to supersonic values. 
In Figs.~\ref{Fig:velocities_angwid_IKTau} and \ref{Fig:velocities_angwid_RDor} we plot the half-line width at zero intensity as a measurement of the wind velocities as a function of the radial distance as indicated by half of the measured spatial FWHM. These plots immediately show that the retrieved wind velocities are much larger than the combined effect of previously determined gas terminal and turbulent velocity\footnote{The term `terminal velocity' is defined in Sect.~\ref{Sec:Introduction}. The `turbulent velocity' is defined as the local velocity dispersion, equal to the local line width at 1/$e$ relative intensity \citep{Morris1985A&A...142..107M}}. Using the 300\,mas aperture spectra as reference, the maximum velocity traced for IK~Tau is 36\,km/s  and for R~Dor 23\,km/s (in both cases the SiO v=0 J=8-7 line, with an upper state energy of $\sim$52\,cm$^{-1}$ (or 75\,K), displays the largest velocity) with the wind being most strongly accelerated around 100-200\,mas for both stars. Previous studies using single-beam telescopes derive a terminal wind velocity, $v_\infty$, of 17.7\,km/s and a turbulent wind velocity, $v_{\rm{turb}}$, of $\sim$1.5\,km/s for IK~Tau \citep{Decin2010A&A...516A..69D}, while for R~Dor $v_\infty$ is $\sim$5.5\,km/s and $v_{\rm{turb}}$ around 1\,km/s  \citep{KhouriPhD, VandeSande2017}.
In stellar winds, velocity profiles are often parametrised using the simplified so-called $\beta$-type law \citep{Lamers1999isw..book.....L}
\begin{equation}
v(r) = v_0 + (v_\infty - v_0) \left( 1 - \frac{r_{\rm{dust}}}{r}\right)^\beta\,,
\end{equation}
or, alternatively,
\begin{equation}
v(r) \approx v_\infty \left(1-\frac{r_0}{r}\right)^\beta\,
\end{equation}
with
\begin{equation}
r_0 = \Rstar \left\{1-\left(\frac{v_0}{v_\infty}\right)^{1/\beta}\right\}
\end{equation}
with $r$ being the distance to the star and $v_0$ the velocity at the dust condensation radius $r_{\rm{dust}}$. A slower acceleration until larger distances is indicated by a larger value of $\beta$.
The parametrised function derived by \citet{Decin2010A&A...516A..69D} in the case of IK~Tau ($\beta$\,=\,1) and by \citet{KhouriPhD} and \citet{VandeSande2017} for R~Dor ($\beta$\,=\,5) are shown in Figs.~\ref{Fig:velocities_angwid_IKTau} and~\ref{Fig:velocities_angwid_RDor}, respectively. Since, as we will see below, the data suggest a slower acceleration, we have plotted, just to guide the eye, a model with a large but arbitrary value of $\beta$\,=\,10. 
How are we to understand these large velocities? Do the ALMA data really indicate that the terminal wind velocity is larger than previously determined? And/or do the high values for the (logarithmic) velocity gradient infer a radiation driven wind mechanism more complex than generally deduced from the momentum equation?

Some lines detected in our survey of IK~Tau were also observed by \citet{Velilla2017A&A...597A..25V} using the IRAM-30m telescope. The velocities derived using the IRAM data are much lower compared to our values (see blue crosses in Fig.~\ref{Fig:velocities_angwid_IKTau}). However, the IRAM data have a much higher rms noise value of 0.0128\,K (or 0.14\,Jy). If we were to determine the line widths of the brightest lines using the 3$\sigma$-values of IRAM as the limit, the velocities would shift down considerably (see black crosses in Fig.~\ref{Fig:velocities_angwid_IKTau}) and would be more in line with the previously determined wind velocity profiles and gas terminal velocity. It is clear that ALMA's increased sensitivity allows us to detect emission over a greater velocity range than IRAM.

\paragraph{Blue wing line profile and stochastic velocity:}
Looking at the line profiles of some molecular transitions that exhibit clearly large velocities, we can notice some distinct features (see Fig.~\ref{Fig:Blue_hole}), such as asymmetric profiles with a blue wing absorption (see also Sect.~\ref{Sec:spatial}) and broad wings (both in the red and blue, but better visible in the blue due to the blue wing absorption). The effect is more pronounced for R~Dor than for IK~Tau, and for the latter is mainly visible in the brightest lines. Even for a smooth 1D stellar wind, it is well known that the line profiles for optically thick lines are often asymmetric, the effect often being dubbed blue wing self-absorption\footnote{We note that the term `blue wing self-absorption' is different from the term `blue wing absorption' discussed in Sect.~\ref{Sec:spatial}.}. The physical cause for this asymmetry is that non-local effects enhance the intensity in the red wing of the profile depressing the intensity in the blue wing since the ($\tau$\,=\,1)-layer is reached in warmer regions for the red-shifted part of the wind as compared to the blue-shifted regions \citep{Morris1985A&A...142..107M}. Another characteristic is the appearance of an emission feature in the blue wing of an optically thick line with extended scattering zone in cases that the telescope beam can spatially resolve the outer line-excitation regions \citep{Schonberg1988A&A...195..198S}\footnote{Calculations done for resolution on arcsec scale}. The broader the line scattering zone, that is,\ the larger the stochastic velocity, the larger  the emission feature. Other physical causes such as the temperature gradient and opacity have an effect on the strength of the feature as well.

For both stars, the blue wing feature has a width at 3$\sigma_{\rm{rms}}$ of around 15$-$20\,km/s for the lines shown in Fig.~\ref{Fig:Blue_hole}. Under the simplifying assumption of a 1D spherically symmetric wind, purely thermal motions result in a  stochastic velocity distribution of a molecule with mass similar to CO of $\sim$1\,km/s in the warm inner wind, reducing to  to $\sim$0.5\,km/s in the cool outer wind regions. For a line with peak strength around 0.5\,Jy/beam (see Fig.~\ref{Fig:Blue_hole}) this stochastic velocity distribution induces a widening of the line profile (at 3$\sigma_{\rm{rms}} \sim$\,12\,mJy/beam) of a factor of  $\sim$4 or $\sim$4\,km/s. While this thermal velocity dispersion contributes to the line broadening, it can not explain the fact that various lines have widths far above the canonical terminal wind velocities.

\paragraph{Pulsations:}
Another cause for velocity disturbance in the stellar atmosphere and inner wind regions are convection-induced pulsations in Mira-type and semi-regular variables \citep{Bowen1988ApJ...329..299B, Liljegren2017arXiv170608332L}. 
Simultaneous spectroscopic monitoring of various spectral features originating in different layers and for several instances of time during a pulsation period enables one to trace the evolution of the global velocity field throughout the outer layers of an AGB star and thereby the mass loss process.
The detailed study by \citet{Nowotny2010A&A...514A..35N} shows that for Miras the radial velocity (RV) amplitude (difference between minimum and maximum values) in the CO $\Delta v$\,=\,3 line amounts to 20--30\,km/s. With an estimated conversion factor between gas velocities and measured RV around 1.2--1.5, the difference between post-shock outflow velocity and pre-shock infall velocity can reach values values of $\sim$35\,km/s. The high-excitation vibrational CO $\Delta v$\,=\,3  sample the regularly pulsating layers of the deep photosphere.
The study by \citet{Nowotny2010A&A...514A..35N} shows that the lower excitation vibrational CO $\Delta v$\,=\,2 lines trace the slow but quite steady outflow velocity pointing towards a line formation in layers where the material is already accelerated and variations due to shock waves are low ($\Delta$RV$\le$10\,km/s). The CO $\Delta v$\,=\,1 lines\footnote{Upper state energies are above 2140\,cm$^{-1}$.} in the data of \citet{Nowotny2010A&A...514A..35N} are blue-shifted for every instance with a measured RV lower than the observed terminal wind velocity $v_\infty$ and display a $\Delta$RV of only $\sim$1--2\,km/s, hence they trace the outflowing material. Most ALMA data have upper state energies lower than those of the CO $\Delta v$\,=\,1 lines  for which reason the velocity variation due to pulsations is estimated to be negligible. 

\medskip

These considerations allow us to conclude that the origin of the large velocities in both stars is a genuine physical mechanism not linked to thermal motions of the gas or pulsation behaviour of the atmospheric layers. We have put forward two different scenarios to explain the observed velocity profile: (1)~a fraction of the inner wind region does not follow the smooth radially outflowing wind but is part of a more complex morphology, and (2)~the global  velocity of the isotropic wind is determined by a more gradual but ultimately more forceful acceleration than is generally assumed. We consider the merits of both of scenarios regarding the dynamical structure deduced from the ALMA data, and their impact on our understanding of stellar evolution.

\subsubsection{Scenario 1: complex 3D morphology}

\begin{figure*}[!htbp]
\includegraphics[width=\textwidth,angle=0]{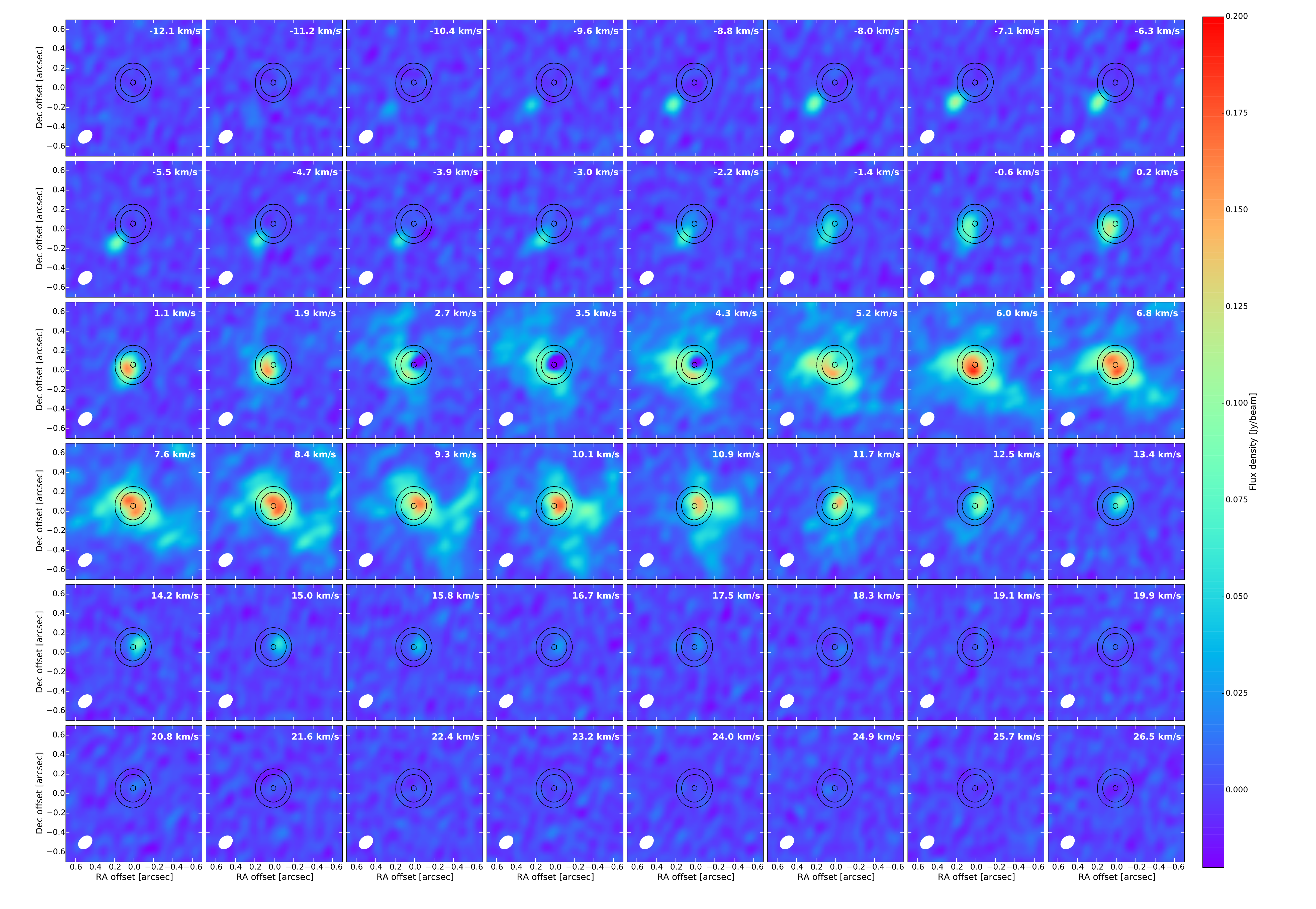}
\caption{Channel map of the SO$_2$ $(13_{4,10}-13_{3,11})$ emission in R~Dor. The circle denotes the place of maximum dust emissivity (taking the contours at 1\%, 10 and 90\% of the total flux). An extra blob appears at $\sim$0.3\arcsec\ from the central position in the blue channel maps.}
\label{Fig:channel_map_blue_blob}
\end{figure*}

Zooming into the inner wind region within 400\,mas, one can notice that the velocity spread is much greater in R~Dor than in IK~Tau. In the case of R~Dor, the cause for this large spread is partly due to a blue emission feature at 0.3\arcsec\ to the south-east (see Fig.~\ref{Fig:channel_map_blue_blob}). This `blue blob' is seen in almost all lines, with the main exception being the HCN lines. A direct comparison of the blue profiles between both stars also elucidates that the wings in R~Dor have a more gradual decrease towards the extreme velocities (Fig.~\ref{Fig:Blue_hole}). As discussed by \citet{Homan2018} these wings are reminiscent of a small disk residing in the inner wind of R~Dor with a tangential velocity of $\sim$12\,km/s at the inner rim residing at $\sim$6\,AU. The blue blob is tentatively postulated to be a potential companion with mass of at least 2.5 earth masses. These morphologies explain the large velocity spread around 200\,mas in Fig.~\ref{Fig:velocities_angwid_RDor}.

On the other hand, the position-velocity diagrams of the ALMA CO data of IK~Tau and the arc-like structures that can be discerned in some other bright molecular lines suggest the presence of semi-concentric shells or perhaps even a (broken) spiral-like structure (Decin et al.\ \textit{in prep.}). The source of these density inhomogeneities might disturb the velocity structure in the inner wind. If the spiral structure arises from the interaction with a binary companion, then wind-binary interaction models show that the velocity of a parcel of gaseous material in the wind can be modified due to both the reflex motion of the mass-losing AGB star and the wake induced by the companion's gravitational potential \citep{Kim2012ApJ...759...59K}. Their results show that the wind velocity $v_w$ (in the centre-of-mass frame) cannot exceed $v^{\prime}_w + v^{\prime}_p$ in any direction, with $v^{\prime}_w$ being the wind speed of the intrinsically isotropic wind and $v^{\prime}_p$ the orbital velocity of the AGB star. If the wind of IK~Tau is indeed subject to such a process, then for a measured $v_w$ of 36\,km/s and $ v^{\prime}_w$ of 17.7\,km/s, this situation would lead to an orbital velocity $v^{\prime}_p$ larger than 18.3\,km/s. For a star of 1\,\Msun\ this would yield an orbital radius of the mass-losing AGB star lower than 2.64\,AU (or 1.7\,\Rstar\ in the case of IK~Tau). As remarked by \citet{Kim2012ApJ...759...59K}, the velocity (modulated by the orbital velocity motion of the binary system) is larger than the intrinsic wind model, which could be considered as mimicking a wind acceleration mechanism. 
However, no binary companion has been detected as yet, albeit the 22\,GHz water masers and OH masers give indirect support to this hypothesis, since they show evidence for multiple shells and/or possibly a biconical outflow more-or-less face-on \citep{Richards2012A&A...546A..16R}.

\subsubsection{Scenario 2: enforced dynamics in an isotropic wind  }

\begin{figure*}[!htbp]
\begin{minipage}[t]{.48\textwidth}
        \centerline{\resizebox{\textwidth}{!}{\includegraphics[angle=0]{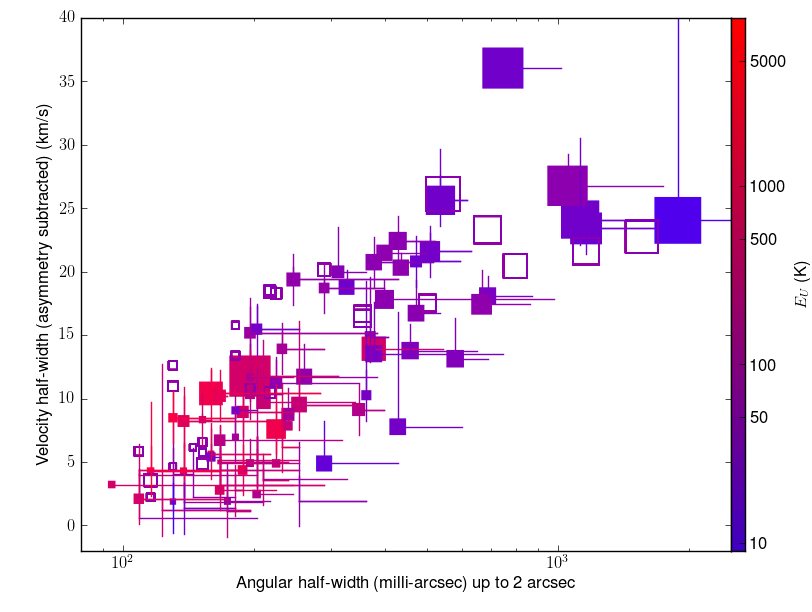}}}
    \end{minipage}
    \hfill
\begin{minipage}[t]{.48\textwidth}
        \centerline{\resizebox{\textwidth}{!}{\includegraphics[angle=0]{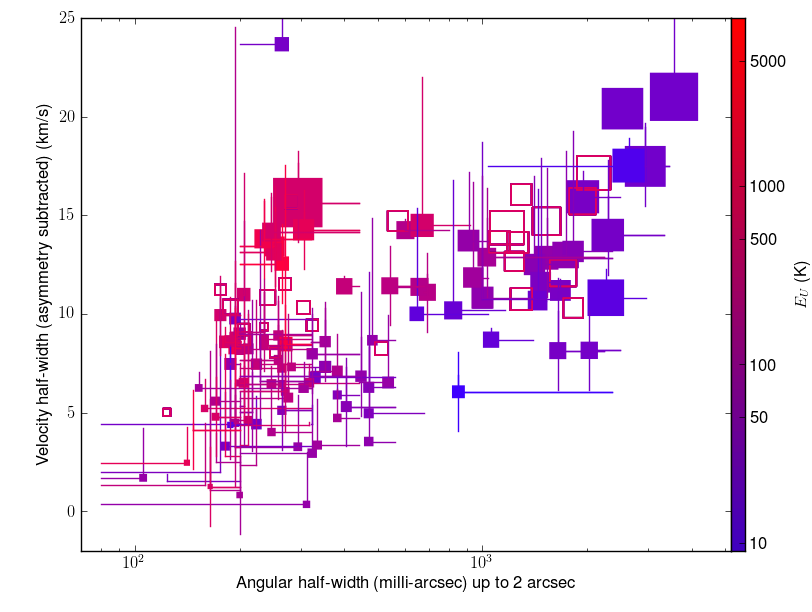}}}
    \end{minipage}
\caption{Relationship between the derived velocity and the angular size (IK~Tau at the left, R~Dor at the right). The solid (open) symbols designate single (one-sided) blends respectively. The error bars for the single lines show the velocity asymmetry, while for the one-sided blended lines the stochastic error is used.  The angular size error bars show the difference between the contour-derived and annular measurements. The shade shows the excitation temperature; the symbol size is proportional to integrated flux.}
\label{Fig:velocity_Anita}
\end{figure*}

While first modelling suggests that (part of) the broad wings in R Dor might be explained by a disk, an explanation for the broad wings in IK~Tau in terms of complex 3D morpho-kinematics is far more tentative. Correcting for the velocity asymmetry\footnote{By taking the velocity side with lowest value with respect to\ v$_{\rm{LSR}}$ as the total velocity half-width. Using columns (5) and (6) in Tables~\ref{Tab:IKTau}-\ref{Tab:RDor}, this is obtained by the simple arithmetic (\textit{`velocity width'}$-$abs(\textit{`velocity asymmetry'}))/2.}, the velocity structure displays a relationship with upper state excitation energy for both stars (in the region beyond 400\,mas; see Fig.~\ref{Fig:velocity_Anita}) indicating that there is no evidence that all high-velocity emission comes from inner wind regions. Fig.~\ref{Fig:velocities_angwid_IKTau} and Fig.~\ref{Fig:velocities_angwid_RDor} suggest a more gradual (larger $\beta$-values) but ultimately more effective acceleration than has been derived from the momentum equation. 

Focusing first on the region within 400\,mas, one indeed notices that the solution to the momentum equation systematically overestimates the derived wind velocities in IK~Tau (Fig.~\ref{Fig:velocities_angwid_IKTau}). A first reason for that overestimation is that \citet{Decin2010A&A...516A..69D} have used the opacities of Fe-rich silicates to model the wind structure. It is well known that small Fe-poor silicates or aluminium-oxides cannot trigger the onset of a stellar wind \citep{Woitke2006A&A...460L...9W}. Fe-rich grains have higher near-infrared opacities facilitating their importance in the photon momentum transfer, but also making them sensitive to sublimation. In addition, theoretical considerations suggest that most of the refractory material should condense into dust grains within a few stellar radii \citep[e.g.][]{Hofner2008A&A...491L...1H, Hofner2016A&A...594A.108H} and that grain growth and destruction processes in that inner wind region are dominated by collision processes resulting in a grain size distribution heavily skewed towards small grains. Therefore, \citet{Decin2006A&A...456..549D} adopt a grain size distribution $n_d(a,r)\,da = A(r)\,a^{-3.5} n_{\rm{H}}(r)\,da$, with a $n_{\rm{H}}$ being the hydrogen number density and the quantity $A(r)$ representing an abundance scale factor giving the number of dust particles in units of particles per H atom. This equation bears the implicit assumption of no further grain growth when the dust particles travel through the wind. However, an interesting finding by \citet{Paquette2011ApJ...732...62P} is that silicate grain nucleation occurs over a wide range in stellar radius --- from approximately sevento more than 20 stellar radii in their models. This additional nucleation would provide more grains to capture photons and accelerate dust (and hence gas) farther from the star. 
We also mention that clumpiness will have an effect on the drift velocity, since the gas might be more effectively coupled to the dust in denser clumps.

Another point of interest is that the logarithmic velocity gradient calculated from the momentum equation is around zero in the region beyond $\sim$30\,\Rstar\ (Fig.~\ref{Fig:velocities_angwid_IKTau}). This arises when the only significant forces determining the wind dynamics are gravity and radiation pressure on grains and for dust opacities $\chi$ being independent of radius \citep{Goldreich1976ApJ...205..144G}. A possible cause of the high velocity gradient at large radial distances can be the increase of $\chi(r)$ with radius so that the radiation pressure becomes increasingly effective with distance from the star \citep{Chapman1986MNRAS.220..513C}. One might suggest that the relative motions of the dust particles in the radially outflowing wind leads to grain growth such that the dust particles have fractal dimensions significantly lower than three. This suggestion is based on the theoretical calculations done by \citet{Dominik2016arXiv161100167D} and the laboratory studies of \citet{Krause2004PhRvL..93b1103K}.
The nucleation models by \citet{Paquette2011ApJ...732...62P} and the dynamical models by \citet{Fleischer1994PhDT.......101F} suggest a narrow grain size distribution with a strong peak at small grain sizes in the inner wind region\footnote{Calculations by, for example, \citet{Dominik1989A&A...223..227D} and \citet{Gobrecht2016A&A...585A...6G} show a broader distribution of grain size, but all of them are heavily skewed towards small grain size. It is that large ensemble of small grains which is most important for the further growth of aggregates further in the wind.}. Further out in the wind, in the region beyond 400\,mas, the relative motions of the dust particles will be governed by thermal (Brownian) motions caused by the random collisions with the gas molecules and turbulence caused by the relative difference in stopping time (captured in the Stokes number) for various dust particles. As discussed by \citet{Dominik2016arXiv161100167D}, subsequent collisions between two aggregates of similar small size can be described by cluster-cluster aggregation (CCA) of which the outcome is the growth of grains that have low fractal dimensions. The details of the outcome depend on the structure of the collision trajectory of the particles and on the rotational state of the colliding aggregates. If so, these fluffy aggregates formed beyond 400\,mas might have a large impact on the radiation pressure efficiency, since particularly in the case that the fractal dimension is  approximately two their shape maximises the interaction with the arriving photons. The main challenge for this scenario is that the dust densities are low at 400\,mas so that the time between collision of two dust grains can be in the order of years (depending on velocity and area of the dust grain) for a smooth wind. However, if the wind is inhomogeneous, one gets clumps which may be 10 to 50 times more dense than the average density, decreasing the collision time with the same factor.
\citet{Chapman1986MNRAS.220..513C} discussed the possibility that $\chi(r)$ might increase at larger radii due to the growth of icy mantles on the surface of the already existing grains as the temperature drops. These new, and more absorptive, coatings should however be in the order of a monolayer. If multiple monolayers are added and the initial fractal monomers are only a nanometre or so in diameter, then one starts to close in the holes and increase the overall fractal dimension, thus decreasing the radiative coupling efficiency. At a distance of 400\,mas, the gas temperature is around 300--600\,K, too high for water to nucleate. However, molecules which are more refractory than water might condense or react if they hit a grain.

One might wonder if this behaviour of a wind velocity larger than expected is also seen in other data of evolved AGB stars and supergiants. Indeed, the ALMA data of the two supergiants $\alpha$ Ori and VY~CMa display velocities of $\sim$24\,km/s and $\sim$105\,km/s, much larger than the canonical terminal wind velocities of 14\,km/s and 35\,km/s \citep{DeBeck2015A&A...580A..36D, Decin2016A&A...592A..76D, Kervella2017arXiv171107983K}, and also in the case of the O-rich AGB stars Mira and EP~Aqr a similar behaviour is noticed \citep[][Homan et al.\ \textit{in prep.}]{Wong2016A&A...590A.127W}. Although one can make the argument that the wind of each of these stars might be subject to some particular systematic dynamical pattern caused by, for instance, direction-dependent outflows, disks, spirals, or the impact of binarity, it is intriguing that the high sensitivity of ALMA is the key to cognise these high velocity tails. Specifically, the line sensitivity in these ALMA data is around 1.6-5\,mJy/beam, for the Mira observations it was around 2.3 mJy/beam. In contrast, the ALMA sensitivity of the observations of the carbon-rich AGB star CW~Leo was only around 200-300\,mJy/beam \citep{Decin2015A&A...574A...5D}, and albeit a spiral is detected in the inner wind of that target, the velocities derived from the line widths do not exceed the canonical terminal velocity of $\sim$14.5\,km/s.

The impact of higher wind velocities cannot be underestimated.  The mass-loss rate is linearly dependent on the (terminal) wind velocity, as can be seen in the equation of mass conservation
$dM(r)/dt = \Mdot(r) = 4 \pi r^2 \rho(r) v(r)$. Hydrogen gas densities are often derived from CO line observations under the assumption of a [CO/H$_2$] ratio determined from stellar evolution models. If wind velocities are indeed higher than previously determined, this would yield an increase of \Mdot\ by the same factor. These higher wind velocities imply a shortening of the AGB phase since the mass of outer atmospheric hydrogen-envelope will be reduced at higher pace, yielding a reduction of the pulsation amplitudes and hence one of the main triggers for the wind will come to an end.

\subsection{Isotopic ratios} \label{Sect:disc_isotopes}

We refrain from performing a population diagram analysis of the molecules observed in the ALMA survey. The underlying assumption of a population diagram is a Boltzmann distribution of states that applies to an isothermal gas in local thermodynamic equilibrium (LTE). The zero-moment maps prove that all molecular transitions are formed over a substantial part of the circumstellar envelope and hence the isothermal assumption does not hold, rather a temperature gradient should be taken into account. Also, molecules with a large dipole moment are mainly radiatively excited by stellar and/or dust photons for which reason LTE is not a valid assumption. As demonstrated by \citet{Kaminski2013ApJS..209...38K} deviations from these assumptions are the reason that one cannot reproduce many molecular population diagrams by a single linear fit and hence any conclusions in terms of excitation temperature or column densities should be considered with care. We intend to analyse the molecular excitation and retrieve the proper molecular abundances under the assumption of statistical equilibrium in forthcoming papers (Danilovich et al.\ \textit{in prep.}).

It is tempting, though, to have a first look at isotopic ratios as derived from the observed emission of isotopologues. These ratios yield constrains on the AGB nucleosynthesis, stellar mass and age, and the Galactic environment in which the star was born. Analogous limitations as in the case of a population diagram analysis hold here as well. Observed line intensity ratios provide us only with a valid indicator of an isotopic ratio in the case of \textit{(i.)} no or the same correction for line optical thickness, \textit{(ii.)} no or the same effect of radiative pumping, and \textit{(iii.)} given high-spatial resolution one should trace the same excitation region.
These limitations are often formulated together such that the method of line intensities ratios is only valid under the assumption of optically thin emission in LTE, although they can be relaxed a bit under the conditions summed up before. The high spatial resolution of ALMA offers us a possibility to properly address point \textit{(iii.)}. The most difficult point to assess, without in-depth modelling, is  point \textit{(ii.)}; it is also the reason to exclude potential masers (such as the numerous $^{28}$SiO, $^{29}$SiO, and $^{30}$SiO lines in our data) as isotope tracers. To address point \textit{(i.)}, we will only consider isotope ratios of the minor isotopologues, specifically of $^{29}$Si, $^{30}$Si, $^{33}$S, $^{34}$S, and $^{37}$Cl. The latter one will be compared to the main stable isotope $^{35}$Cl. Isotopologues probing these ratios are $^{29}$SiS,$^{30}$SiS, Si$^{33}$S, Si$^{34}$S, $^{33}$SO\footnote{The hyperfine components of $^{33}$SO have been summed when comparing to the non-hyperfine molecule $^{34}$SO.}, $^{34}$SO, Na$^{35}$Cl, and Na$^{37}$Cl. For all species, with the exception of NaCl, intensity ratios of lines with the same quantum numbers are considered to be compliant with point \textit{(iii.)}. For NaCl this was not possible, and the Na$^{35}$Cl J=27-26  and J=26-25 lines are compared to the Na$^{35}$Cl J=28-27 transition. Line intensity ratios are corrected for the difference in Einstein A-coefficient, or in terms of the quantities given by the CDMS database the predicted integrated intensity (at 300\,K), the frequency, the rotational-spin partition function, and the lower state energy. In the case of NaCl, also a correction for the upper state degeneracy is taken into account. Since neither SiS or NaCl are detected in R~Dor, only the $^{34}$S/$^{33}$S ratio can be derived. Often, only one corresponding transition between both isotopologues is available since we also exclude lines that are blended or resolved out. As a consequence, no formal uncertainty on the derived ratio can be calculated.
For IK~Tau, we derive that $^{29}$Si/$^{30}$Si$\approx$4.1, $^{34}$S/$^{33}$S$\approx$5.6$\pm$0.36, and $^{35}$Cl/$^{37}$Cl$\approx$1.1$\pm$0.21, while for R~Dor $^{34}$S/$^{33}$S$\approx$7.8$\pm$0.7. 

The large dipole moment of NaCl ($\mu$\,=\,9\,D)\footnote{The dipole moments of SiS and SO are only 1.735\,D and 1.535\,D, respectively.} implies a large uncertainty on the derived $^{35}$Cl/$^{37}$Cl fraction, since these lines might be highly sensible to radiative pumping via, for example, line overlap even in other transitions. This might explain why the derived ratio is much lower than what has been derived for the Sun, interstellar medium, some dense interstellar clouds, and the carbon-rich AGB star IRC\,+10216 which all have values around 3.1 \citep{Cernicharo2010A&A...518L.136C, Cernicharo2010A&A...518L.115C, Agundez2011A&A...533L...6A}.  $^{35}$Cl and $^{37}$Cl are thought to be formed during explosive oxygen-burning in supernovae \citep{Jaschek2009bces.book.....J} and should not be modified during the red giant and asymptotic giant phase evolution of low mass stars. As such, it is expected that the $^{35}$Cl/$^{37}$Cl ratio in IK~Tau should be close to the value of its parental cloud.

The isotopes $^{33}$S and $^{34}$S are mainly products of oxygen burning \citep{Chin1996A&A...305..960C, Mauersberger1996A&A...313L...1M} with $^{34}$S being quite sensitive to metallicity (Woosley and Weaver 1995). These isotopes and their relative ratio are not expected to be modified during the evolution of a low-mass star. Indeed, the $^{34}$S/$^{33}$S-values derived for IK~Tau and R~Dor are close to values for the interstellar medium \citep[6.27$\pm$1.01,][]{Chin1996A&A...305..960C} and solar system \citep[$\approx 5.5$, ][]{Kahane1988A&A...190..167K}; the same conclusion was reached in the case of the carbon-rich AGB star IRC\,+10216 \citep[5.7$^{+0.8}_{-0.6}$,][]{Kahane1988A&A...190..167K}.

The silicon isotope fraction $^{29}$Si/$^{30}$Si of IK~Tau is slightly higher than the value derived by \citet{Decin2010A&A...516A..69D}, being $^{29}$Si/$^{30}$Si\,=\,3 with an uncertainty of a factor of 2. We refer to the in-depth discussion on the silicon isotope ratios, including the ratio w.r.t.\ $^{28}$Si, by \citet{Decin2010A&A...516A..69D}. The specific $^{29}$Si/$^{30}$Si ratio is considered to be quite high. Compared to the solar isotopic ratio, ($^{29}$Si/$^{30}$Si)$_\odot$\,=\,1.52, IK~Tau seems underabundant in neutron-rich Si isotopologues.
\citet{Zinner2006ApJ...650..350Z} studied the change in silicon isotopic ratios when stars evolve along the asymptotic giant branch, but they found that no noticeable changes occur in the Si isotopes when the star is still O-rich. Consequently, for these O-rich AGB stars the silicon isotope ratios reflect the interstellar cloud out of which the star was born. As such, \citet{Decin2010A&A...516A..69D} suggested that IK~Tau was born in an interstellar medium with a mixture enriched by X-type grains of which supernovae type~II are thought to be main contributors.


\section{Summary}\label{Sec:conclusions}

In this paper we have presented the first ALMA spectral survey of two oxygen-rich AGB stars. The high mass-loss rate star IK~Tau (\Mdot$\sim$5$\times 10^{-6}$\,\Msun/yr) and low mass-loss rate star R~Dor (\Mdot$\sim$1$\times 10^{-7}$\,\Msun/yr) were surveyed between 335 and 362\,GHz at a spatial resolution of $\sim$150\,mas. This spatial resolution corresponds to the locus of the main dust formation in both stars.
Some 200 molecular features were detected in each source, arising from 34 molecules (including isotopologues). Careful analysis of the spectra, total intensity maps and radially averaged flux density profiles lead to the compilation of a spectral atlas for both sources, which is visually represented in Figs.~\ref{Fig:atlas_IKTau}--\ref{Fig:atlas_RDor} and tabulated in Tables~\ref{Tab:IKTau}--\ref{Tab:RDor}. This atlas can be used to retrieve the molecular abundance structure in the envelope, deduce the local and global morphology patterns, analyse the  kinematics, etc. 
Moreover, the atlas is very useful to shape future observing programmes for ALMA, SMA, PdBI, etc.

Specifically, the ALMA data cover transitions from AlCl, AlO, AlOH, CO, CS, H$_2$O, HCN, NS, NaCl, SO, SO$_2$, SiO, SiS, TiO, and TiO$_2$ with some of their isotopologues, including rotational lines in both the ground state and vibrational excited states. SiO lines are detected up to $v$\,=\,5. A clear dichotomy is seen in the sulphur chemistry of both stars: whereas (almost) all sulphur seems to be locked up in SO and SO$_2$ in R~Dor, while CS and SiS are also prominently present in IK~Tau. NaCl and NS are detected in IK~Tau, but not in R~Dor. As discussed by \citet{Decin2017arXiv170405237D}, the gas-phase aluminium chemistry between both targets is clearly different. TiO and TiO$_2$ are detected well beyond the main dust condensation region in both stars, indicating that some fraction does not partake in dust nucleation and growth.

The channel maps and total intensity maps testify to a complex wind morphology and show the presence of blobs, arcs, and/or a small disk in the inner wind region. This non-homogeneous density distribution will have a profound impact on the ongoing chemical processes, since lower density regions will allow energetic interstellar UV photons to penetrate deeper into the envelope, resulting in a more active photochemistry in these inner winds. This process might explain the detection of molecules, such as HCN, SiS and CS.  These molecules are indicators of a carbon-rich chemistry, and their detection in an oxygen-rich environment points towards  processes such as photochemistry and/or pulsation-induced shocks yielding a non-equilibrium chemistry.

The high sensitivity of ALMA allows us to study the weak wings of the line profiles and from that deduce the wind velocity structure in the transition region from sub-sonic to supersonic values. Surprisingly, a large fraction of lines display line widths, and hence velocities, well beyond the canonical wind terminal velocity, for both stars. These large velocities have remained undetected in previous observations. A (tentative) spiral-like structure in IK~Tau, and a small disk combined with a blue-shifted blob in R~Dor might explain (part of) these large velocities. A correlation between the wind velocity and excitation temperature seems to suggest that a more gradual but finally more forceful acceleration is shaping the wind speed of the intrinsically radially outflowing material. We propose that the further growth of grains in the region beyond $\sim$30\,\Rstar\ leads to fractal grains with fractal dimension around two, hence increasing the radiation pressure efficiency in that part of the wind.

Quite a few species show a `hole' at the stellar position in the blue-shifted channel maps, manifested in the spectrum as a blue wing absorption feature.  This is particularly prominent for brighter lines and for spectra extracted with an aperture comparable to the stellar diameter. The cause of this blue hole is the crossing of impact parameters through the stellar disk and the outflowing material, much in analogy with the classical P-Cygni profiles in massive-star winds. Assuming that the emission at extreme blue and red-shifted velocity arises from (complex) 3D structures which barely contribute to the line-of-sight emission at impact parameters crossing the stellar surface, the sharp edge of the blue absorption peak allows us to determine the terminal plus turbulent velocity of the (classical) isotropic wind.

\begin{acknowledgements}
LD, TD, and JN acknowledge support from the ERC consolidator grant 646758 AEROSOL. LD also acknowledges support from the FWO Research Project grant G024112N
and TD acknowledges from the Fund of Scientific Research Flanders (FWO).
This paper makes use of the following ALMA data: ADS/JAO.ALMA2013.1.00166.S. ALMA is a partnership of ESO (representing 
its member states), NSF (USA) and NINS (Japan), together with NRC 
(Canada) and NSC and ASIAA (Taiwan), in cooperation with the Republic of 
Chile. The Joint ALMA Observatory is operated by ESO, AUI/NRAO and NAOJ.
Credit CASA: International consortium of scientists based at the National
Radio Astronomical Observatory (NRAO), the European Southern
Observatory (ESO), the National Astronomical Observatory of Japan
(NAOJ), the CSIRO Australia Telescope National Facility (CSIRO/ATNF),
and the Netherlands Institute for Radio Astronomy (ASTRON) under the
guidance of NRAO.
\end{acknowledgements}

\bibliographystyle{aa}
\bibliography{ALMA_IKTau_RDor}

\onecolumn

{\small{
\begin{landscape}
\setlength{\tabcolsep}{.6mm}
\begin{longtable}{llrrrrrrrrrrrcp{2cm}}
\caption{\label{Tab:IKTau}Spectral lines observed in IK Tau with ALMA. The asterisks in the penultimate column denote if no fiducial measurements could be made.}\\
\hline \hline
\multicolumn{1}{c}{(1)} & \multicolumn{1}{l}{(2)} & \multicolumn{1}{c}{(3)} & \multicolumn{1}{c}{(4)} & \multicolumn{1}{c}{(5)} & \multicolumn{1}{c}{(6)} & \multicolumn{1}{c}{(7)} & \multicolumn{1}{c}{(8)} & \multicolumn{1}{c}{(9)} & \multicolumn{1}{c}{(10)} & \multicolumn{1}{c}{(11)} & \multicolumn{1}{c}{(12)} & \multicolumn{1}{c}{(13)} & \multicolumn{1}{c}{(14)} & \multicolumn{1}{c}{(15)}\\
 \multicolumn{1}{c}{Molecule} & \multicolumn{1}{l}{Quantum} & \multicolumn{1}{c}{$\nu_{\rm{rest}}$} & \multicolumn{1}{c}{$E_{\rm{low}}$} & \multicolumn{1}{c}{Velocity\tablefootmark{a}} & \multicolumn{1}{c}{Velocity\tablefootmark{a}} & \multicolumn{1}{c}{Peak\tablefootmark{b}} & \multicolumn{1}{c}{Integrated\tablefootmark{b}} & \multicolumn{1}{c}{Velocity\tablefootmark{a}} & \multicolumn{1}{c}{Velocity\tablefootmark{a}}  & \multicolumn{1}{c}{Peak\tablefootmark{b}} & \multicolumn{1}{c}{Integrated\tablefootmark{b}} & \multicolumn{1}{c}{Angular\tablefootmark{a}} &\multicolumn{1}{c}{Comment\tablefootmark{c}} & \multicolumn{1}{c}{Comment\tablefootmark{d}}\\
 & \multicolumn{1}{l}{numbers} & & & \multicolumn{1}{c}{width}  & \multicolumn{1}{c}{asymmetry} & \multicolumn{1}{c}{flux} & \multicolumn{1}{c}{flux} & \multicolumn{1}{c}{width} & \multicolumn{1}{c}{asymmetry}  & \multicolumn{1}{c}{flux} & \multicolumn{1}{c}{flux} & \multicolumn{1}{c}{width} & \multicolumn{1}{c}{line} & \multicolumn{1}{c}{mom0}\\
 & & \multicolumn{1}{c}{[GHz]} & \multicolumn{1}{c}{[K]} & \multicolumn{1}{c}{[km s$^{-1}$]} & \multicolumn{1}{c}{[km s$^{-1}$]} &  \multicolumn{1}{c}{[Jy]} & \multicolumn{1}{c}{[Jy km s$^{-1}$}] & \multicolumn{1}{c}{[km s$^{-1}$]} & \multicolumn{1}{c}{[km s$^{-1}$]} &  \multicolumn{1}{c}{[Jy]} & \multicolumn{1}{c}{[Jy km s$^{-1}$]} & \multicolumn{1}{c}{}[mas] & \multicolumn{1}{c}{identification} & \multicolumn{1}{c}{map}\\
 \cmidrule(r){5-8} 
 \cmidrule(l){9-12}
 & & & & \multicolumn{4}{c}{Aperture 320\,mas} & \multicolumn{4}{c}{Aperture 800\,mas} & & & \\
 \hline
\endfirsthead
\caption{continued.}\\
\hline \hline
\multicolumn{1}{c}{(1)} & \multicolumn{1}{l}{(2)} & \multicolumn{1}{c}{(3)} & \multicolumn{1}{c}{(4)} & \multicolumn{1}{c}{(5)} & \multicolumn{1}{c}{(6)} & \multicolumn{1}{c}{(7)} & \multicolumn{1}{c}{(8)} & \multicolumn{1}{c}{(9)} & \multicolumn{1}{c}{(10)} & \multicolumn{1}{c}{(11)} & \multicolumn{1}{c}{(12)} & \multicolumn{1}{c}{(13)} & \multicolumn{1}{c}{(14)} & \multicolumn{1}{c}{(15)} \\
 \multicolumn{1}{c}{Molecule} & \multicolumn{1}{l}{Quantum} & \multicolumn{1}{c}{$\nu_{\rm{rest}}$} & \multicolumn{1}{c}{$E_{\rm{low}}$} & \multicolumn{1}{c}{Velocity\tablefootmark{a}} & \multicolumn{1}{c}{Velocity\tablefootmark{a}} & \multicolumn{1}{c}{Peak\tablefootmark{b}} & \multicolumn{1}{c}{Integrated\tablefootmark{b}} & \multicolumn{1}{c}{Velocity\tablefootmark{a}} & \multicolumn{1}{c}{Velocity\tablefootmark{a}}  & \multicolumn{1}{c}{Peak\tablefootmark{b}} & \multicolumn{1}{c}{Integrated\tablefootmark{b}} & \multicolumn{1}{c}{Angular\tablefootmark{a}} &\multicolumn{1}{c}{Comment\tablefootmark{c}} & \multicolumn{1}{c}{Comment\tablefootmark{d}}\\
 & \multicolumn{1}{l}{numbers} & & & \multicolumn{1}{c}{width}  & \multicolumn{1}{c}{asymmetry} & \multicolumn{1}{c}{flux} & \multicolumn{1}{c}{flux} & \multicolumn{1}{c}{width} & \multicolumn{1}{c}{asymmetry}  & \multicolumn{1}{c}{flux} & \multicolumn{1}{c}{flux} & \multicolumn{1}{c}{width} & \multicolumn{1}{c}{line} & \multicolumn{1}{c}{mom0}\\
 & & \multicolumn{1}{c}{[GHz]} & \multicolumn{1}{c}{[K]} & \multicolumn{1}{c}{[km s$^{-1}$]} & \multicolumn{1}{c}{[km s$^{-1}$]} &  \multicolumn{1}{c}{[Jy]} & \multicolumn{1}{c}{[Jy km s$^{-1}$}] & \multicolumn{1}{c}{[km s$^{-1}$]} & \multicolumn{1}{c}{[km s$^{-1}$]} &  \multicolumn{1}{c}{[Jy]} & \multicolumn{1}{c}{[Jy km s$^{-1}$]} & \multicolumn{1}{c}{}[mas] & \multicolumn{1}{c}{identification} & \multicolumn{1}{c}{map}\\
 \cmidrule(r){5-8} 
 \cmidrule(l){9-12}
 & & & & \multicolumn{4}{c}{Aperture 320\,mas} & \multicolumn{4}{c}{Aperture 800\,mas} & & & \\
 \hline
\endhead
\hline
\endfoot
\hline
\endlastfoot
              SO$_2$ &            $20_{4,16} \to 20_{3,17}$ ($\nu_{2}=1$) & 335.1285 &  963 & 10.48 &   -0.4 &    0.043  &   0.261 &  5.24 &    1.2 &    0.030  &   0.054 &   478 & single                         &                  -              \\
                 SiO &                                  $8 \to 7$ (v = 5) & 335.2820 & 8735 & 13.97 &   -5.4 &    0.059  &   0.578 & 12.22 &   -3.6 &    0.044  &   0.349 &   333 & single                         &                  Compact        \\
          Si$^{34}$S &                                $19 \to 18$ (v = 0) & 335.3419 &  145 & 41.90 &    0.4 &    0.393  &   8.200 & 38.41 &    0.4 &    0.374  &   7.888 &  1028 & single                         &                  -              \\
          SO$^{18}$O &                          $42_{5,37} \to 41_{6,36}$ & 335.4791 &  837 &  5.24 &    3.0 &    0.033  &   0.205 &  $-$  &   $-$  &     $-$   &   $-$   &   391 & single                         &                  -              \\
          $^{29}$SiO &                                  $8 \to 7$ (v = 3) & 335.8807 & 5282 & 22.66 &   -1.8 &    5.055  &  35.268 & 20.92 &   -3.6 &    5.017  &  34.258 &   565 & single                         &                  Compact        \\
           TiO$_{2}$ &                         $51_{9,43} \to 50_{10,40}$ & 336.0286 & 1050 & 13.94 &  -11.5 &   (0.014) &   $-$   &  5.23 &   -6.3 &   (0.009) &   $-$   &   391 & uncertain                      &                  Compact        \\
\hline
\end{longtable}
\tablefoot{
\tablefoottext{a}{Values for the velocity width, velocity asymmetry, and angular width are given only in the case of clear detections.}\\
\tablefoottext{b}{Values for the integrated flux are given only for non-blended lines or lines with a minor blend contribution. Peak-flux values are listed for all detected species, but might not reflect the real peak flux in the case of strongly overlapping blends. In the latter case, the values are put in parentheses.}\\
\tablefoottext{c}{Comments based on appearance of line in the spectrum for the 320\,mas extraction aperture.
`Single' denotes that the detected line is not blended.
`Low/high blend' means that there is contamination on the low/high velocity side of a line
`Mid blend' means that a line has a blend on both sides.
`Minor' means that the line is a minor contributor.
`Hyper' indicates that a line is part of a hyperfine structure line.
`Uncertain' reflects that the line identification is unsure.
}\\
\tablefoottext{d}{Comments based on the morphology of the zero-moment map. 
`Compact' means that the emission is not significantly different from the
beam shape and the edge contours are closely spaced, suggesting that
the emission is unresolved and the extent could be much smaller than
the imaged extent. 
`-' means that the shape is fairly regular and that the
edge contours are not very closely spaced.
`Irregular' means that the emission appears irregular and that there are
often additional faint clumps outside the measured contour. Hence, the measured extent might be a lower limit.
`Arc-off' means that the transition is concentrated on one side of the stellar
position.
`Resolved out' is emission which can be measured to the maximum extent
sampled and appears to have strong artefacts further out, i.e.\ the measured extent is an
underestimate of the true extent. 
`Faint scatter' is emission from  clumps scattered around. In
these latter two cases the angular width is measured from the azimuthally averaged emission.
`Core-halo' means that, in the case of two high-excitation lines, they have a compact peak but a (surprisingly) large total extent.}
`Arc' means that an arc-like feature is detected in the channel map, the direction of which is given.
`S': south, `E': east, `SW': south-west, `NW': north-west, `SE': south-east\\
}
\end{landscape}
}}

{\small{
\begin{landscape}
\setlength{\tabcolsep}{.5mm}
\begin{longtable}{llrrrrrrrrrrrcp{2cm}}
\caption{\label{Tab:RDor}Spectral lines observed in R Dor with ALMA. The asterisks in the penultimate column denote if no fiducial measurements could be made.}\\
\hline \hline
\multicolumn{1}{c}{(1)} & \multicolumn{1}{l}{(2)} & \multicolumn{1}{c}{(3)} & \multicolumn{1}{c}{(4)} & \multicolumn{1}{c}{(5)} & \multicolumn{1}{c}{(6)} & \multicolumn{1}{c}{(7)} & \multicolumn{1}{c}{(8)} & \multicolumn{1}{c}{(9)} & \multicolumn{1}{c}{(10)} & \multicolumn{1}{c}{(11)} & \multicolumn{1}{c}{(12)} & \multicolumn{1}{c}{(13)} & \multicolumn{1}{c}{(14)} & \multicolumn{1}{c}{(15)}\\
 \multicolumn{1}{c}{Molecule} & \multicolumn{1}{l}{Quantum} & \multicolumn{1}{c}{$\nu_{\rm{rest}}$} & \multicolumn{1}{c}{$E_{\rm{low}}$} & \multicolumn{1}{c}{Velocity\tablefootmark{a}} & \multicolumn{1}{c}{Velocity\tablefootmark{a}} & \multicolumn{1}{c}{Peak\tablefootmark{b}} & \multicolumn{1}{c}{Integrated\tablefootmark{b}} & \multicolumn{1}{c}{Velocity\tablefootmark{a}} & \multicolumn{1}{c}{Velocity\tablefootmark{a}}  & \multicolumn{1}{c}{Peak\tablefootmark{b}} & \multicolumn{1}{c}{Integrated\tablefootmark{b}} & \multicolumn{1}{c}{Angular\tablefootmark{a}} &\multicolumn{1}{c}{Comment\tablefootmark{c}} & \multicolumn{1}{c}{Comment\tablefootmark{d}}\\
 & \multicolumn{1}{l}{numbers} & & & \multicolumn{1}{c}{width}  & \multicolumn{1}{c}{asymmetry} & \multicolumn{1}{c}{flux} & \multicolumn{1}{c}{flux} & \multicolumn{1}{c}{width} & \multicolumn{1}{c}{asymmetry}  & \multicolumn{1}{c}{flux} & \multicolumn{1}{c}{flux} & \multicolumn{1}{c}{width} & \multicolumn{1}{c}{line} & \multicolumn{1}{c}{mom0}\\
 & & \multicolumn{1}{c}{[GHz]} & \multicolumn{1}{c}{[K]} & \multicolumn{1}{c}{[km s$^{-1}$]} & \multicolumn{1}{c}{[km s$^{-1}$]} &  \multicolumn{1}{c}{[Jy]} & \multicolumn{1}{c}{[Jy km s$^{-1}$}] & \multicolumn{1}{c}{[km s$^{-1}$]} & \multicolumn{1}{c}{[km s$^{-1}$]} &  \multicolumn{1}{c}{[Jy]} & \multicolumn{1}{c}{[Jy km s$^{-1}$]} & \multicolumn{1}{c}{}[mas] & \multicolumn{1}{c}{identification} & \multicolumn{1}{c}{map}\\
 \cmidrule(r){5-8} 
 \cmidrule(l){9-12}
 & & & & \multicolumn{4}{c}{Aperture 320\,mas} & \multicolumn{4}{c}{Aperture 800\,mas} & & & \\
 \hline
\endfirsthead
\caption{continued.}\\
\hline \hline
\multicolumn{1}{c}{(1)} & \multicolumn{1}{l}{(2)} & \multicolumn{1}{c}{(3)} & \multicolumn{1}{c}{(4)} & \multicolumn{1}{c}{(5)} & \multicolumn{1}{c}{(6)} & \multicolumn{1}{c}{(7)} & \multicolumn{1}{c}{(8)} & \multicolumn{1}{c}{(9)} & \multicolumn{1}{c}{(10)} & \multicolumn{1}{c}{(11)} & \multicolumn{1}{c}{(12)} & \multicolumn{1}{c}{(13)} & \multicolumn{1}{c}{(14)} & \multicolumn{1}{c}{(15)} \\
 \multicolumn{1}{c}{Molecule} & \multicolumn{1}{l}{Quantum} & \multicolumn{1}{c}{$\nu_{\rm{rest}}$} & \multicolumn{1}{c}{$E_{\rm{low}}$} & \multicolumn{1}{c}{Velocity\tablefootmark{a}} & \multicolumn{1}{c}{Velocity\tablefootmark{a}} & \multicolumn{1}{c}{Peak\tablefootmark{b}} & \multicolumn{1}{c}{Integrated\tablefootmark{b}} & \multicolumn{1}{c}{Velocity\tablefootmark{a}} & \multicolumn{1}{c}{Velocity\tablefootmark{a}}  & \multicolumn{1}{c}{Peak\tablefootmark{b}} & \multicolumn{1}{c}{Integrated\tablefootmark{b}} & \multicolumn{1}{c}{Angular\tablefootmark{a}} &\multicolumn{1}{c}{Comment\tablefootmark{c}} & \multicolumn{1}{c}{Comment\tablefootmark{d}}\\
 & \multicolumn{1}{l}{numbers} & & & \multicolumn{1}{c}{width}  & \multicolumn{1}{c}{asymmetry} & \multicolumn{1}{c}{flux} & \multicolumn{1}{c}{flux} & \multicolumn{1}{c}{width} & \multicolumn{1}{c}{asymmetry}  & \multicolumn{1}{c}{flux} & \multicolumn{1}{c}{flux} & \multicolumn{1}{c}{width} & \multicolumn{1}{c}{line} & \multicolumn{1}{c}{mom0}\\
 & & \multicolumn{1}{c}{[GHz]} & \multicolumn{1}{c}{[K]} & \multicolumn{1}{c}{[km s$^{-1}$]} & \multicolumn{1}{c}{[km s$^{-1}$]} &  \multicolumn{1}{c}{[Jy]} & \multicolumn{1}{c}{[Jy km s$^{-1}$}] & \multicolumn{1}{c}{[km s$^{-1}$]} & \multicolumn{1}{c}{[km s$^{-1}$]} &  \multicolumn{1}{c}{[Jy]} & \multicolumn{1}{c}{[Jy km s$^{-1}$]} & \multicolumn{1}{c}{}[mas] & \multicolumn{1}{c}{identification} & \multicolumn{1}{c}{map}\\
 \cmidrule(r){5-8} 
 \cmidrule(l){9-12}
 & & & & \multicolumn{4}{c}{Aperture 320\,mas} & \multicolumn{4}{c}{Aperture 800\,mas} & & & \\
 \hline
\endhead
\hline
\endfoot
\hline
\endlastfoot
           TiO$_{2}$ &                          $24_{8,16} \to 24_{7,17}$ & 335.0670 &  273 & 15.73 &    5.6 &   (0.037) &   $-$   & 15.73 &    5.6 &   (0.036) &   $-$   &   435 & uncertain                      & -                               \\
              SO$_2$ &            $20_{4,16} \to 20_{3,17}$ ($\nu_{2}=1$) & 335.1285 &  963 & 18.35 &   -0.5 &    0.137  &   1.399 & 16.60 &    1.2 &    0.138  &   1.200 &   517 & single                         & -                               \\
                 SiO &                                  $8 \to 7$ (v = 5) & 335.2820 & 8735 & 26.20 &   -1.1 &    0.470  &   5.622 & 24.45 &    0.6 &    0.388  &   4.619 &   529 & single                         & Compact                         \\
          SO$^{18}$O &                          $42_{5,37} \to 41_{6,36}$ & 335.4791 &  837 & 18.33 &    2.9 &    0.128  &   1.266 &  9.60 &   -2.2 &    0.129  &   0.795 &   517 & single                         & -                               \\
           TiO$_{2}$ &                          $29_{3,27} \to 29_{2,28}$ & 335.5054 &  303 & 20.07 &   -9.3 &   (0.032) &   $-$   &  2.62 &    2.8 &   (0.010) &   $-$   &   411 & single                         & -                               \\
              SO$_2$ &                          $29_{5,25} \to 30_{2,28}$ & 335.7732 &  447 & 13.08 &   -3.7 &   (0.024) &   $-$   &  9.59 &   -0.3 &   (0.027) &   $-$   &   376 & uncertain                      & -                               \\
\hline
\end{longtable}
\tablefoot{
\tablefoottext{a}{Values for the velocity width, velocity asymmetry, and angular width are given only in the case of clear detections.}\\
\tablefoottext{b}{Values for the peak flux and integrated flux are given only for non-blended lines or lines with a minor blend contribution.}\\
\tablefoottext{c}{Comments based on appearance of line in the spectrum for the 300\,mas extraction aperture. For more details: see Table~\ref{Tab:IKTau}.}\\
\tablefoottext{d}{Comments based on the morphology of the zero-moment map. For more details: see Table~\ref{Tab:IKTau}.}
}
\end{landscape}
}}

\begin{appendix}
\section{Channel maps} \label{App:channel_maps}
In this section, we present the HCN(4-3) channel maps of R~Dor and IK~Tau, and the NaCl(26-25) channel map of IK~Tau. These channel maps serve as support of the discussion in Sect.~\ref{Sec:spatial}.

\begin{figure*}
\includegraphics[width=\textwidth]{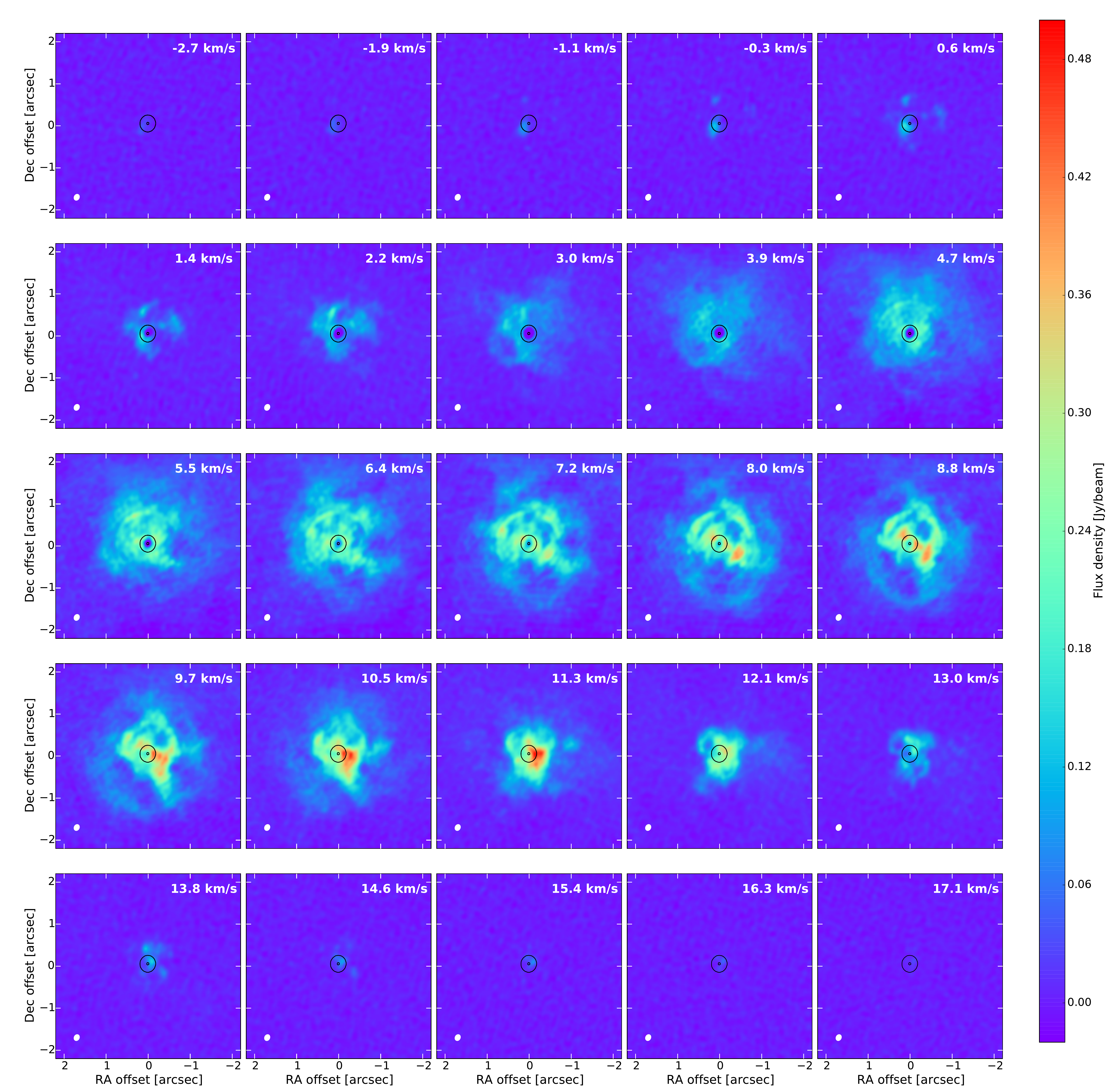}
\caption{Channel map of the HCN(4-3) emission in R~Dor. The circle denotes the place of maximum dust emissivity (taking controus at 1\% and 99\% of the total flux). The contrast is best visible on screen.}
\label{Fig:RDor_HCN_channel}
\end{figure*}

\begin{figure*}
\includegraphics[width=\textwidth]{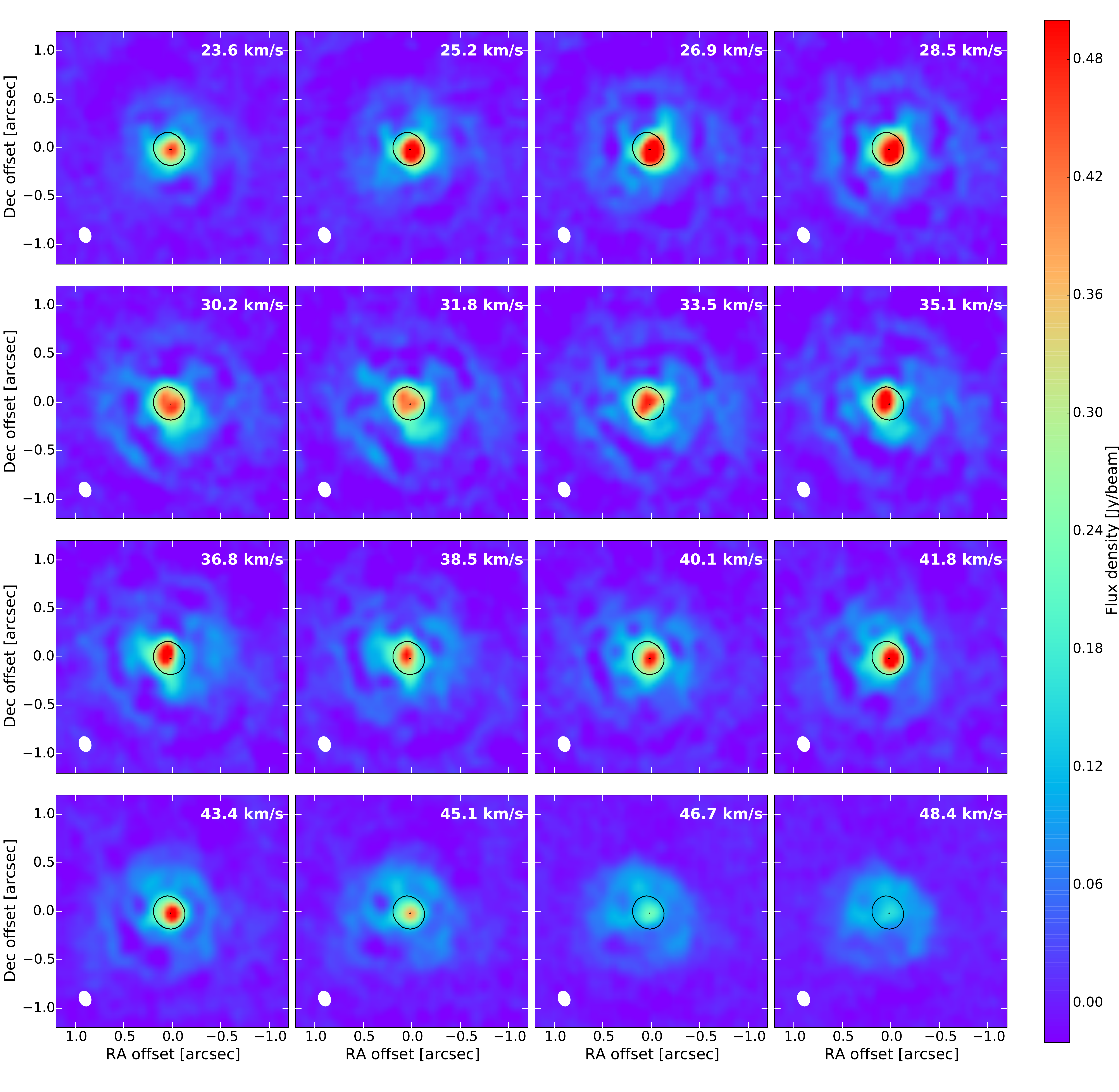}
\caption{Channel map of the HCN(4-3) emission in IK~Tau. The circle denotes the place of maximum dust emissivity (taking controus at 1\% and 99\% of the total flux). The contrast is best visible on screen.}
\label{Fig:IKTau_HCN_channel}
\end{figure*}

\begin{figure*}
\includegraphics[width=\textwidth]{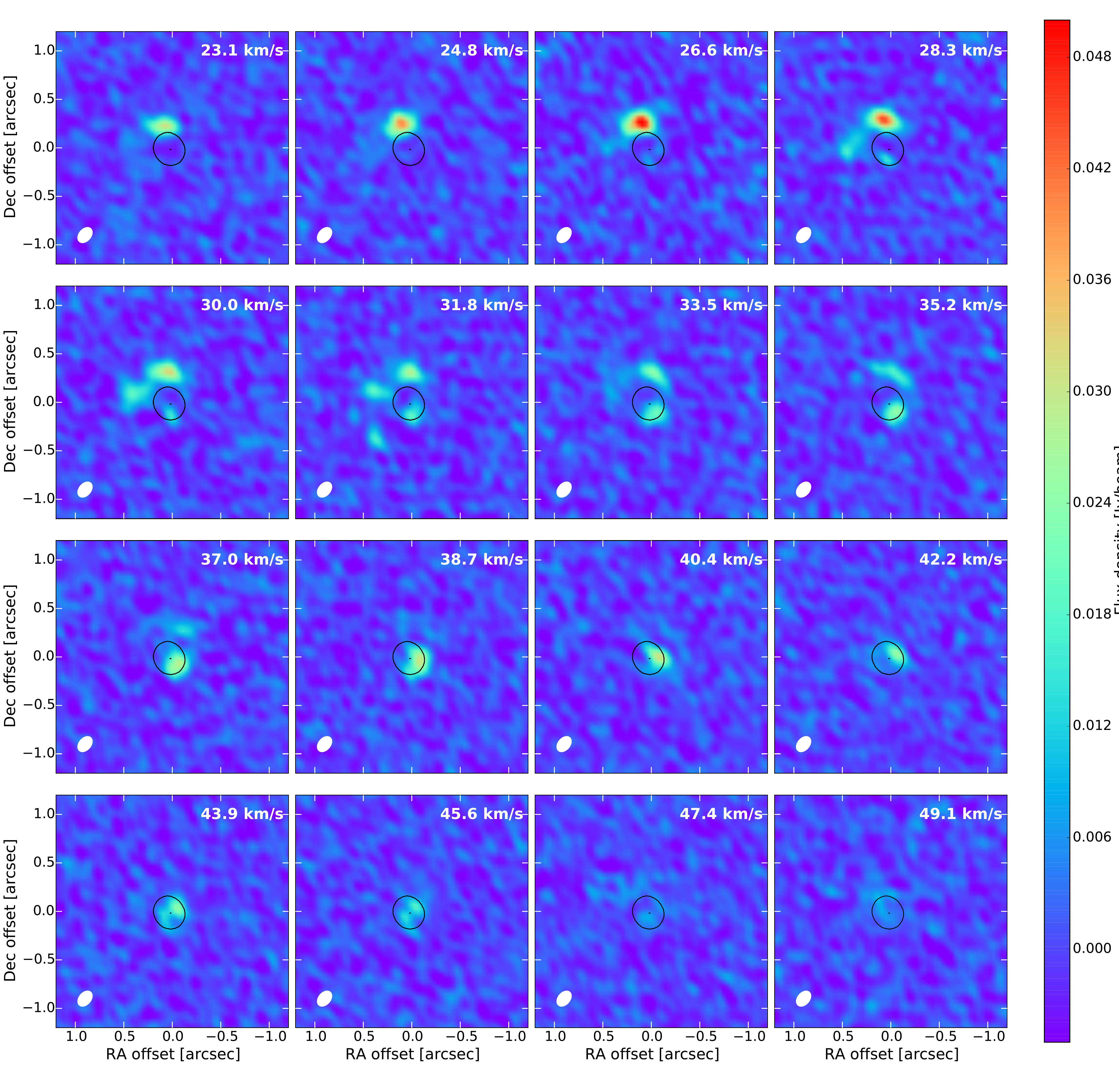}
\caption{Channel map of the NaCl(26-25) emission in IK~Tau. The circle denotes the place of maximum dust emissivity (taking contours at 1\% and 99\% of the total flux). The contrast is best visible on screen.}
\label{Fig:IKTau_NaCl_channel}
\end{figure*}

\end{appendix}

\end{document}